\documentclass[aps,prx,showpacs,twoside,twocolumn,10pt,floatfix]{revtex4-2}
\usepackage{times,epsfig,amssymb,amsfonts,amsmath,bm,subfigure,mathtools,amsthm,braket,soul,enumitem,color} 
\usepackage[normalem]{ulem}
\usepackage{graphicx}
\usepackage{float}
\usepackage{natbib}

\newcommand{\stkout}[1]{\ifmmode\text{\sout{\ensuremath{#1}}}\else\sout{#1}\fi}
\definecolor{magenta}{rgb}{1.0, 0.0, 0.56}

\begin{document}
\title{Experimental quantum pattern recognition in IBMQ and diamond NVs}
\author{Sreetama Das$^{1,2}$, Jingfu Zhang$^{3}$, Stefano Martina$^{1,2}$, Dieter Suter$^3$, Filippo Caruso$^{1,2,4}$}
\affiliation{$^1$Department of Physics and Astronomy, University of Florence, Via Sansone 1, Sesto Fiorentino, I-50019, Italy}
\affiliation{$^2$European Laboratory for Non-Linear Spectroscopy (LENS), University of Florence, Via Nello Carrara 1, Sesto Fiorentino, I-50019, Italy}
\affiliation{$^3$Fakultaet Physik, Technische Universitaet Dortmund, D-44221 Dortmund, Germany}
\affiliation{$^4$QSTAR and CNR-INO, Largo Enrico Fermi 2 - 50125 Firenze - Italy}

\begin{abstract}

One of the most promising applications of quantum computing is the processing of graphical data like images. 
Here, we investigate the possibility of realizing a quantum pattern recognition protocol based on swap test, and use the IBMQ noisy intermediate-scale quantum (NISQ) devices to verify the idea. We find that with a two-qubit protocol, swap test can efficiently detect the similarity between two patterns with good fidelity, though for three or more qubits the noise in the real devices becomes detrimental. To mitigate this noise effect, we resort to destructive swap test, which shows an improved performance for three-qubit states. Due to limited cloud access to larger IBMQ processors, we take a segment-wise approach to apply the destructive swap test on higher dimensional images. 
In this case, we define an average overlap measure which shows faithfulness to distinguish between two very different or very similar patterns when simulated on real IBMQ processors. As test images, we use binary images with simple patterns, greyscale MNIST numbers and MNIST fashion images, as well as binary images of human blood vessel obtained from magnetic resonance imaging (MRI). We also present an experimental set up for applying destructive swap test using the nitrogen vacancy centre (NVs) in diamond. Our experimental data show high fidelity for single qubit states. Lastly, we propose a protocol inspired from quantum associative memory, which works in an analogous way to supervised learning for performing quantum pattern recognition using destructive swap test.
\end{abstract}

\maketitle

\section{Introduction}
\label{sec:intro}
In the last few decades, the advancement in quantum information processing has significantly impacted the fields of computation and communication technologies. The superiority of a quantum protocol lies in the fact that it can accomplish tasks either impossible by a classical protocol \cite{teleportation, *densecoding}, or in some cases it performs exponentially or polynomially faster compared to its classical analogue, the most prominent examples being Shor's factorization algorithm \cite{shor} and Grover's search algorithm \cite{grover}. These traits of a quantum system, particularly the computational speed-up, has been crucial in developing powerful algorithms for quantum simulation \cite{simulation1, *simulation2, *simulation3, *simulation4, *simulation5, *simulation6, *simulation7, *simulation8}, Boson sampling \cite{bsampling1, *bsampling2, *bsampling3, *bsampling4, *bsampling5, *bsampling6}, solving systems of linear equations \cite{harrow_2009, *Cai_2013, *Pan_2014, *barz_2017, *zheng_2017}, and many other computational problems. Recently, the possibility to use quantum properties to facilitate artificial intelligence and machine learning has been extensively investigated \cite{Manzano_2009, lloyd2013quantum, rebentrost_2014, qmachine1, *qmachine2, *qmachine3, *qmachine4, buffoni_2020}.
We are living in the era of quantum supremacy \cite{qsupremacy_google, *china_supremacy1, *china_supremacy2, *china_supremacy3}. It has been possible to build quantum processors (processing units) with several tens of qubits, some of them being accessible to users from anywhere in the world through cloud-based sharing \cite{ibmq}. However, these machines are still extremely noisy and, for this reason, they are called Noisy Intermediate-Scale Quantum (NISQ) devices \cite{Preskill2018quantumcomputingin}.

This progress is staring to affect the field of image processing. Vision is the most fundamental mechanism of obtaining information about a system. Thus, image processing is a necessary task in a broad range of directions like medical science, space science, automobile technologies etc. With the advancement in quantum technologies, a growing interest has been directed towards implementing quantum systems for improved image processing.  Some quantum image processing protocols have been proposed and tested, which show polynomial and exponential speed up; an example is quantum edge detection \cite{qsobel_2015, suter_2017, Cavalieri2020AQE, Xu:20}. Another widely used image processing task is pattern recognition.
The application of the latter ranges from identifying a string of integers or texts present in an 1D array, to recognizing 2D patterns such as faces of people, handwritten digits, geometric shapes, and other tasks.
In 2002, Trugenberger introduced the idea of quantum associative memory, and put forward a quantum pattern recognition protocol based on that \cite{Trugenberger2002}. Using quantum Fourier transform, Schutzhold in 2003 devised another protocol for pattern recognition which demonstrated exponential speed-up over its classical analogue \cite{schutzhold_2003}. A number of follow-up works attempted to improve the above protocols or presented similar algorithms inspired from them \cite{petruccione_book2014, prousalis_2019, banchi2020}. Except these, quantum pattern recognition protocols based on the framework of classical Hopfield neural network \cite{neigovzen_2009}, the hidden shift problem \cite{montanaro2015}, pixel gradient calculation \cite{Zhang2015LocalFP}, Grover's algorithm \cite{jiang_2016, soni_2020, tezuka_2022} has been proposed.
Some of them adopted a quantum machine learning approach \cite{lloyd2013quantum, qmachinepattern1, *qmachinepattern2, *qmachinepattern3, *qmachinepattern4, schuld_qmachine}. 
However, when tested on real quantum systems, the protocols are inevitably subject to noise, which degrades their efficiency. Thus, one must learn how to control noise, and counteract its effects.

In this work, we consider a quantum pattern recognition algorithm based on swap test. The swap test is used to calculate the closeness between two quantum states. In quantum image processing, the classical images are encoded in quantum states, hence the swap test emerges as a very plausible approach to find similarity between two quantum images. This has been previously mentioned and briefly discussed in \cite{suter_2017} and \cite{Cavalieri2020AQE}. The question remains as what will be the efficiency of this protocol in real quantum systems. This is motivation of our work, where the real systems are the IBMQ devices \cite{ibmq}. We aim to identify a pattern in the target image by comparing it with the patterns present in a large set of reference images, and finding an exact match. We find that, for systems as small as three qubits, and corresponding states which can be prepared using a few Hadamard and controlled NOT gates, the performance of swap test is low due to the strong presence of noise. To curb the effect of noise, we try an alternate but equivalent circuit, often called as `destructive swap test' in the past literature  \cite{garcia_2013}, in which the number of necessary gates is reduced. By using this novel approach, we observe an improved performance of the pattern recognition protocol for three-qubit states in real IBMQ devices (see Section \ref{sec:swap_test_result} for more details). 
We then go on to try the destructive swap test in these systems for higher dimensional binary and greyscale images, including small-dimensional biomedical images obtained in the lab with MRI. We also present an experimental setup and the resulting data where destructive swap test is performed using diamond nitrogen vacancy (NVs) centres. Our results suggest that destructive swap test can be a potential alternative to achieve the goal of swap test, at a reduced level of noise. Lastly, we propose a quantum protocol inspired from the quantum associative memory, by which one can recognize a pattern from a large set of reference patterns.

The paper is organized as follows. In Section \ref{sec:encoding}, we describe the encoding of a classical image in a quantum state. In Section \ref{sec:swap_dswap}, we discuss swap test and its performance in real quantum computers. This leads the way to destructive swap test and its improvements over swap test. In Section \ref{sec:swap_test_result}, we demonstrate the performance of destructive swap test in real quantum systems, when identifying similar geometric patterns encoded in binary images, as well as binary images obtained from MRI of human blood vessel. We also present a short analysis of how the success probability of the protocol changes with varying noise in the real quantum processors. In Section \ref{sec:greyscale}, we present similar results for greyscale images. In Section \ref{experimental} we show the experimental result of destructive swap test in diamond NVs. In Section \ref{sec:qmachine}, we propose a quantum algorithm analogous to supervised learning, for using destructive swap test in pattern recognition. Lastly, Section \ref{sec:conclusion} contains the concluding remarks.


\begin{figure}
    \centering
    \includegraphics[width=0.39\textwidth]{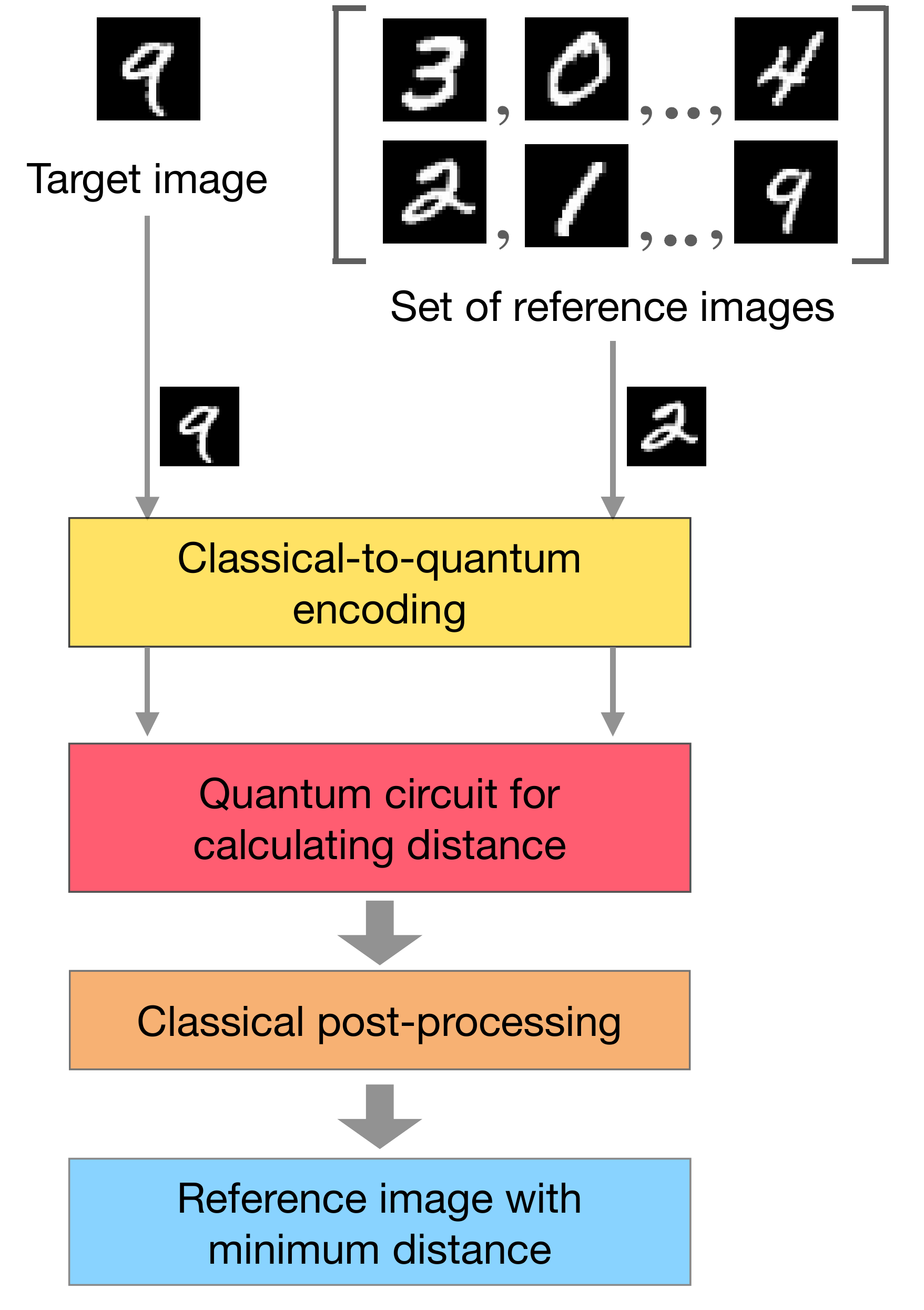}\vspace{0.5em}
    \caption{A schematic diagram of our pattern recognition protocol. In the target image, one aims to detect the handwritten number ``9". There is a set of reference images with handwritten digits from 0 to 9. The target image and an arbitrary chosen reference image, in this case ``2" for demonstration purpose, is encoded as quantum states, passed as the inputs of a quantum protocol to calculate the distance between them. The measurement result undergoes a classical post-processing. The protocol is repeated for all the reference images. The reference image with minimum distance to the target image is assigned as the pattern to be detected.}
    \label{schematic}
\end{figure}

\begin{figure*}
    \centering
    \subfigure[]{\includegraphics[width=0.39\textwidth]{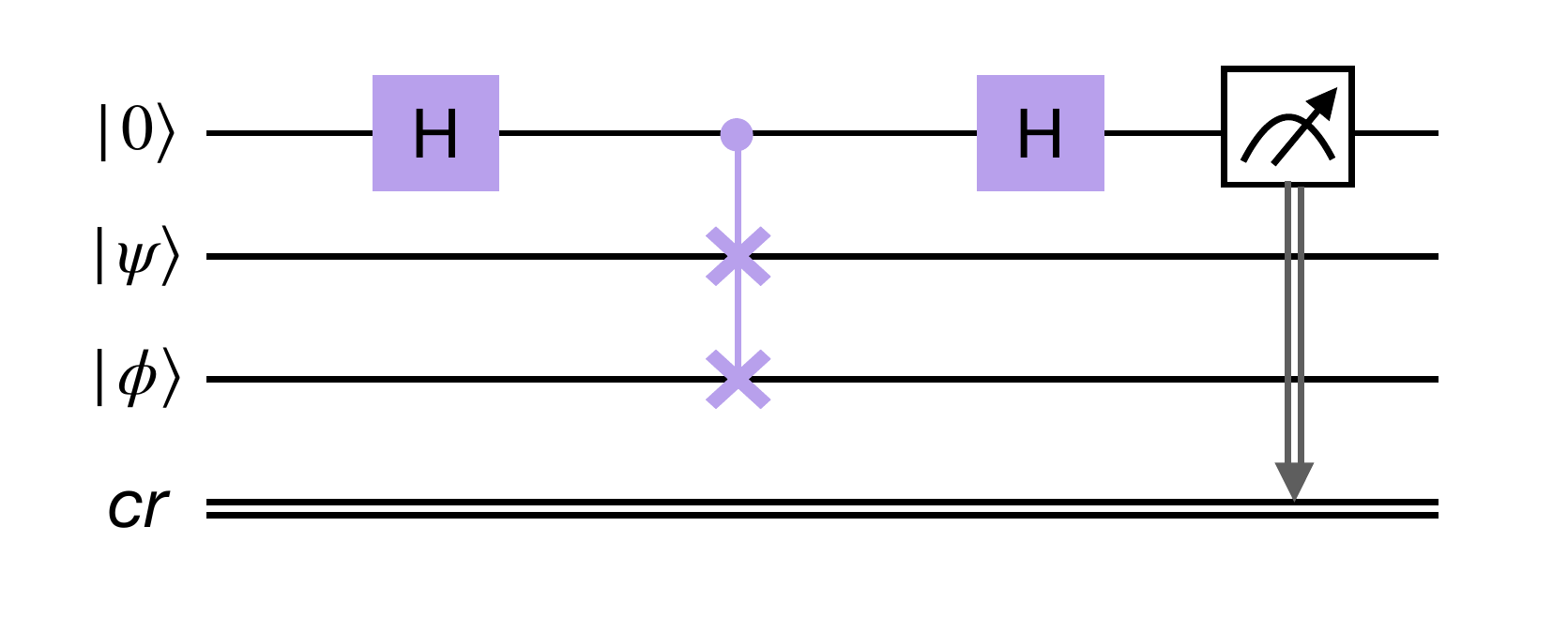}}
    \subfigure[]{\includegraphics[width=0.43\textwidth]{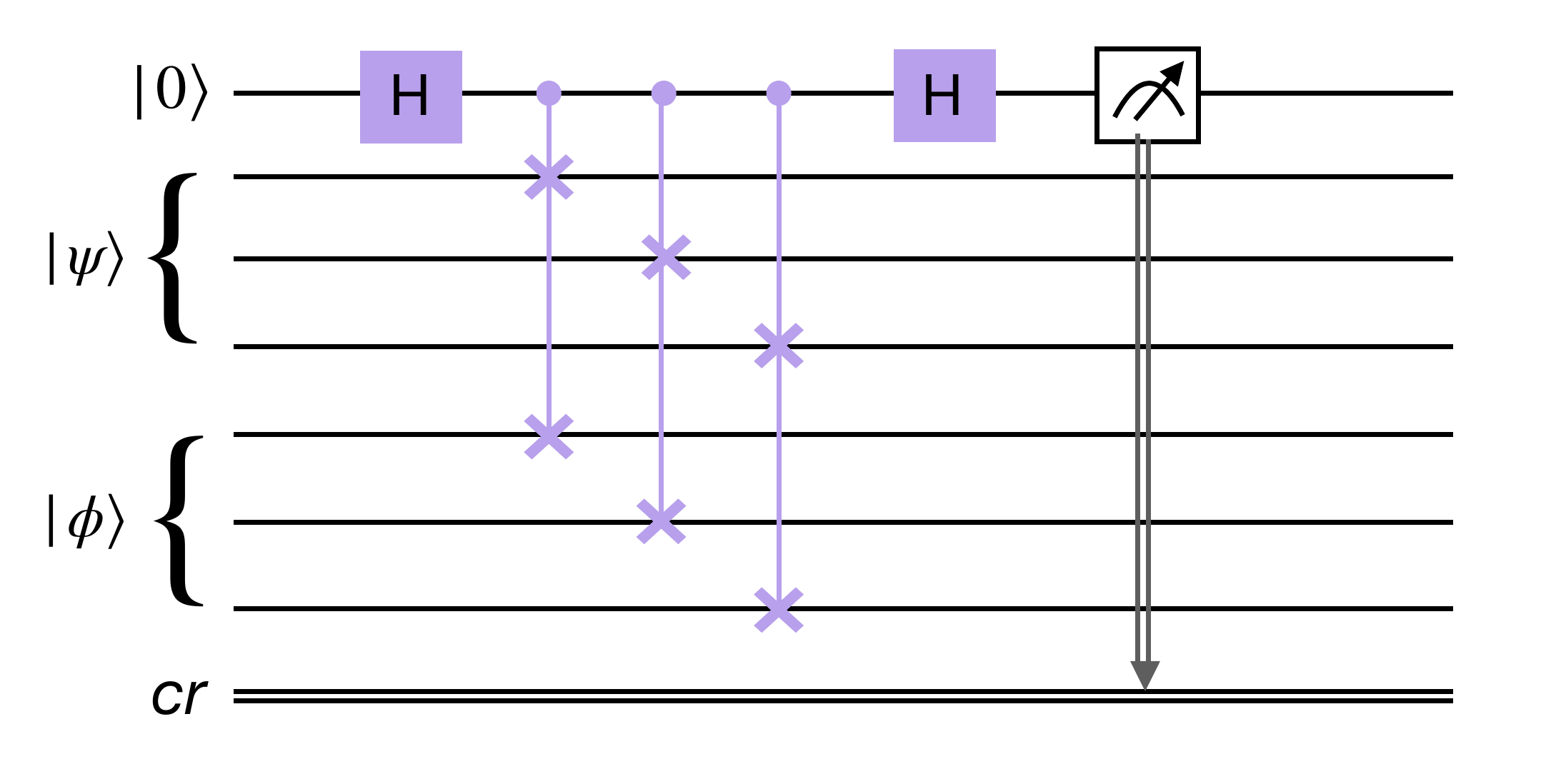}}
    \subfigure[]{\includegraphics[width=1\textwidth]{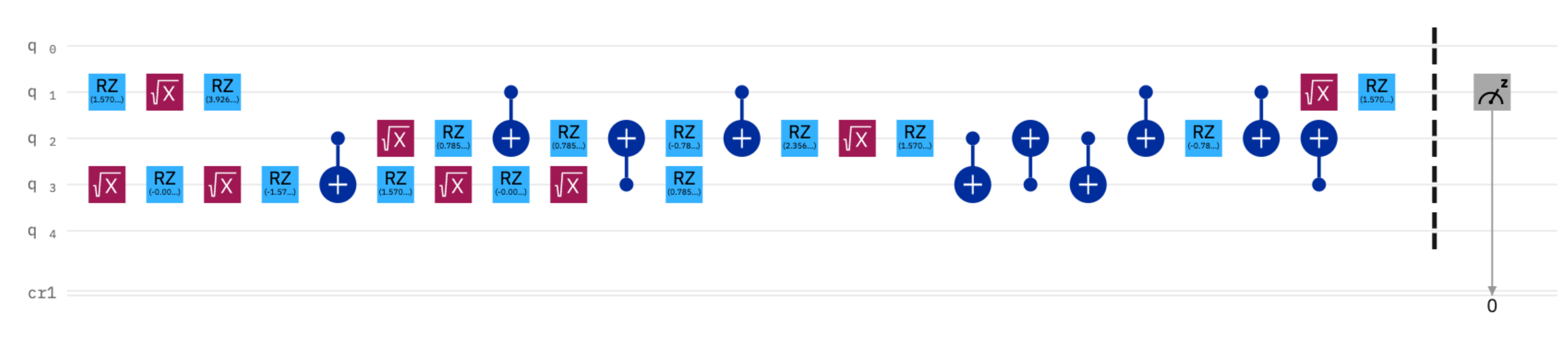}}
    \caption{The circuit diagram of Swap test for (a) single qubit states and (b) three-qubit states. (c) The swap test circuit after transpilation in $\mathrm{ibmq}\_\mathrm{manila}$, for two single qubit states, both being $|0\rangle$. The figure is exported from Qiskit.}
    \label{fig:swapcircuit}
\end{figure*}

\section{Encoding an image in a quantum state}
\label{sec:encoding}
The first task in quantum image processing is to encode the pixel positions and corresponding pixel values of a classical image in a quantum state. A number of encoding processes has been proposed to date, e.g. Flexible Representation of Quantum Images (FRQI) \cite{frqi} and Novel Enhanced Quantum Representation (NEQR) \cite{neqr}. In this work, we will use a different encoding method first used by the authors in \cite{suter_2017}, which is often referred to as Quantum Probability Image Encoding (QPIE).

We work with a 2D image of $N=P\times Q$ pixels, which is classically encoded in a $P\times Q$ dimensional matrix $M$, the matrix element $M(i,j)$ denoting the pixel value at position $(i,j)$. To encode this image in a quantum state, we can use any $m$-qubit system, such that $2^{m}\geq N$. The Hilbert space $\mathcal{H}_{2}$ of each qubit is spanned by the computational basis $\{|0\rangle,|1\rangle\}$. From the $m$-qubit register, we choose $n$ qubits such that $2^{n}=N$. We initialize them in the state $|f\rangle=\sum_{i}c_{i}|i\rangle$ where the basis vectors $|i\rangle$ span the $2^{n}$ dimensional Hilbert space $\mathcal{H}_{2}^{\otimes n}$, encoding the positions of the pixels. The coefficients $c_{i}$ encode the corresponding pixel values. The first $P$ elements of $|f\rangle$ encode the first column of $M$, the next $P$ elements encode the second column, and so on. Thus, each pixel position corresponds to a particular quantum state $|i\rangle$ of $\mathcal{H}_{2}^{\otimes n}$, and its pixel value corresponds to the coefficient $c_{i}$ in $|f\rangle$.  Lastly, the state $|f\rangle$ must be normalized.

Most generally, the images can have a color format e.g. RGB or sRGB, where the pixel values can vary over a range. The images can also be greyscale, where the varying pixel values denote different shades of grey. However, depending on which type of information we want to acquire from the images, converting them to binary images can be operationally advantageous for further image processing tasks, e.g. edge detection. In a binary image, all the pixel values are $0$ or $1$. The conversion can be achieved by choosing a suitable threshold pixel value $p$ in between the minimum and maximum pixel values of the color image. The pixel values equal or higher than the threshold are converted to 1, those below the threshold are converted to 0. Depending on the choice of $p$, the resulting images differ.



Now that we are able to encode the images, we proceed to apply a quantum pattern recognition protocol in the next section. In classical image processing, a brute-force method is to measure the pixel values of all the N pixels. However, once encoded as an $n$-qubit quantum state, the number of measurements necessary to obtain certain information about an image is drastically reduced compared to the classical case. To define our pattern recognition protocol, we assume that we have a target image with an unknown pattern to be detected, and a large set of reference images with different patterns which are known to us. The target pattern can be detected if it is compared with each reference image and an exact match is found. In case an exact match does not exist in the reference set, one can still look for the closest match. Translated into the language of quantum physics, one needs to calculate the closeness or distance between the quantum states representing the images. A schematic diagram of the protocol is presented in Fig. \ref{schematic}. The distance is determined by swap test, as we discuss in the next section.

    

\section{Swap test}
\label{sec:swap_dswap}

Swap test is an widely used protocol for calculating the overlap between two quantum states \cite{buhrman2001,gottesman_2001, kang2019}. Specifically, if $|\psi\rangle$ and $|\phi\rangle$ are two pure states, swap test can be used to calculate $|\langle \psi|\phi\rangle|^{2}$, which is also a valid distance measure between the two states, known as quantum Fidelity \cite{Jozsa_1994, schumacher_1995}. The circuit of swap test for single qubit states is shown in Fig. \ref{fig:swapcircuit}(a). It can be generalized to $n$-qubit states by repeating the controlled-swap gate for all qubit pairs, as shown in Fig. \ref{fig:swapcircuit}(b). Each of the states $|\psi\rangle$ and $|\phi\rangle$ is encoded using three qubit quantum registers $Q_{1}$ and $Q_{2}$ respectively. There is an auxiliary qubit $Q_{a}$ which is initialized in the computational basis state $|0\rangle$. After initializing all the qubits, a Hadamard gate $H$ is applied on $Q_{a}$, which transforms $|0\rangle$ to $|+\rangle=\frac{1}{\sqrt{2}}(|0\rangle + |1\rangle)$. This qubit is then used as the control qubit to apply the controlled swap operation between all consecutive qubit pairs of $Q_{1}$ and $Q_{2}$. In the controlled swap operation, if the control qubit is in state $|1\rangle$, the states of the two target qubits are interchanged. If the control qubit is in state $|0\rangle$, nothing changes. This is followed by another Hadamard operation on $Q_{a}$. Lastly there is a classical register ``cr" which registers the measurement outcome on the auxiliary qubit. The outcome can be $``0"$ or $``1"$, associated with the measurement basis states $|0\rangle$ and $|1\rangle$ respectively. The measurement is taken on a large number of identically simulated circuits. If $|\psi\rangle=|\phi\rangle$, the measurements will return ``$0$" with probability $P(0)=1$. 
If they are orthogonal, $P(0)=P(1)=\frac{1}{2}$. For all other cases, $P(0)$ varies between $\frac{1}{2}$ and $1$. The fidelity between two states is then
\begin{equation}
\label{fidelity}
\mathcal{F}=|\langle \phi|\psi\rangle|^{2}=2P(0)-1,
\end{equation}
and the overlap $\mathcal{I}$ is
\begin{equation}
\label{overlap}
\mathcal{I}=|\sqrt{\mathcal{F}}|.
\end{equation}
A higher value of $\mathcal{I}$ implies higher degree of similarity between two states. Both $\mathcal{F}$ and $\mathcal{I}$ can be equivalently used to measure the degree of similarity between two quantum states. In this work, for most parts (except the diamond NVs), we stick to $\mathcal{I}$ since it is more sensitive to small changes in $P(0)$.

We assume that for our pattern recognition protocol, both the target and reference images have same dimension. The target image  is encoded in a quantum state, and passed as one of the inputs in the swap test circuit. On the other hand, each of the reference images is encoded in a quantum state, and then passed as the other input. If two patterns match exactly, one gets $``0"$ with probability 1 in the measurement of the auxiliary qubit. Otherwise, one can detect the closest pattern of the target image as the one with which it has highest overlap. 

\begin{figure}[t]
\centering

    \includegraphics[width=0.43\textwidth]{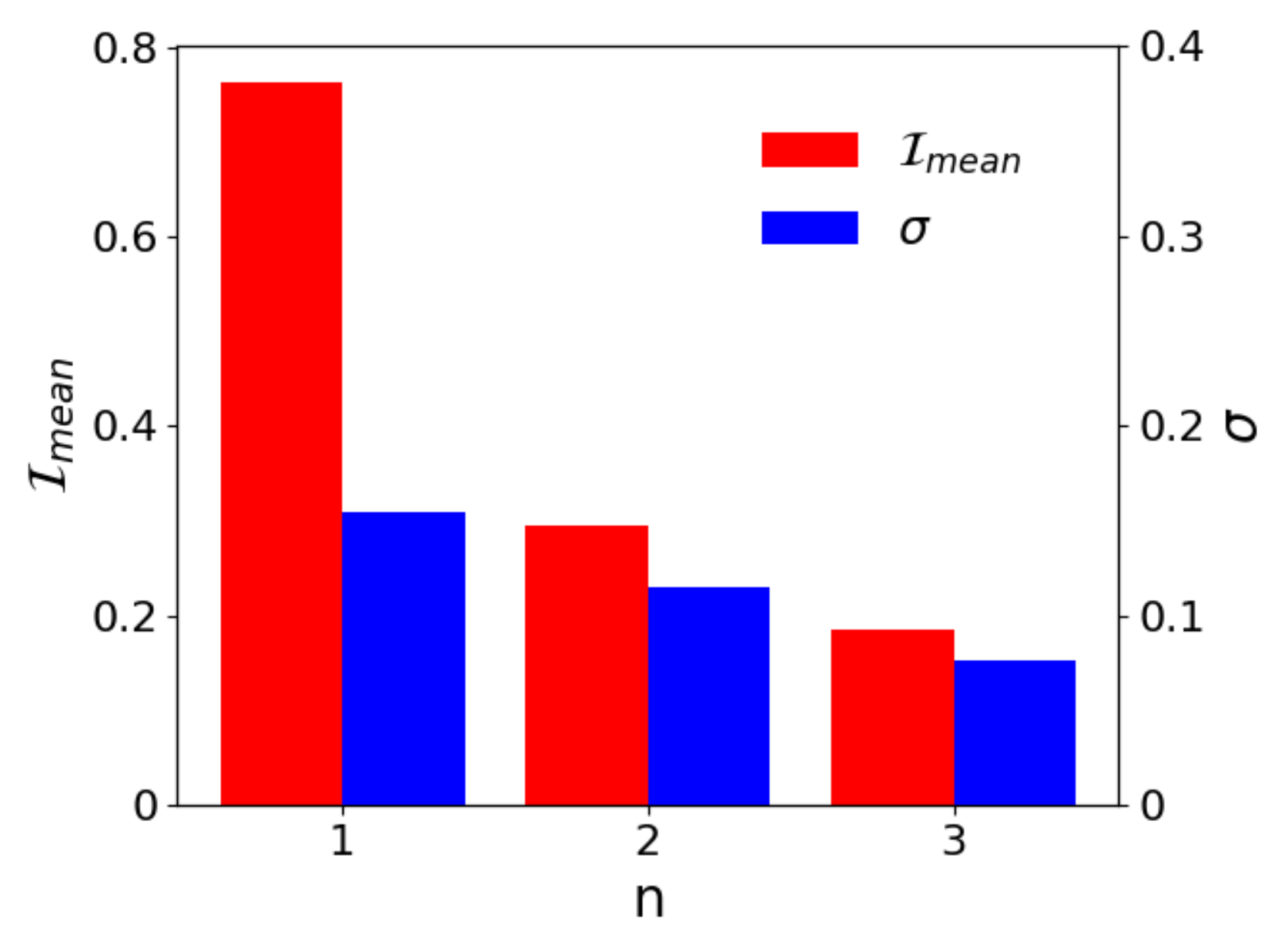}
    \caption{The mean overlap $\mathcal{I}_{mean}$ between two identical states encoded using $n$ qubits and the corresponding standard deviation $\sigma$ for swap test. The data are obtained by performing 100 runs of the circuit in real IBMQ processors.}
    \label{fig:swap_test}
\end{figure}

\begin{figure*}[ht!]
\begin{minipage}{0.425\textwidth}%
    \subfigure[]
    {
        \includegraphics[width=\textwidth]{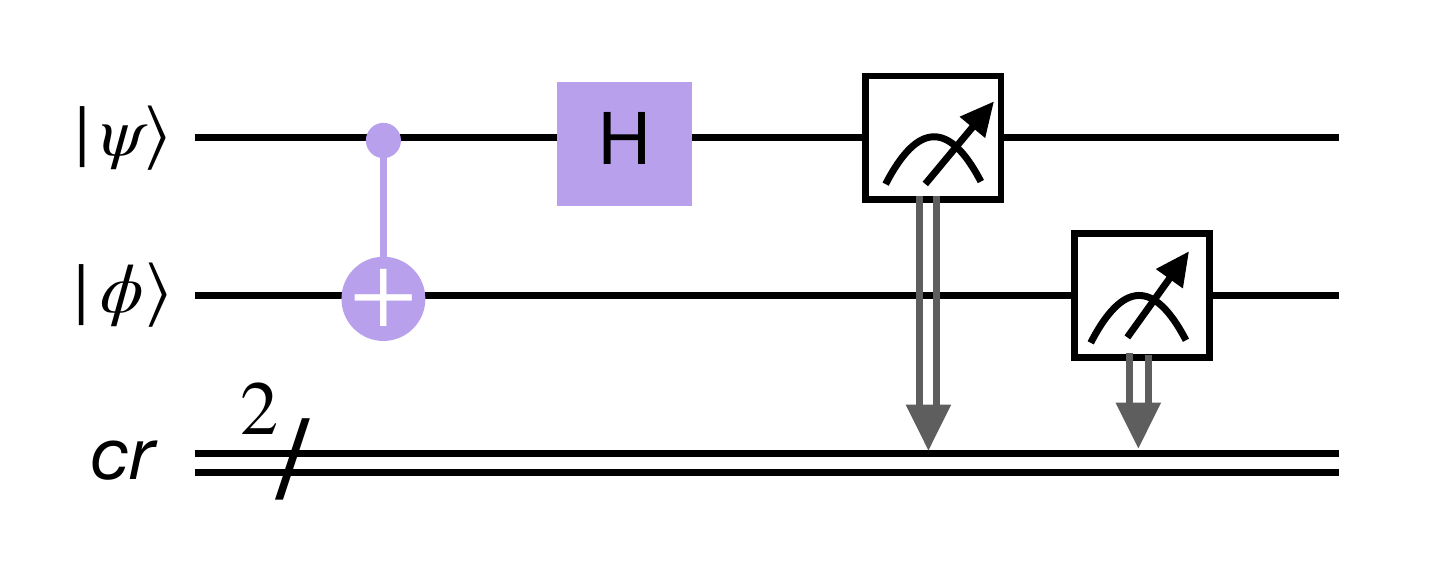}
    }
    
    \subfigure[]
    {
        \includegraphics[width=\textwidth]{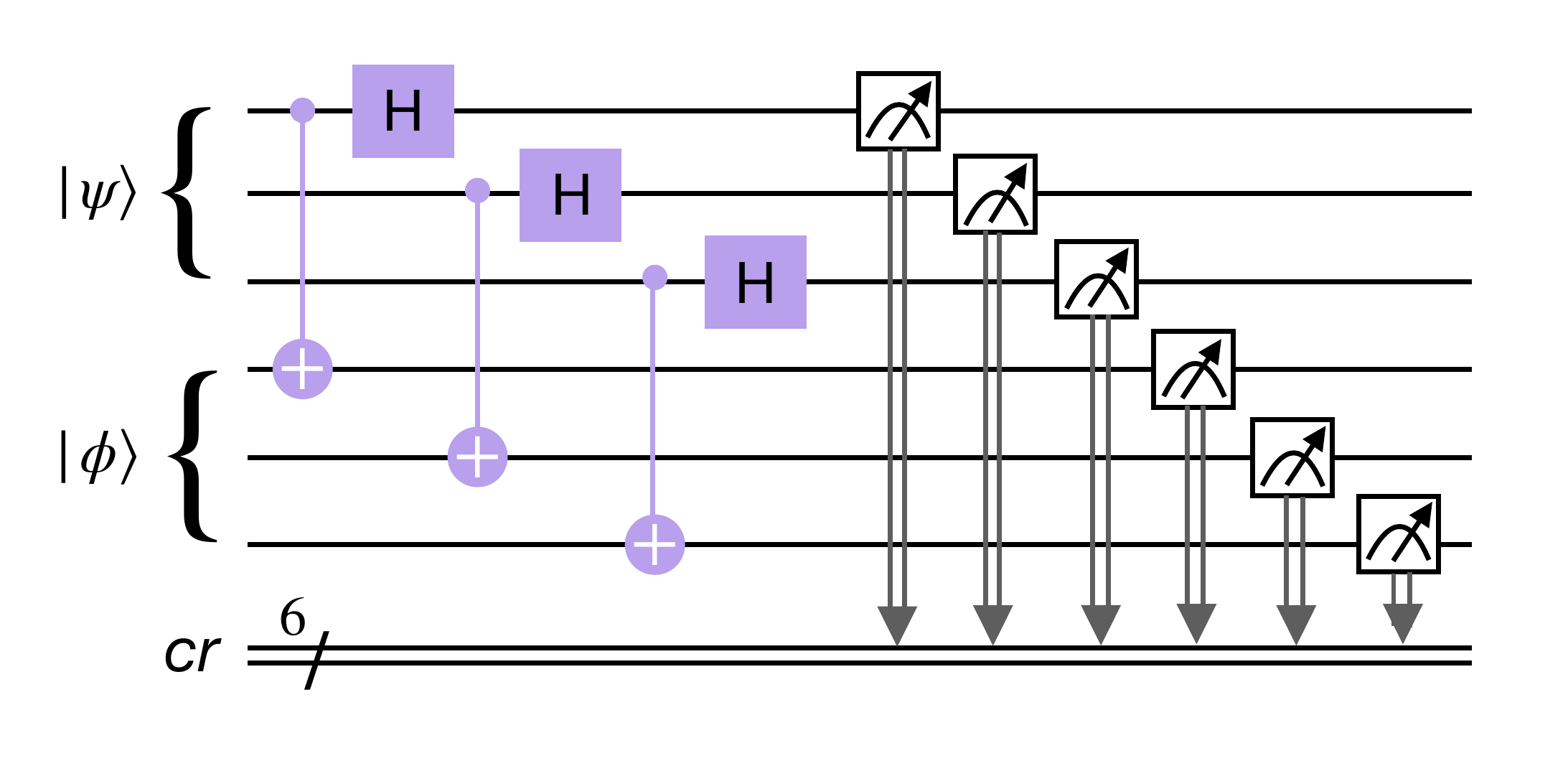}
    }
    \end{minipage}%
    \hspace{0.5mm}
    \begin{minipage}{0.425\textwidth}%
        \subfigure[]
        {
            \includegraphics[width=0.9\linewidth]{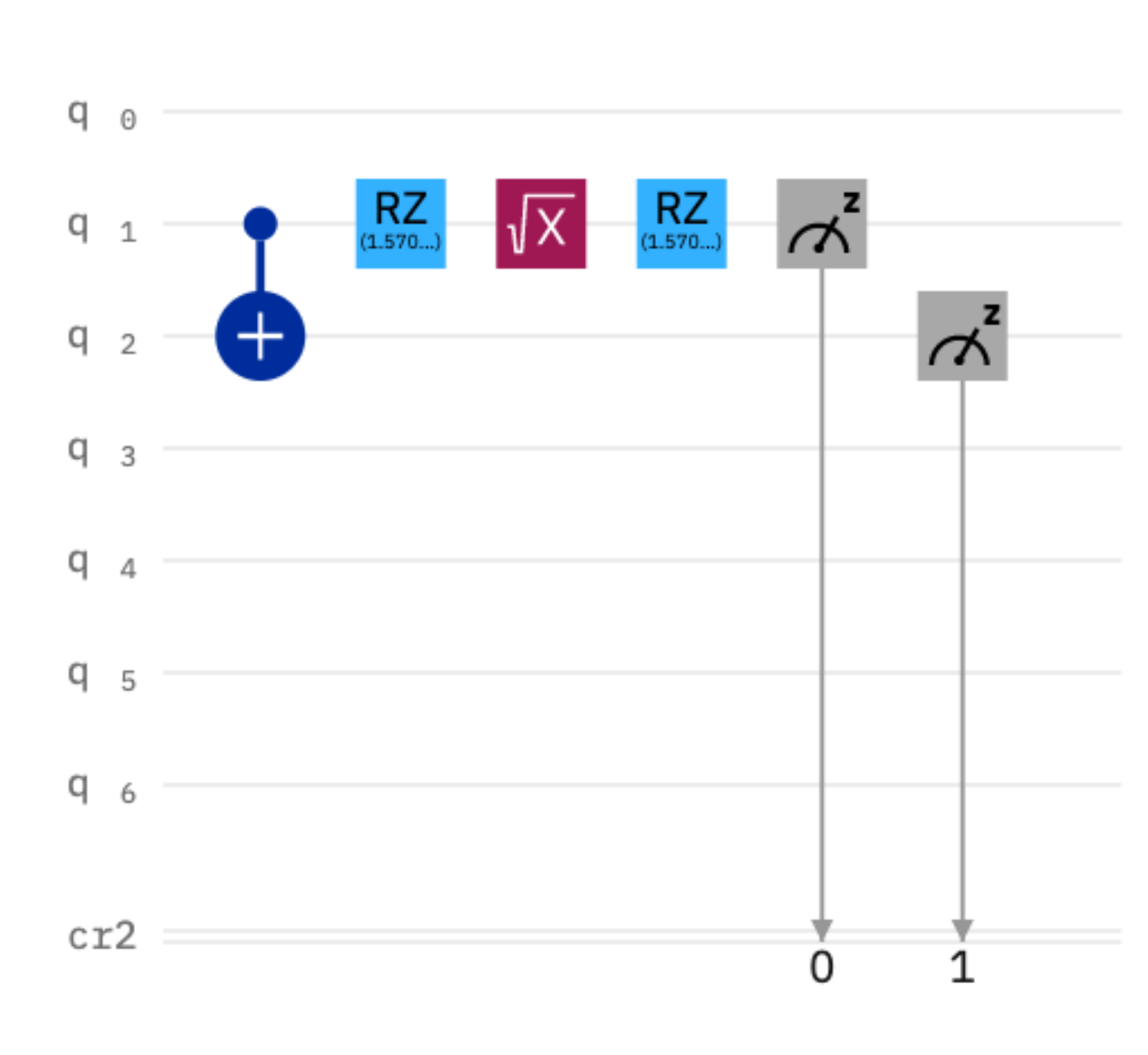}
        }
    \end{minipage}

\caption{The circuit diagram of destructive swap test for (a) single qubit states and (b) three-qubit states. (c) The destructive swap test circuit after transpilation in $\mathrm{ibmq}\_\mathrm{manila}$, for two single qubit states, both being $|0\rangle$. The figure is exported from Qiskit.}
\label{fig:destructive_swap}
\end{figure*}


 In real quantum processors, the performance is unavoidably affected by noise. The latter can appear due to decoherence, interaction among qubits, gate fidelity, state preparation and measurements. The adverse effect of noise on any protocol increases with increasing number of qubits and gates. In IBMQ, each of the real quantum processors has a set of ``basis gates" consisting of some particular single qubit and two-qubit gates. When a circuit is simulated using these, it typically needs to be``transpiled", i.e. the action of all the complex gates are obtained by using the basis gates. Thus in the actual simulation, the number of gates are higher compared to what we add to the circuit. This has been shown in Fig. \ref{fig:swapcircuit}(c) for single qubit states. For higher dimensional states, performance of swap test invariably degrades. We demonstrate this in Fig. \ref{fig:swap_test} where we perform swap test for two states encoded using one qubit, two qubits and three qubits respectively. For all three cases, we take the two states to be same and calculate the overlap between them. The states corresponding to these three cases are $|\psi_{1}\rangle=\frac{1}{\sqrt{2}}\big(|0\rangle + |1\rangle\big)$, $|\psi_{2}\rangle=\frac{1}{\sqrt{2}}\big(|00\rangle + |11\rangle\big)$, and $|\psi_{3}\rangle=\frac{1}{\sqrt{2}}\big(|000\rangle + |111\rangle\big)$. These states are typical exemplary states, and we prepare them by directly adding Hadamard and controlled NOT (CNOT) gates to the circuit. The CNOT gate has a control qubit and a target qubit, the state of the latter undergoes a bit-flip operation (gate corresponding to pauli-x matrix $\sigma_{x}$) if the control qubit is in state $|1\rangle$, otherwise nothing changes. The overlap is calculated for 100 realizations of the identical circuit, and then the mean overlap $\mathcal{I}_{mean}$ and the standard deviation $\sigma$ is calculated. For this simulation, and for all the subsequent sections, we have used the 5 and 7-qubit IBMQ systems which are- $\mathrm{ibmq}\_\mathrm{manila}$, $\mathrm{ibmq}\_\mathrm{jakarta}$, $\mathrm{ibm}\_\mathrm{lagos}$, $\mathrm{ibm}\_\mathrm{perth}$, $\mathrm{ibmq}\_\mathrm{casablanca}$ and $\mathrm{ibmq}\_\mathrm{quito}$.
 We see that for single qubit case, in spite of the presence of noise, the performance of swap test is quite good. For two qubits, the noise becomes significant, and for the three-qubit case the value of the overlap worsen. Due to the cloud inaccessibility to higher dimensional systems, we could not check the swap test for four qubit states. However the results are sufficient to suggest that the swap test quickly loses its relevance for states having dimension higher than 3, and this poses a primary challenge to utilize it for pattern recognition in even fairly small images. Even in the NV setup, the experimental results deviate because of imperfect realization of the pulse sequences, as will be discussed in Section \ref{experimental}. However, the quantum simulators do not suffer this problem. One can check that for two same images, the mean inner product and standard deviation will always return values $1$ and $0$ respectively. This is because the IBMQ simulators are classical computers that simulates the results of an ideal quantum evolution. Hence they can still be used to check the performance of swap test irrespective of the dimension, while being within their size limit, which is 32 qubits for the $\mathrm{qasm}\_\mathrm{simulator}$.

\subsection{Destructive swap test}
In \cite{garcia_2013}, the authors showed that an equivalent circuit for swap test can be built without using the auxiliary qubit, at the cost of measuring all the qubits used to encode the states. The controlled swap gate used in the swap test, is a three-qubit gate which can be realized by two CNOTs and one CCNOT (controlled controlled NOT) gate. In the new circuit, all the controlled swap gates are replaced by same number of two-qubit CNOT gates. Also, only one Hadamarrd gate is required. Thus, compared to the original swap test, in this case the number of gates in the transpiled circuit decreases significantly, hence reducing the noise in the output. This protocol is referred to as ``destructive swap test", because any superposition in the output is destroyed on measuring all the qubits. Later in \cite{cincio_2018}, the authors used machine learning approach to find the optimized circuit for calculating overlap between two states, and they rediscovered the destructive swap test as the optimum algorithm. Their results, simulated using 5-qubit IBMQ and 19 qubit Rigetti's quantum computer, shows a huge improvement of destructive swap test over the auxiliary qubit-assisted swap test, in terms of reducing the noise in the output. The circuit for destructive swap test for single qubit states is shown in Fig. \ref{fig:destructive_swap}(a). For $n$-qubit states, the gate sequence of (CNOT+Hadamard) is applied to the $n$ pairs of qubits belonging to the two states, as shown in Fig. \ref{fig:destructive_swap}(b). The circuit of Fig. \ref{fig:destructive_swap}(a) after transpilation is shown in Fig. \ref{fig:destructive_swap}(c), which clearly uses a lot less number of gates compared to Fig. \ref{fig:swapcircuit}(c). For each shot of circuit simulation, the classical bit string corresponding to measurement outcomes from the qubits of $|\psi\rangle$ is denoted by $O^{\psi}$, and that for $|\phi\rangle$ is denoted by $O^{\phi}$. Each of $O^{\psi}$ and $O^{\phi}$ can have $2^{n}$ different possibilities. The joint string $O^{\psi}O^{\phi}$ constitutes the measurement outcomes of the total system, which has $2^{2n}$ possibilities. Each element $O^{\psi/\phi}_{i}$ $(i=1,2,...,n)$ is a bit, taking values either $``0"$ or $``1"$. Suppose $O^{\psi}_{i}$ and $O^{\phi}_{i}$ are the $i^{\mathrm{th}}$  bits of the two strings respectively. As shown in \cite{garcia_2013}, an equivalent situation of obtaining $``1"$ in the auxiliary qubit measurement in the standard swap test occurs when $O^{\psi}$ and $O^{\phi}$ are such that $\sum_{i}\mathrm{AND}(O^{\psi}_{i},O^{\phi}_{i})$ has odd parity. For all such sequences $O^{\phi}$ and $O^{\psi}$, we take summation over the probabilities $P(O^{\psi}O^{\phi})$ of the joint outcomes $O^{\psi}O^{\phi}$. A nonzero value of this sum implies that $|\psi\rangle$ and $|\phi\rangle$ are not same. For this reason, we call this sum as the ``probability of failure". By subtracting it from unity, we get the ``success probability" which we denote by $P(0)$. A higher value of $P(0)$ implies higher overlap between two states. When $|\psi\rangle$ and $|\phi\rangle$ are same, the success probability is 1. The fidelity and the overlap can again be calculated using Eq. (\ref{fidelity}) and Eq. (\ref{overlap}).\\
Example: for two-qubit states, the probability of failure is obtained by summing over the probabilities of outcomes 0101, 0111, 1010, 1011, 1101 and 1110. For each of them, the first two bits correspond to the measurement outcome of qubits from one state, and the last two bits from the other state.
In ideal systems the swap test and destructive swap test will give same success/failure probability to distinguish between two images. It's not the case for noisy systems.




\begin{figure*}
    \centering
    \subfigure[]{\includegraphics[width=0.85\textwidth]{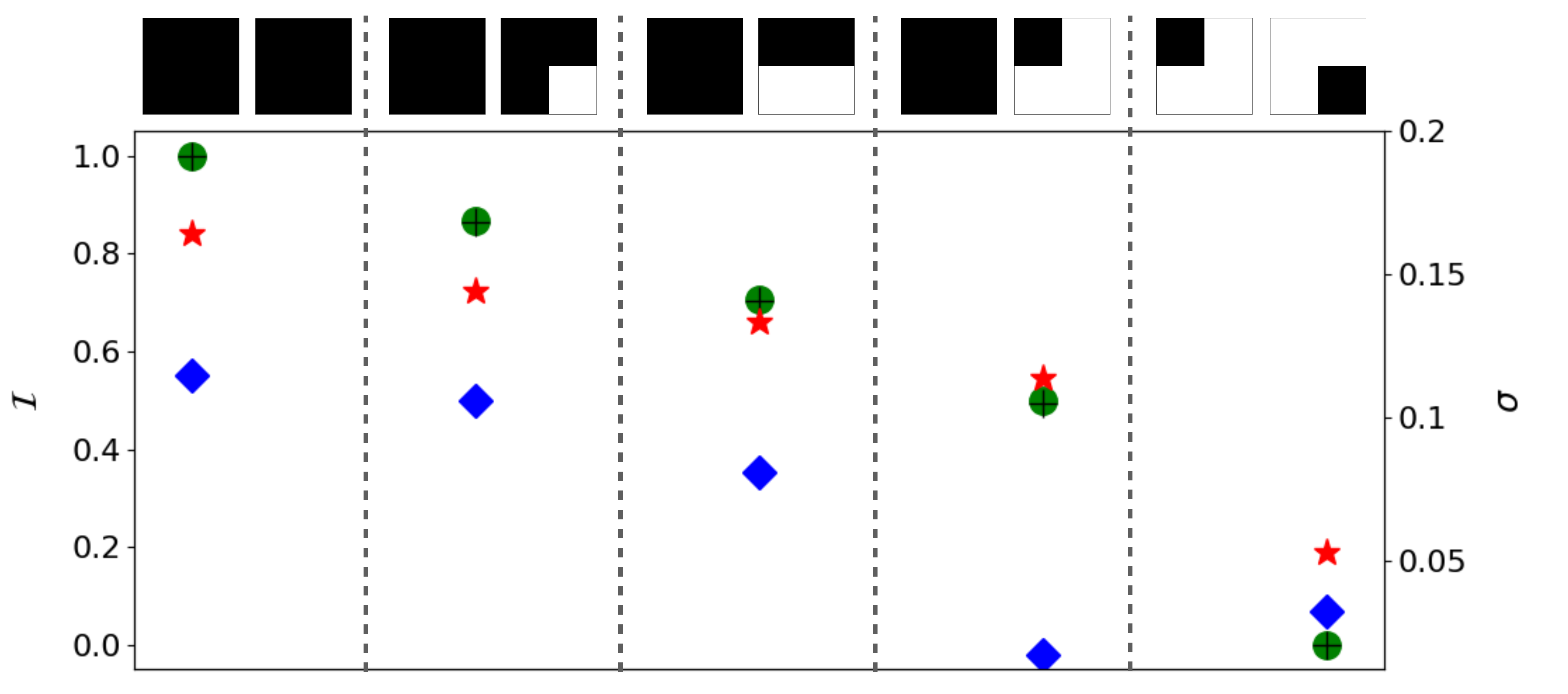}}
    \subfigure[]{\includegraphics[width=0.85\textwidth]{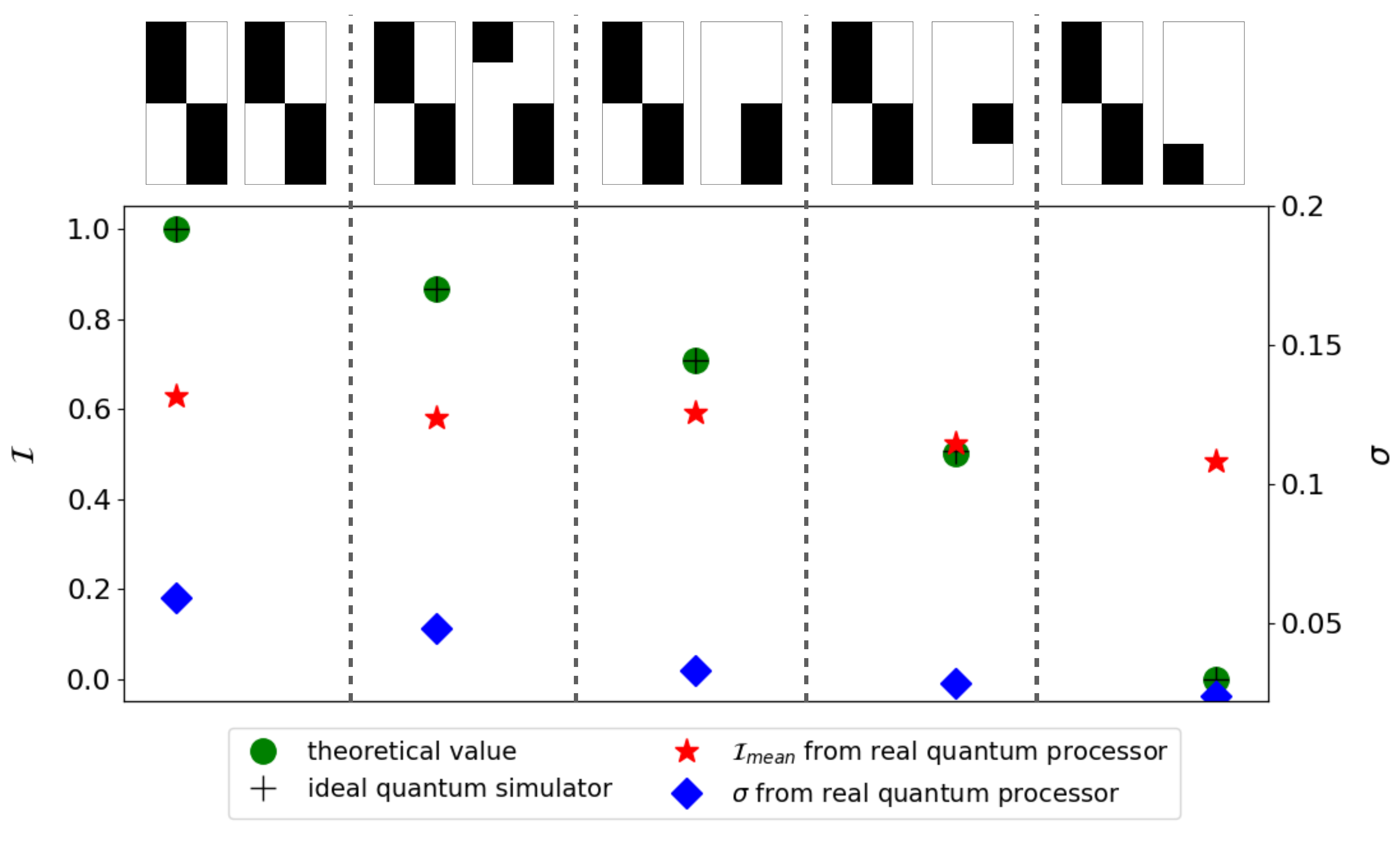}}


    \caption{Destructive swap test for pattern recognition in (a) two-qubit and (b) three-qubit images. In each vertical panel, the image on the left is the target image, and the one in the right is the reference image. The $y$-axis in the left denotes the overlap and the one in the right denotes the standard deviation corresponding to real quantum processors. The legendbox in (b) applies to both the figures. The mean and standard deviation are calculated over 100 runs of the circuit.
    }
    \label{fig:2_3qubit_dswap}
\end{figure*}


\begin{figure}
    \hspace{2.8em}
    \includegraphics[width=0.47\textwidth]{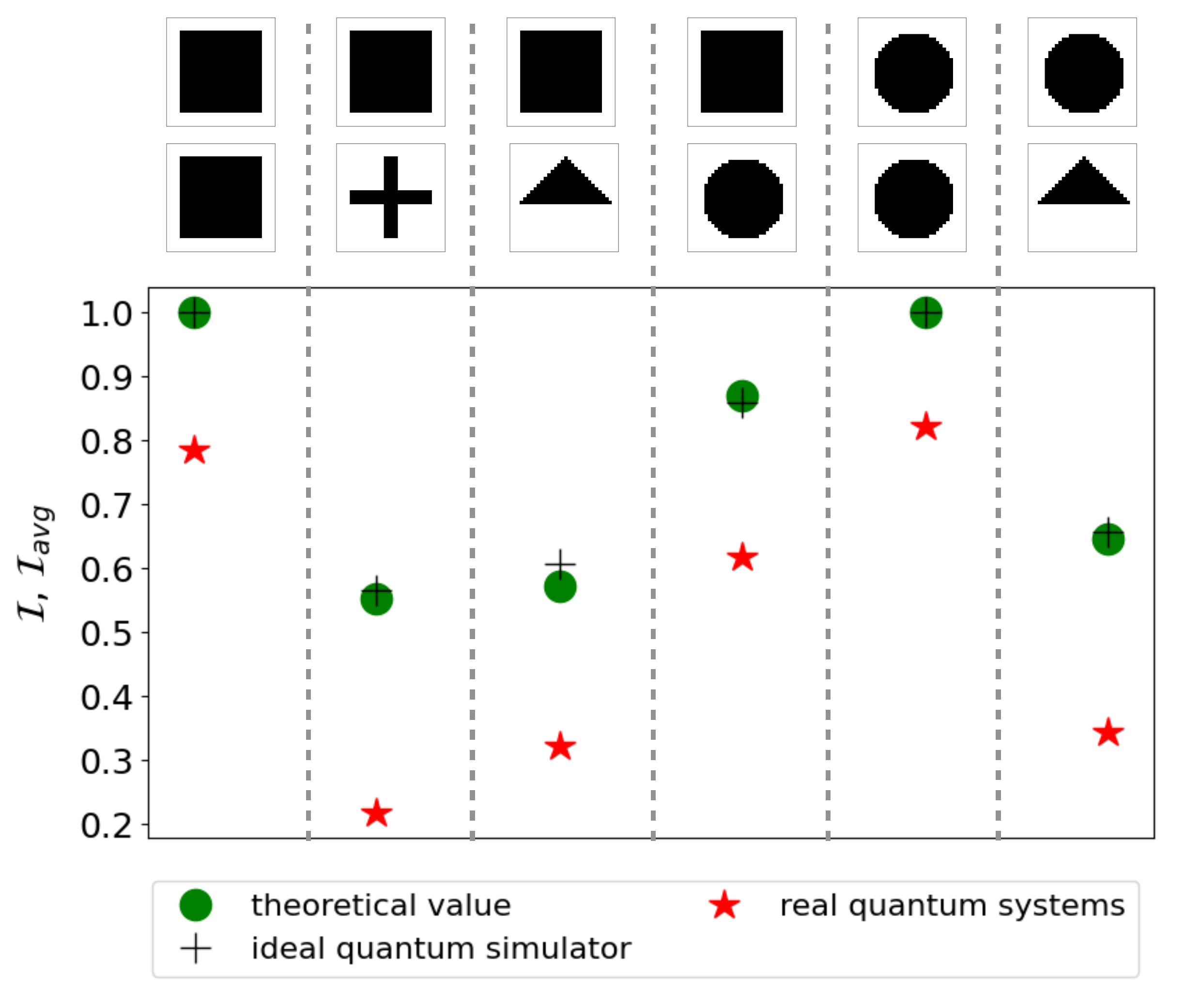}
    \caption{The destructive swap test applied to compare the overlap between different geometric patterns in binary images. Each image is $32 \times 32$ pixels. The two images placed next to each other in a vertical column are compared. In each column, the red square denotes the ideal overlap, the blue cross denotes the overlap obtained from a quantum simulator and the green triangle is the average overlap obtained from real quantum processors by dividing the original image into $2\times 2$ segments. The vertical axis is used to plot both the overlap for full images, and the average overlap over the segments.}
    \label{64patterns}
\end{figure}


\begin{figure}[t]
    \centering
    \includegraphics[width=0.47\textwidth]{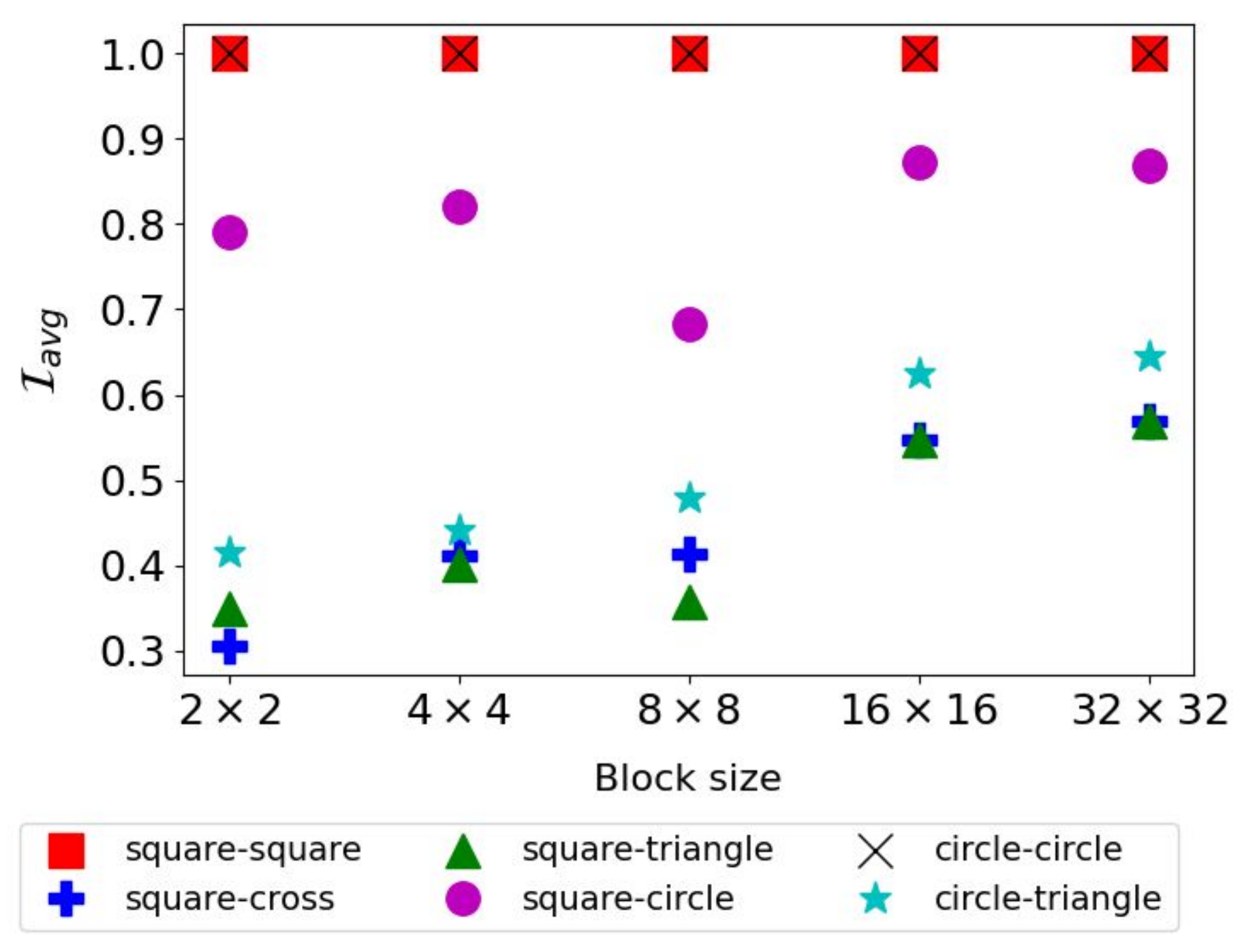}
    \caption{The variation of the average overlap $\mathcal{I}_{avg}$ with varying dimension of the blocks used to divide the original $32\times 32$ images. The results are obtained by using IBMQ simulator. Different markers and colors corresponds to comparison between different pairs of geometric shapes as indicated in the legend.}
    \label{32patterns}
\end{figure}

\section{Destructive swap test for pattern recognition}
\label{sec:swap_test_result}

In this section, we present a comparative study between the swap test and destructive swap test in real quantum processors as well as in the ideal quantum simulators provided by IBMQ. We use binary images where the black pixels compose a `pattern' against the background of white pixels. Our goal is to identify this pattern in a certain image. We start with small images with $2\times 2$ and $2\times 4$ pixels, as shown in Fig. \ref{fig:2_3qubit_dswap}(a) and Fig. \ref{fig:2_3qubit_dswap}(b). The black and white pixels correspond to pixel values ``1" and ``0" respectively. The images are encoded using 2 and 3 qubits respectively. As shown in the previous section, the swap test in these systems is highly noisy. We want to check whether destructive swap test can reduce the effect of noise. For 100 simulations of the same circuit in real quantum processors, we calculate the mean overlap $\mathcal{I}_{mean}$ and the standard deviation $\sigma$. For comparison, we also show in Fig. \ref{fig:2_3qubit_dswap}(a) and Fig. \ref{fig:2_3qubit_dswap}(b) the theoretical value of the overlap, and that obtained from IBMQ simulator.
For the two-qubit images, the result shows $7\%$ error to distinguish between same images, which is a significant improvement over Fig. \ref{fig:swap_test}. For the three-qubit images, this error is around $31\%$, which is still an improvement over the swap test. Also the relative behaviour of $\mathcal{I}_{mean}$ for different reference images are consistent with the actual overlap, and hence the destructive swap test can be faithfully used for pattern recognition. This also demonstrates that, even though the number of measurements increase, the noise in the output is much less compared to swap test. However, the destructive swap test needs a more complicated classical post-processing of the measurement outcomes. Thus, there is a trade-off between swap test and destructive swap test in terms of classical post processing, and number of quantum gates used in the circuit. The reader may note that, even when comparing two same images, the success/failure probability can vary little bit depending on the particular quantum image states being considered. This is because the number of gates used to prepare different initial states can be different, hence the effect of noise can vary.

To perform destructive swap test for larger images, we need access to higher dimensional real quantum processors. The largest system we have access to are the 7 qubit IBMQ processors. A way out is to divide the original image into smaller segments each of which can be encoded using the available few-qubit systems. Then one can apply destructive swap test on the pair of segments belonging to the same coordinates from the target and the reference image. In case of binary images, there can exist segments on which all the pixels have pixel values 0. Using our encoding method, we cannot encode these white blocks as valid quantum states because $c_{i}=0$, $\forall i$. Hence, we run the circuit only when both the blocks corresponding to target and reference image have at least one pixel with non-zero pixel value. For each such pair of blocks, we calculate the overlap. One drawback of this segment-wise approach is that it does not anymore calculate the distance between the full quantum states encoding the images. Also by taking this approach, we exclude those cases when one among the pair of blocks is white and another has black pixels. This exclusion induces error in the results.

In other words, a particular pattern can have same number of pixels in common with two or more different patterns, e.g. in Fig. \ref{64patterns}, the triangle has the same pixels overlapping with the square, the circle, and with itself. Hence all those reference patterns will correspond to exactly the same number of blocks contributing to the overlap with the triangle. Thus to detect the closest pattern, one should also take into account the white blocks which are not common between the target and reference pattern. For this, we take summation of all segment-wise overlaps, and divide it by $\mathrm{max}\{N_{1}, N_{2}\}$, where $N_{1}$ and $N_{2}$ are the number of blocks having non-zero pixel values corresponding to the patterns in the target and reference image respectively. We call this ``average overlap" $\mathcal{I}_{avg}$, and use this to capture the closeness between two patterns. We show the results in Fig. \ref{64patterns}, where we divide the original $32\times 32$ pixels images containing simple geometric patterns into $2\times 2$ pixels segments, and apply destructive swap test on the $16\times 16$ pairs of segments with same coordinates from the target and reference image. We also have presented the actual value of the overlap between the full quantum image states and those obtained from the quantum simulator.
Clearly, the average overlap is highest when the patterns from both the images match exactly. Moreover, despite the noise, the comparative behaviour of $\mathcal{I}_{avg}$ is consistent with that of the actual overlap between the patterns.
This segmenting process indeed needs significantly higher number of measurements, and computational time. However, as the computational power of quantum computers improves more and more over the years, we will be able to select bigger segments from original images, eventually being able to encode the whole images using the available qubits. 
To check the validity of this segmenting process for bigger than $2\times 2$ blocks, we increase the block size step by step, and compare $\mathcal{I}_{avg}$ between the same pair of patterns as in Fig. \ref{64patterns}. Due to the size limitation of accessible real quantum processors, we use IBMQ simulator to simulate the circuits, and the results have been presented in Fig. \ref{32patterns}. For every block size, the comparative behaviour between different pairs of patterns remains consistent with the ideal situation. There exist small anomalies, e.g. for the $\mathcal{I}_{avg}$ of square-circle pair when the blocks are $8\times 8$ pixels. This may arise because, as stated before, $I_{avg}$ is not entirely equivalent to the overlap $\mathcal{I}$ between full quantum images. The results again indicate that the destructive swap test is faithful for pattern recognition, and $\mathcal{I}_{avg}$ can be used for large dimensional images with simple binary patterns.


\begin{figure}[t]
\centering
\includegraphics[width=0.48\textwidth]{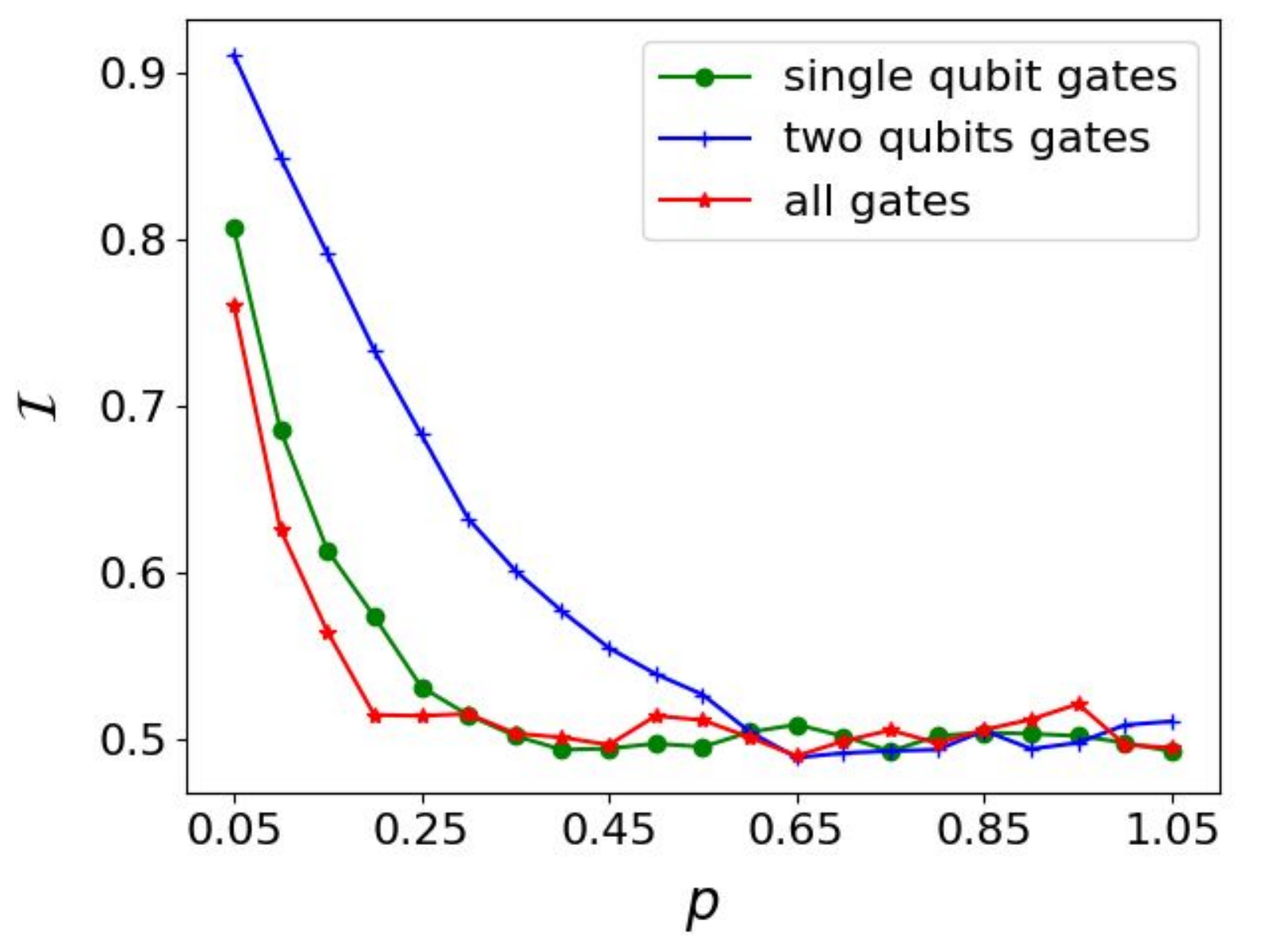}
\caption{The behaviour of the overlap $\mathcal{I}$ between two identical two-qubit image states, with varying depolarizing noise strength in the circuit gates. The blue curve corresponds to the case when the noise in the two CNOT gates is varied, keeping all the single qubit gates at very low noise. The green one is for the case when the CNOT gate noise is low, but all the single-qubit gate noise varies. The red curve is when all the gate noises are varied together.}
\label{gate_error}
\end{figure}

\subsection{Noise robustness}

Due to the impossibility of completely removing the presence of noise, the real IBMQ processors are subject to gate errors, state preparation and measurement errors, and readout errors. The former is due to imperfect implementation of a gate, and the later is due to imperfect measurement. As all the IBMQ processors are calibrated on certain time intervals, the values of these errors fluctuate. To understand the robustness of our results against these fluctuating errors, it is important to check how the success probability of destructive swap test changes with changing noise. Since we cannot manually change the errors in IBMQ processors, we resort to the simulation using ``NoiseModel" class provided by IBMQ. The simulator copies the error values as well as other system parameters like topology, coupling and basis gates of a user-specified real quantum processor, so that the simulation result mimics what is obtained from that real experiment. It is also possible to customize the noise for all the basis gates, and then use the `$\mathrm{qasm}\_\mathrm{simulator}$' to simulate the result. For example, one can choose among depolarizing noise, Pauli noise etc., vary the noise strength, and select gates to which the noise is to be applied. With this approach, we study three cases, i.e. (i) the single qubit gates are significantly noisy, (ii) the two-qubit gates are significantly noisy, and (iii) all the gates are equally noisy. In all three cases, we consider a depolarizing noise. A quantum state $\rho$ subject to depolarizing noise evolves to a mixture of itself and the maximally mixed state, i.e. $\rho \rightarrow (1-p)\rho + p\frac{I}{4}$. Here $p$ is the noise strength which satisfies the bound $0 \leq p \leq 1+\frac{1}{d^2 - 1}$, $d$ being the dimension of $\rho$, and $I$ is the identity operator. We vary $p$ from 0.05 to 1.05, and calculate the overlap between two identical $2\times 2$ image by destructive swap test. Our results are shown in Fig. \ref{gate_error}. For this study, we keep the readout error at the constant value of 0.01, which is of the same order of magnitude as the typical readout errors in real quantum processors. As we can see from the figure, the noise in single qubit gates is more detrimental than the noise in two-qubit gates for destructive swap test. When all the gates are noisy, then the circuit performance worsen, as expected. The fluctuation reduces for higher values of $p$, but in that regime the value of $\mathcal{I}$ is also small. However, for the IBMQ systems we have used, the single-qubit gate errors are always well below 0.01, and the two-qubits gate errors are below 0.1, which implies that $\mathcal{I}>0.55$ for two-qubit states in this noise range.
Since we use only depolarizing noise to model the errors, the actual results from real quantum processors are not exactly same as shown in Fig. \ref{gate_error}. However, the noise simulation gives an idea about the overall behaviour of the protocol with changing noise at different gates.

\begin{figure}[t]
\includegraphics[width=0.47\textwidth]{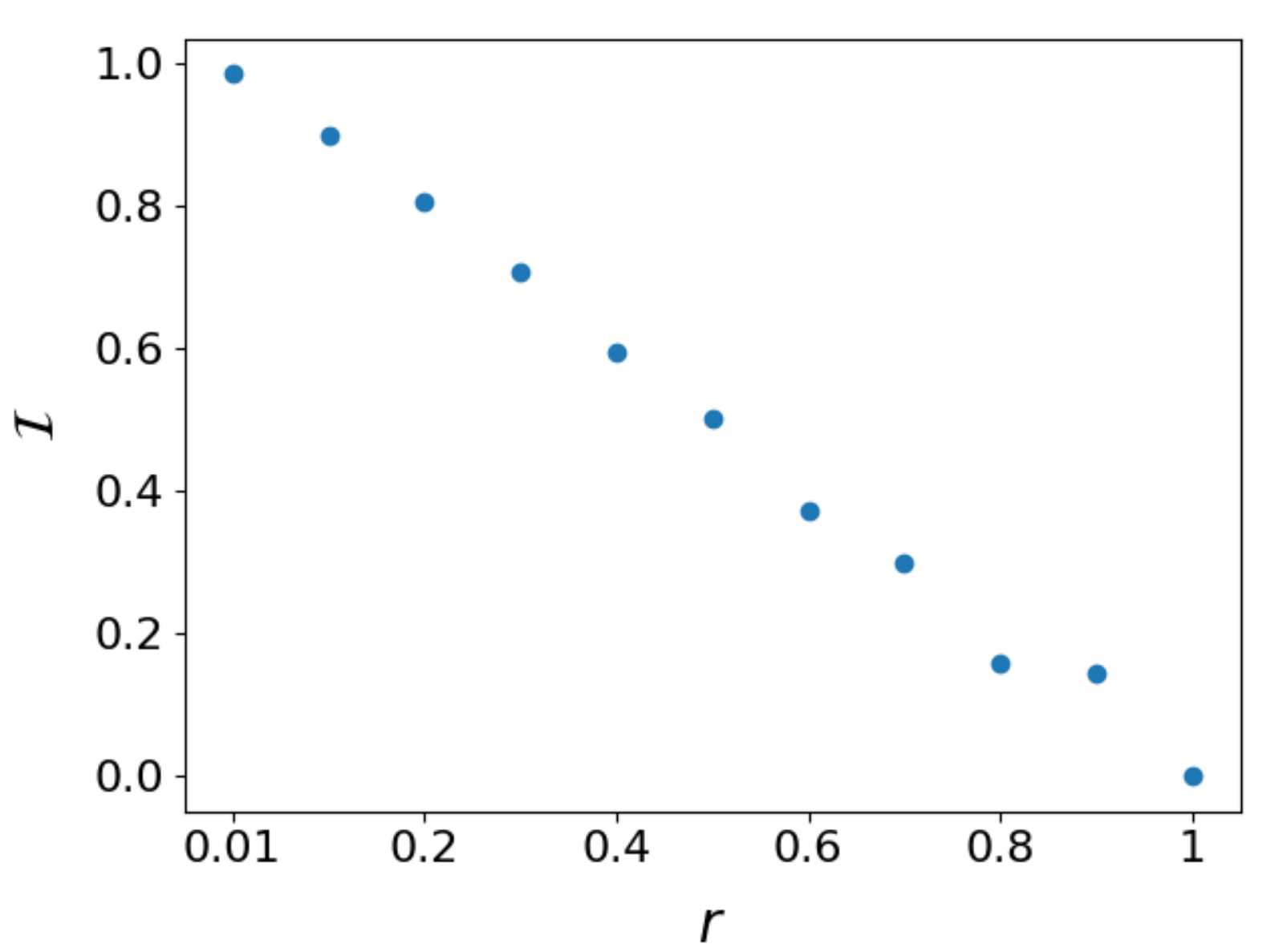}
\caption{The behaviour of the overlap $\mathcal{I}$ between two identical two-qubit states with varying readout error strength, when simulated using destructive swap test.}
\label{readout_error}
\end{figure}

We also study the success probability by varying the readout error $r$. The readout error is the fraction indicating the number of times a $``0"$ or $``1"$ in the output is measured erroneously as $``1"$ or $``0"$. The result is presented in Fig. \ref{readout_error}. The gate errors corresponding to all the gates has been kept at $10^{-3}$ which is of the same order of magnitude as the lowest gate errors in real processors. Typically the readout error in these processors are of the order of $10^{-2}$.

\subsection{Feature recognition in MRI images}
\label{subsec:MRI}
\begin{figure}
    \centering
    \subfigure[]{\includegraphics[width=0.30\textwidth]{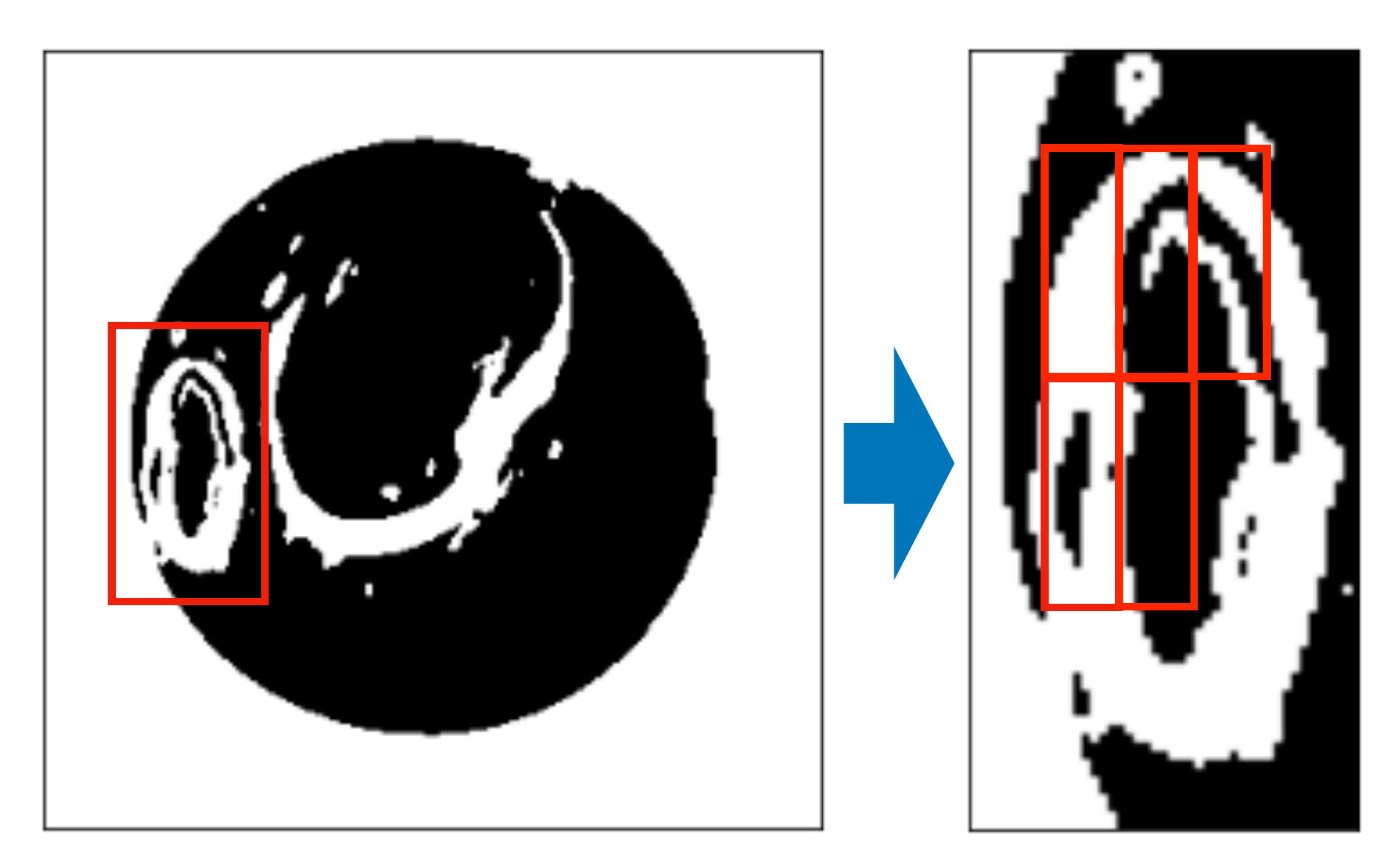}}\hspace{3em}
    \subfigure[]{\includegraphics[width=0.055\textwidth]{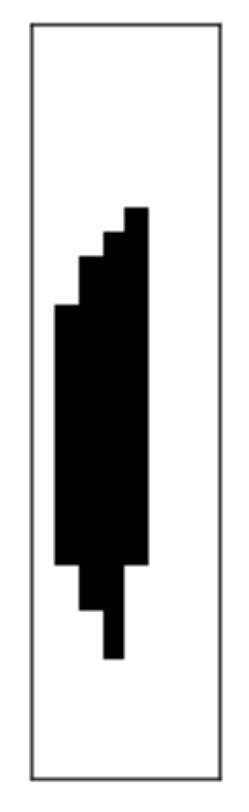}}
    \subfigure[]{\includegraphics[width=0.48\textwidth]{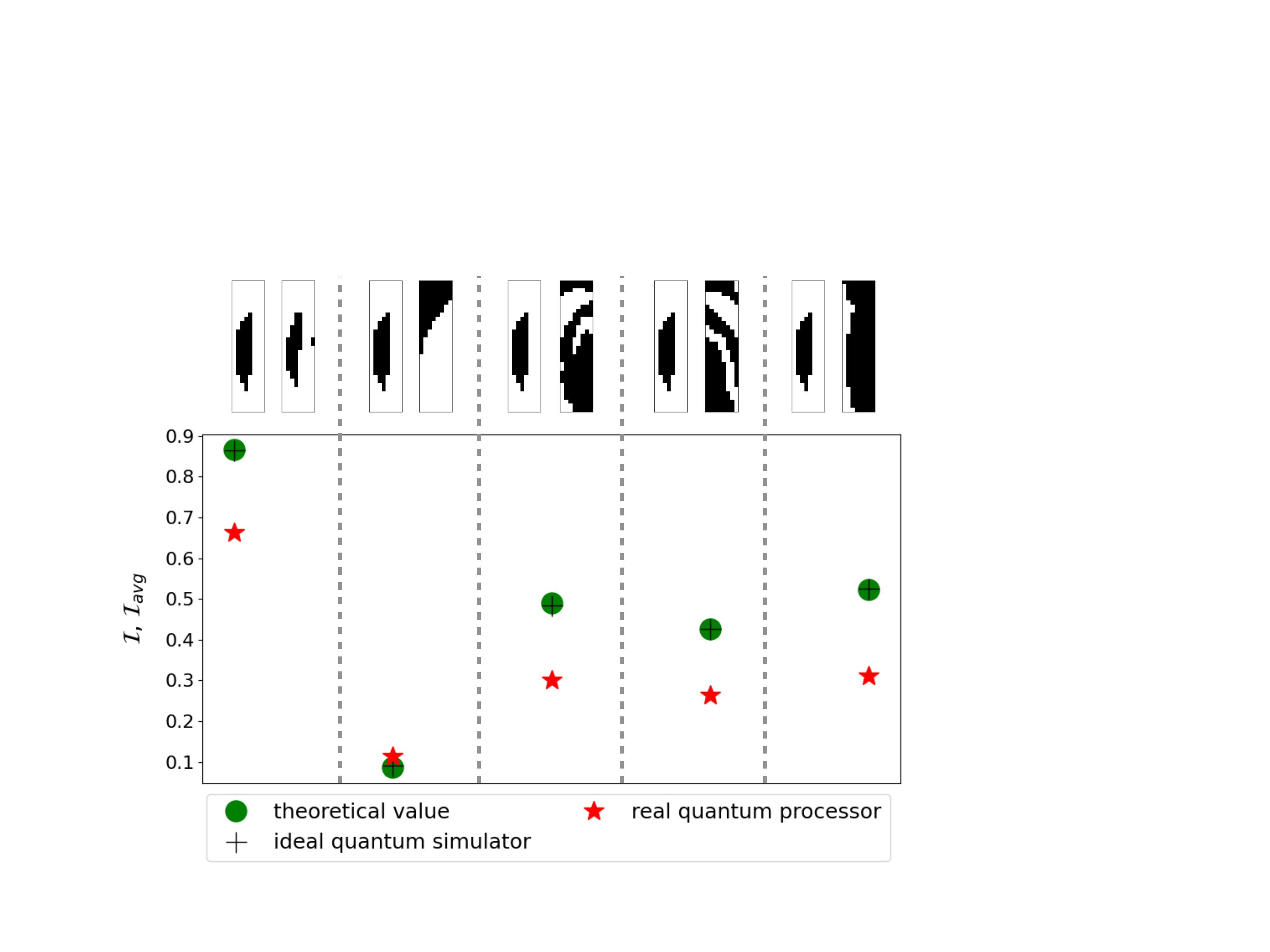}}
    \caption{Destructive swap test performed in real quantum processors to detect a cavity in the MRI image of human blood vessel. (a) (In the left) The full image after conversion from colored to binary. (In the right) A zoomed-in segment of the smaller vessel containing the cavity to be identified. The five rectangular areas with red outline is compared with the reference cavity. (b) The reference cavity, obtained from another similar MRI image. (c) The theoretical overlap $\mathcal{I}$, the overlap from quantum simulator, and the average overlap $\mathcal{I}_{avg}$ obtained by comparing (b) with the six segments in (a).}
    \label{vessel}
\end{figure}

To consider a more practical application e.g. in biomedicine, we repeat the destructive swap test to detect a particular feature in an image of human blood vessel obtained from MRI in our lab. The original images are colored images. We choose a threshold value to binarize the image so as to reveal all the details of the vessel against the background of Formaldehyde. In this particular case, we want for instance, to identify a cavity present in the tissue of the smaller vessel in the left of the image, see Fig. \ref{vessel}(a). To do so, we compare different sections of that blood vessel with a sample image of a cavity. The later is extracted from another MRI image of human blood vessel. In Fig. \ref{vessel}(a), we have highlighted a few of the sections of the blood vessel which are compared with the reference cavity in Fig. \ref{vessel}(b). We select $2\times 2$ blocks from each image, belonging to the same coordinates, and calculate $\mathcal{I}_{avg}$. In Fig. \ref{vessel}(c), we present the comparison of $\mathcal{I}_{avg}$ with the theoretical overlap and that obtained from quantum simulator. In this case as well, $\mathcal{I}_{avg}$ has highest value when the target and reference cavities match very closely, proving again the utility of destructive swap test, as well as the average overlap measure defined by us. This also opens up the possibility of pattern recognition in medical images using destructive swap test.


\begin{figure*}
    \subfigure[]{\includegraphics[width=1\textwidth]{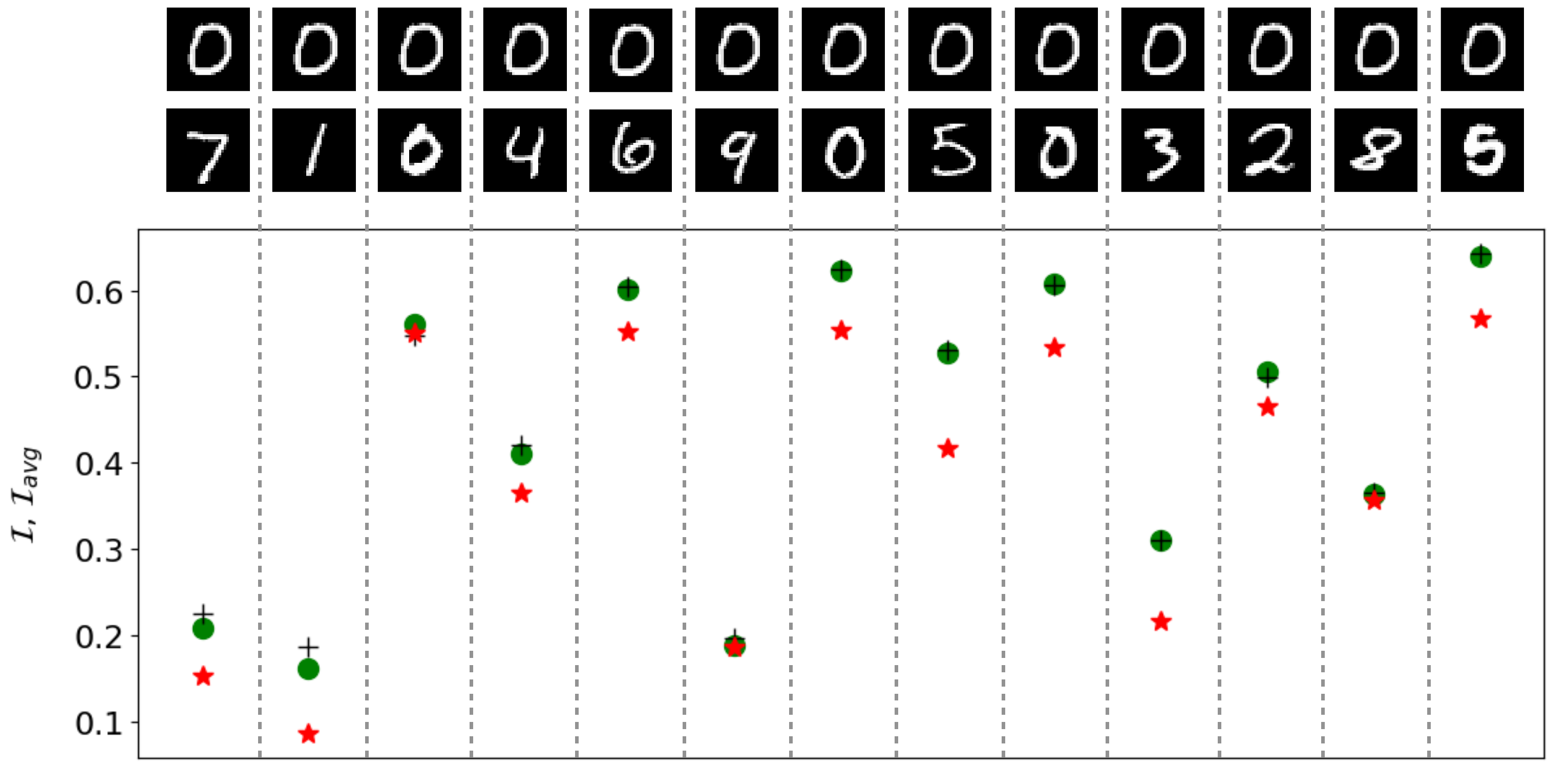}}
    \subfigure[]{\includegraphics[width=1\textwidth]{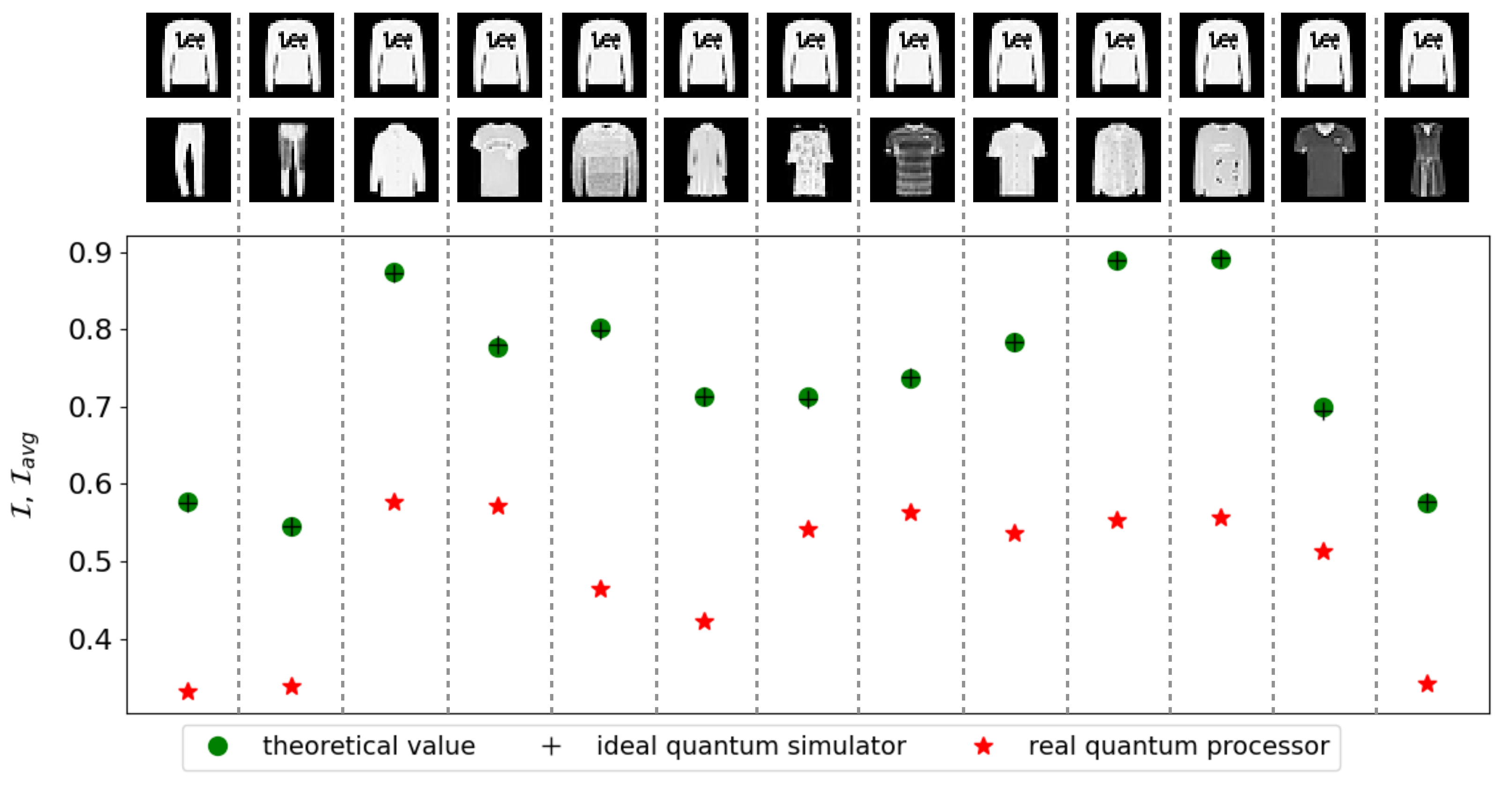}}
    \caption{Comparison between the actual overlap $\mathcal{I}$ between two quantum images, the overlap obtained from a quantum simulator, and the average overlap $\mathcal{I}_{avg}$ obtained from real IBMQ processors. The plots are for (q) MNIST numbers and (b) MNIST Fashion images. Two images along a vertical column are compared.}
     \label{MNIST}
\end{figure*}



\section{Pattern recognition in greyscale images}
\label{sec:greyscale}
While binary images are the simplest image prototypes, in practice the images acquired for quantum image processing tasks are mostly greyscale or with an RGB color profile. Hence it is important to investigate whether the destructive swap test can efficiently identify patterns in greyscale/RGB images. Irrespective of the pixel values, we can encode these images in a quantum state using QPIE method, and then apply the destructive swap test to measure the overlap between them. First, we use MNIST images as the prototype for greyscale images. MNIST images is a large set of images, frequently used in machine learning and pattern recognition algorithms \cite{deng2012mnist}. 
Each image is $28\times 28$ pixels, containing a handwritten digit between 0 and 9. The dimension of these images is not an integer power of 2, hence they cannot be encoded using QPIE method. To tackle this, we add rows and columns all having black pixels to the original images, to increase the dimension to $32\times 32$ pixels, which can be encoded using 10 qubits. As our target image, we pick a particular image containing ``0" as the pattern, and compare it with a set of other images with different numbers. We simulate the circuit in IBMQ simulator to obtain the overlap  $\mathcal{I}$  between full quantum images. Then we divide them into $2\times 2$ segments to calculate $\mathcal{I}_{avg}$ from real quantum processors. The results have been presented in Fig. \ref{MNIST}(a), which shows that the behaviour of $\mathcal{I}_{avg}$ for different reference images is consistent with the theoretical overlap or the simulated one. It suggests that the destructive swap test can identify images with similar greyscale patterns by picking the images with highest $\mathcal{I}$ or $\mathcal{I}_{avg}$. For example, the overlap of the target image with the reference image containing ``6", is close to the overlap with reference images containing ``0". This is expected because the images are handwritten and in some cases ``6" can resemble the circular nature in ``0". Of course, if the reference images are enough noisy, the destructive swap test results can indicate a wrong digit for the target image. 

We also simulate the same circuit for MNIST fashion images. The result is presented in Fig. \ref{MNIST}(b). In this case, $\mathcal{I}_{avg}$ is not as efficient as for MNIST numbers. Though the reference images with comparatively higher $\mathcal{I}_{avg}$ matches well with those having highest overlap, choosing one particular closest pattern based on $\mathcal{I}_{avg}$ may not exactly match with the actual closest pattern. This is also due to the higher complexity of the images, as it is clear from the pictures that the pixels in these images encode more shades of grey compared to MNIST numbers. However, the $\mathcal{I}_{avg}$ in this case can still be used to distinguish between widely different categories of the fashion images, e.g. shirts and dresses.




\section{Experimental demonstration of destructive swap test in diamond NVs}
\label{experimental}
\begin{figure*}
    \includegraphics[width=0.95\textwidth]{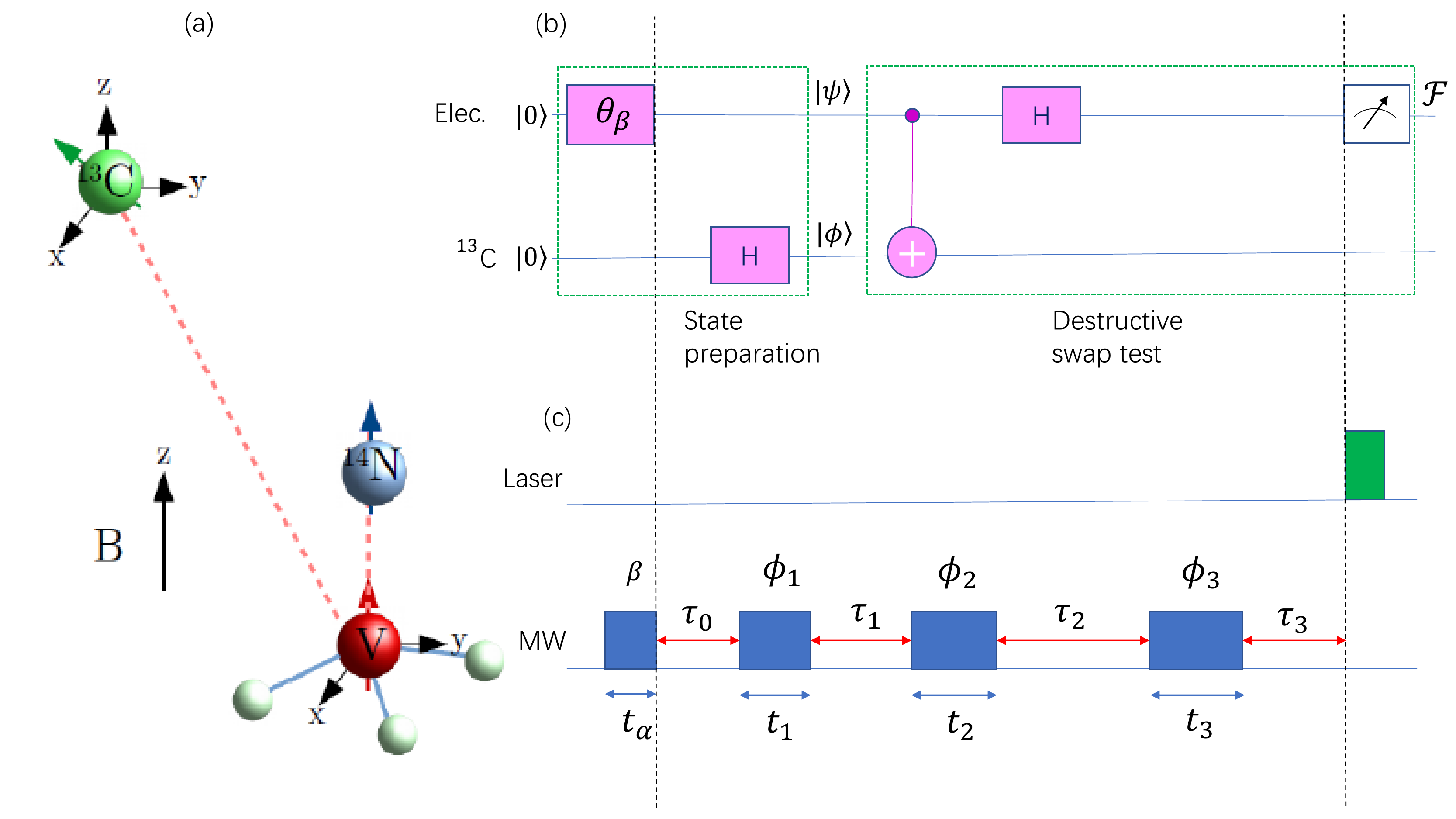}
    \caption{Experiment scheme to demonstrate the destructive swap test in NVs. (a) Structure of the NV center with a coupled $^{13}$C spin. (b) Quantum circuit with the steps of  the state preparation and destructive swap test, indicated by the green dotted boxes. The operation $\theta_{\beta}$ denotes an operation as $\theta_{\beta}=e^{\mp i\theta I_y}$, corresponding $\beta=\pi/2$ and $\beta=3\pi/2$, respectively. The observable $\mathcal{F}  = |\langle \phi |\psi\rangle|^2$ is the population of the electron state $|0\rangle$. (c) The microwave (MW) pulse sequence for implementing the circuit in (b). The dashed lines indicate the correspondence between the operations in the circuit (b) and pulse sequence (c),   where the first MW pulse is for $\theta_{\beta}$, and the other three MW pulses are for the two Hadamard gates sandwiched by the CNOT gate, and the laser pulse detects $\mathcal{F}$. The Rabi frequency of the MW pulses is 2 MHz. The pulse durations are $t_1=0.206$, $t_2=0.122$, $t_3=0.182$ $\mu$s, the pulse phases are $\phi_1=317^{\circ}$, $\phi_2=273^{\circ}$, $\phi_3=90^{\circ}$, and delays $\tau_0=4.336$, $\tau_1=3.478$, $\tau_2=2.368$, $\tau_4=4.988$ $\mu$s. }
    \label{exp}
\end{figure*}

\begin{figure*}
    \includegraphics[width=0.95\textwidth]{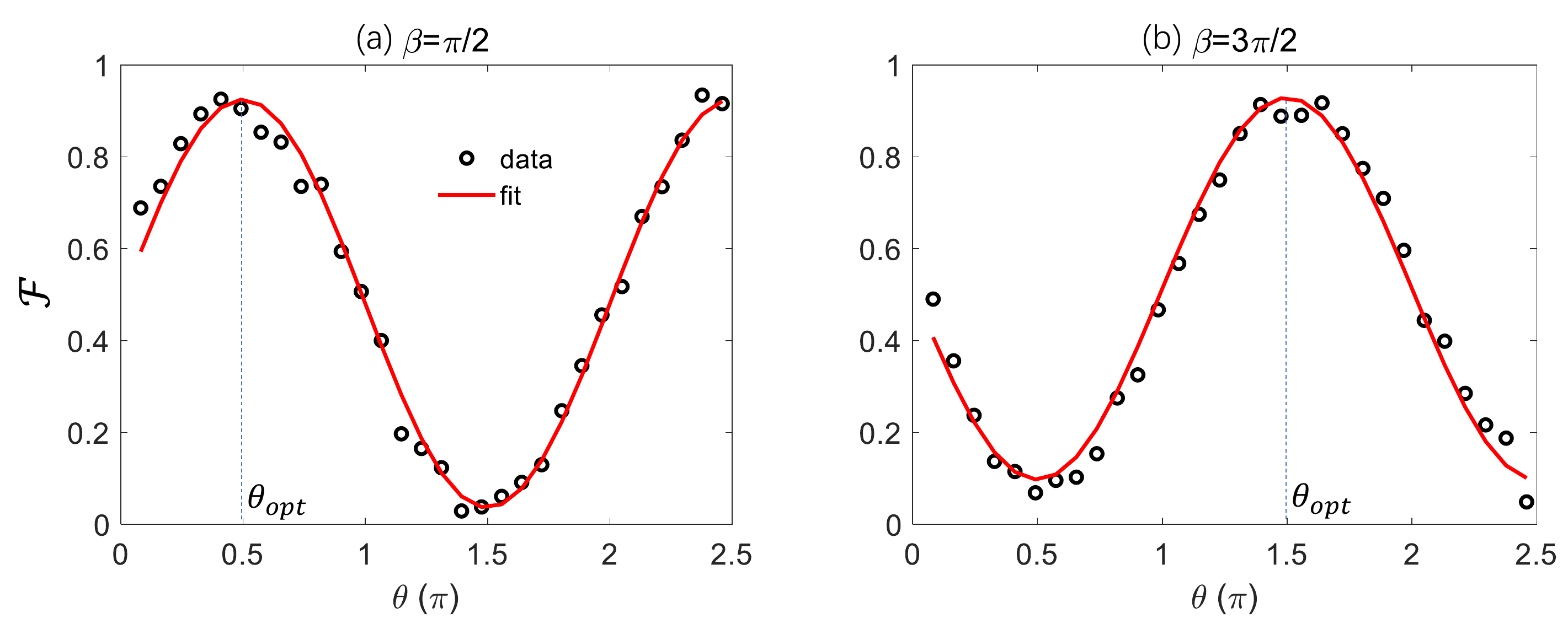}
    \caption{Experiment results for $\mathcal{F}$ vs $\theta_{\beta}$ to demonstrate the swap test in NVs, where the flip angles and the phases in operation $\theta_{\beta}$ are indicated as the horizontal axis and in the panels. The vertical dashed lines indicate the angles corresponding the maximums fidelity in fit.}
    \label{expres}
\end{figure*}

The IBMQ processors are based on superconducting qubit technology. To test the efficiency of destructive swap test in a different quantum system, we use two qubits in diamond NVs \cite{6991,Suter201750} to demonstrate the performance of destructive swap test.  The experiment scheme is shown in Fig. \ref{exp}. The  demonstration starts with the pure state $|00\rangle$, which is prepared by the schemes proposed in the previous works \cite{zhang18,zhang19,swathi19,arXiv:2108.13738}. In Fig. \ref{exp} (a), we show the structure of the NV center with a coupled $^{13}$C spin. Here we choose the electron spin in states with $m_S =0$ and $m_S =-1$ as qubit 1, and $^{13}$C spin as qubit 2, where $m_S$ denotes quantum number for the electron spin.  Fig. \ref{exp} (b) shows the  quantum circuit to demonstrate the destructive swap test. 

We choose $|\phi\rangle=\frac{1}{\sqrt{2}}(|0\rangle+|1\rangle)$, generated by applying a Hadamard gate $H$ to state $|0\rangle$. We use the destructive swap test to measure the fidelity between state $|\phi\rangle$ and various states $|\psi\rangle$, which is generated by applying an operation $\theta_{\beta}=e^{\mp i\theta I_y}$ to state $|0\rangle$, where  $\beta=\pi/2$ or $\beta=3\pi/2$, respectively, and $I_y$ denotes $y$ component of the  spin $1/2$ operator. The fidelity $\mathcal{F}$ is the population of the electron spin in state $|0\rangle$. The theoretic $\mathcal{F}$ is calculated as
\begin{equation}
    \mathcal{F}_{th}  = \frac{1}{2}(1\pm \sin\theta)
\end{equation} 
with maximums at $\theta_{opt}=\pi/2$ or $3\pi/2$, for $\beta=\pi/2$ or $\beta=3\pi/2$, respectively.   
In Fig.  \ref{exp} (c), we show  the microwave (MW) pulse sequence for implementing the circuit in (b). The first pulse is for the operation $\theta_{\beta}$, and the other three MW pulses are for the two Hadamard gates sandwiched by the CNOT gate. The laser pulse detects the observable $\mathcal{F}$.

The experiment results are shown in Fig. \ref{expres}, where figures (a-b) correspond to $\beta = \pi/2$ and $3\pi/2$, respectively. The experiment data $\mathcal{F}$ can be fitted by the theory $\mathcal{F}_{th}$ as    
\begin{equation}
    \mathcal{F} = a + b \mathcal{F}_{th}
\end{equation}
where parameters $a$ and $b$ are obtained in fit  as $a = 0.04$, $b = 0.89$ in Fig. \ref{expres} (a), and $a = 0.10$, $b = 0.83$ in Fig. \ref{expres} (b).
The deviation between
experiment and theory can be mainly attributed to the theoretical imperfection of the pulse sequence with theory fidelity $0.9$ in the optimizing the pulse parameters, dephasing effects of the electron spin since the total pulse duration is up to 16.3 $\mu$s and is comparable  to $T_2^* \approx 35$ $\mu$s and the statistics of the photon detection. 

\begin{figure*}
    \includegraphics[width=0.75\textwidth]{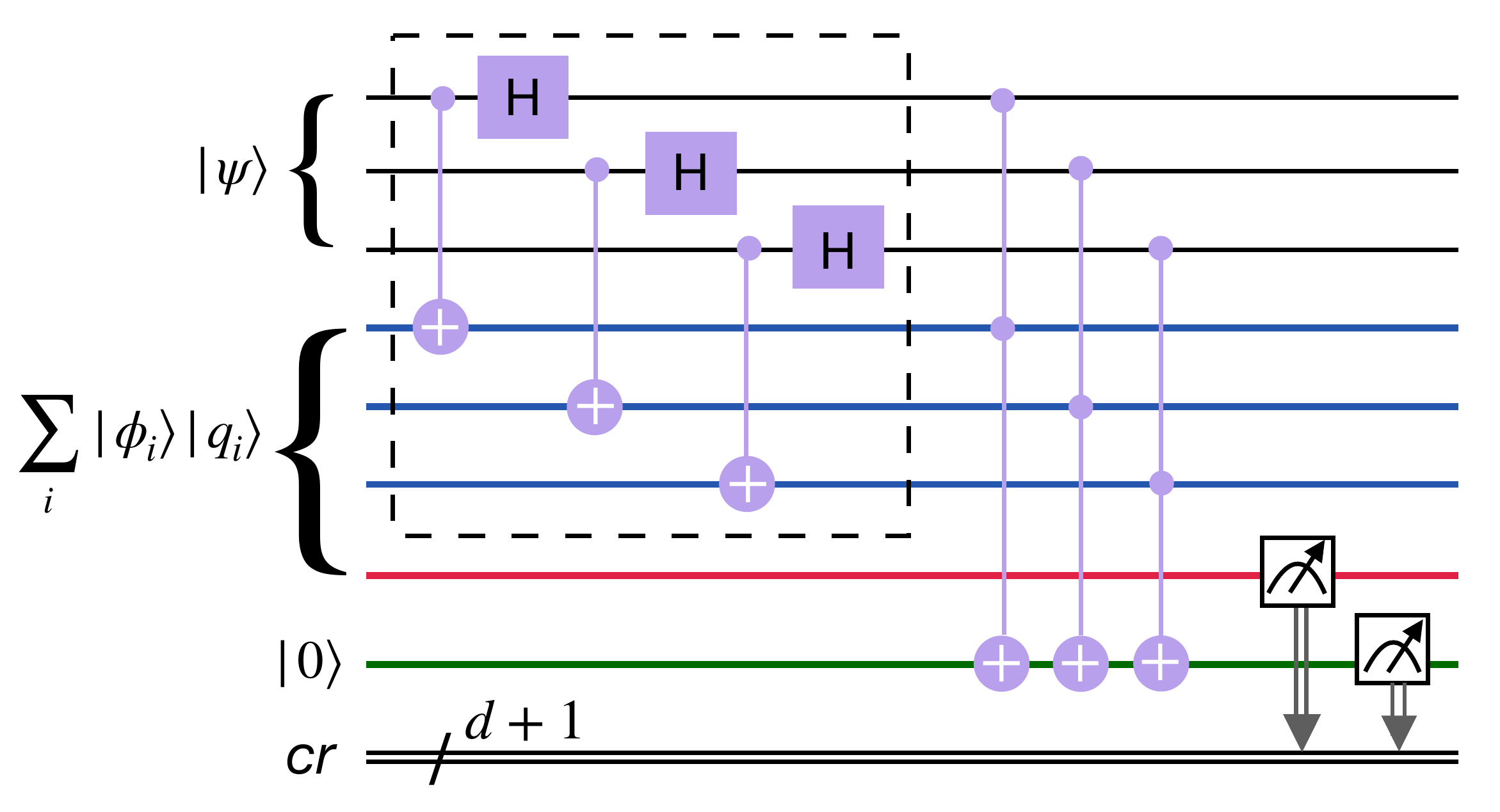}
    \caption{The quantum circuit of supervised learning algorithm based on destructive swap test, to detect the closest pattern to a given pattern.}
    \label{machinelearn_qc}
\end{figure*}

\section{Supervised learning with destructive swap test}
\label{sec:qmachine}
In classical computers, information is stored in a particular memory address of the RAM. These computers recognize a given input by knowing its exact location in the memory. The term \textit{associative memory} is used to describe instances when an object is recognized not by accessing its location in the memory, but by associating the partial knowledge about it with the already stored information in the memory. This is similar to the functioning of human brain, e.g. while solving a crossword puzzle. A computational application of associative memory is artificial neural networks \cite{muller2012neural}. Associative memory is advantageous over RAM while solving complex problems, but on the other hand requires huge capacity to store information for an efficient performance. In \cite{Trugenberger2002}, C. A. Trugenberger introduced the idea of quantum associative memory to be used for quantum pattern recognition. Besides exploiting the quantum speed-up, the protocols based on quantum associative memory can overcome the limitation of short storage capacity in classical systems. Inspired by \cite{Trugenberger2002}, a number of previous works have proposed quantum pattern recognition protocols which work in a similar way to classical supervised learning. In \cite{lloyd2013quantum}, the authors proposed a machine learning protocol for assigning a particular class to a vector, by optimizing the mean distance between the target and reference vectors. In \cite{petruccione_book2014}, the authors propose a similar protocol for pattern recognition, by taking the $k$-nearest neighbour approach to optimize the distance between target and reference states. In this section, we devise a quantum machine learning protocol for pattern recognition, which uses destructive swap test. The corresponding circuit for a three-qubit state is shown in Fig. \ref{machinelearn_qc}. The target state $|\psi\rangle$ has been encoded using three quantum registers, which are represented by the black horizontal lines in the figure. Let us assume that there are $d$ reference states $\phi_{i}$ $(i=1,2,...,d)$, each having the same dimension as $|\psi\rangle$, from which we have to detect the closest pattern to $|\psi\rangle$. The three quantum registers which can be used to encode these states are denoted by the blue horizontal lines in the figure. We take a $d$ dimensional quantum system (qudit), the basis vectors of which can be associated with the $d$ reference states. In Fig. \ref{machinelearn_qc}, the qudit is represented as the red horizontal line. Now we prepare the following superposition,
\begin{equation}
    |\Phi\rangle=\frac{1}{N^{\prime}}\sum_{i}|\phi_{i}\rangle|q_{i}\rangle,
\end{equation}
where $q_{i}$ $(i=0,1,..., d-1)$ is the $i^{\mathrm{th}}$ basis vector in the Hilbert space $\mathcal{H}^{d}$ of the qudit, and $N^{\prime}$ is the normalization constant. The blue and the red quantum registers are initialized in the joint state $|\Phi\rangle$. There is an auxiliary qubit which is initialized in the state $|0\rangle$, represented by the green horizontal line in Fig. \ref{machinelearn_qc}. Thus the initial state of the total system is,
\begin{equation}
    |\Psi_{0}\rangle=\frac{1}{N^{\prime}}|\psi\rangle \otimes \sum_{i}|\phi_{i}\rangle|q_{i}\rangle \otimes |0\rangle.
\end{equation}
Now we apply the gates in accordance to the destructive swap test, i.e. CNOT gates on each pair of qubits from target and reference states, followed by Hadamard on the qubits of the target state. After this, the total state is,

\begin{eqnarray}
    &|\Psi_{1}\rangle=\frac{1}{N^{\prime}}\sum\limits_{i}(C^{i}_{1}|000000\rangle+C^{i}_{2}|000001\rangle+... \nonumber \\
    &+C^{i}_{64}|111111\rangle)\otimes |q_{i}\rangle \otimes |0\rangle.
\end{eqnarray}

\begin{figure}
    \includegraphics[width=0.45\textwidth]{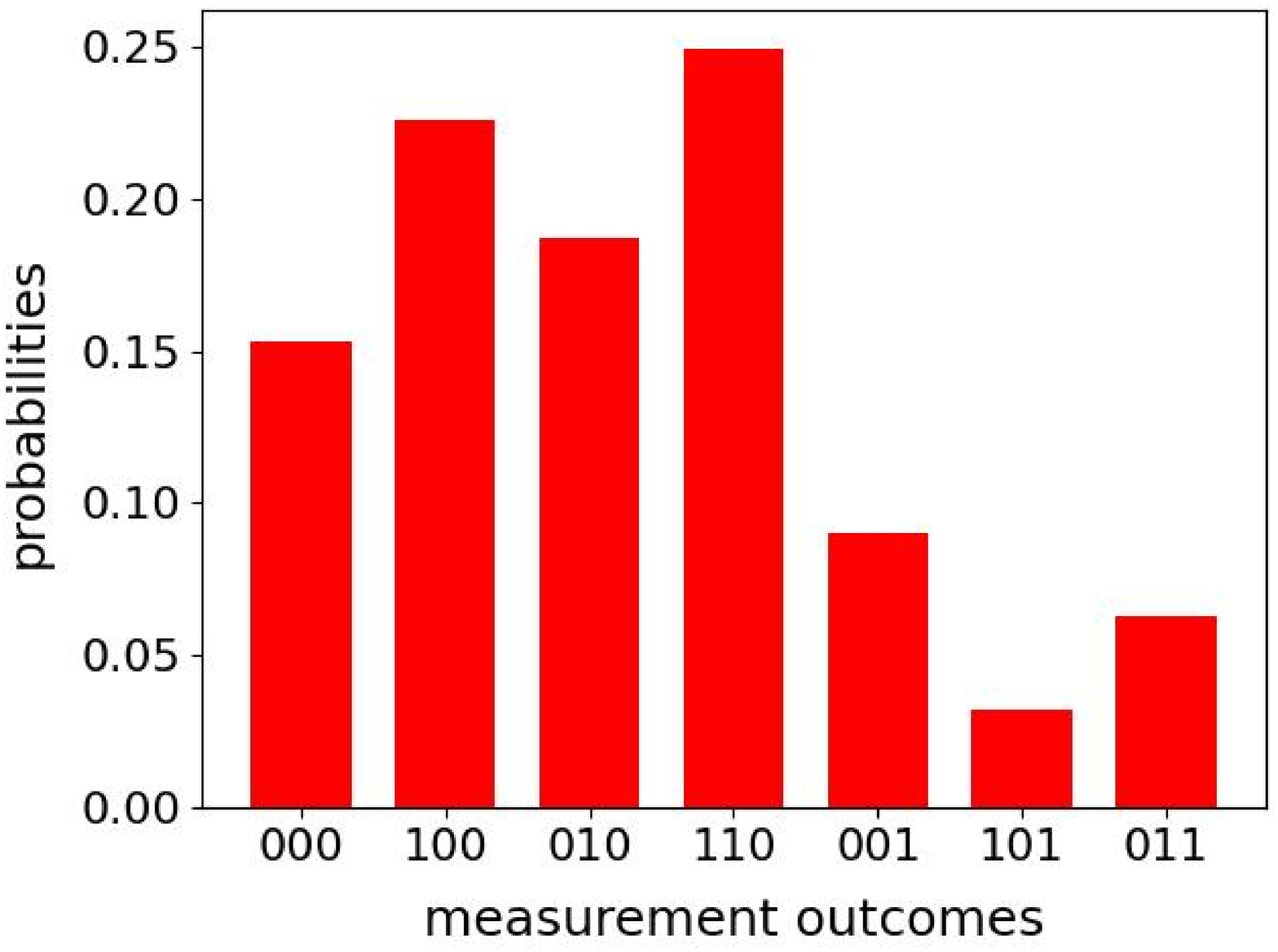}
    \caption{The supervised learning protocol based on destructive swap test performed to detect the closest pattern of a two-qubit target image from a set of four reference images, also shown in Fig. \ref{fig:2_3qubit_dswap}. The horizontal axis shows the classical bits corresponding to the measurement outcomes from the joint system composed of the qudit (in this case four dimensional) and the auxiliary qubit. The vertical axis denotes the corresponding probabilities. The success probability corresponding to the basis $|110\rangle$ is the highest, which corresponds to the case of equal target and reference images.}
    \label{machinelearn_barplot}
\end{figure}

Here $C_{j}^{i}$ $(j=1, 1,..., 64)$ are the coefficients of the vector in the joint Hilbert space $\mathcal{H}^{6}$ of target and reference states associated with $|q_{i}\rangle$. We remind that the basis vectors in this space for which the pairwise AND of the bits have odd parity, contributes to the failure probability. To differentiate between these basis vectors and the rest, we apply CCNOT gates such that each pair of the target and reference state qubit acts as the control qubits, and the auxiliary qubit flips its state when both of the former are in state $|1\rangle$. Thus, the additional qubit keeps a count of the parity. If the basis vector has odd parity, the auxiliary qubit has final state $|1\rangle$. In case of even parity, the qubit state remains $|0\rangle$. The total state after applying the CCNOT gate can be written as,

\begin{eqnarray}
    &|\Psi_{2}\rangle=\frac{1}{N^{\prime}}\sum\limits_{i}\big[\big( C^{i}_{1}|000000\rangle + C^{i}_{1}|000001\rangle +.. \big)\otimes |q_{i}\rangle \otimes |0\rangle \nonumber \\
    &+\big( C^{i}_{16}|001111\rangle + C^{i}_{19}|010010\rangle +... \big)\otimes |q_{i}\rangle \otimes |1\rangle\big]
\end{eqnarray}

Thus, the qudit keeps the memory of the reference images, whereas the auxiliary qubits keeps record of the basis vectors contributing to success and failure probabilities. Now, repeated joint measurement on the qudit and the auxiliary qubit can reveal the success and failure probabilities corresponding to each reference image. The one having the largest success probability is closest to the target image. In Fig. \ref{machinelearn_barplot}, we show the measurement outcomes as a histogram when a two-qubit target image is identified by comparing it with four reference images. For this task, we use the same images as in the first four columns of Fig. \ref{fig:2_3qubit_dswap}. We denote the target image as $|S_{1}\rangle$, and the set of reference patters as $\{|S_{1}\rangle, |S_{2}\rangle, |S_{3}\rangle, |S_{4}\rangle\}$. To label the reference states, we use the basis states $\{|00\rangle, |01\rangle, |10\rangle, |11\rangle\}$ of a two-qubit system, and prepare the following superposition,
\begin{equation}
    |\Phi\rangle=\frac{1}{N^{\prime}}\big( |S_{1}\rangle|11\rangle + |S_{2}\rangle|01\rangle + |S_{3}\rangle|10\rangle + |S_{4}\rangle|00\rangle\big)
\end{equation}
Since the target image is $|S_{1}\rangle$, we expect that at the end of the protocol described above, the joint measurement on the two-qubit system and the auxiliary qubit will give highest probability corresponding to the state $|110\rangle$. Here the first two bits correspond to the two-qubit system, whereas the last bit corresponds to the auxiliary qubit. The height of the bar plots in Fig. \ref{machinelearn_barplot} are consistent with the expectation. The circuit in this case has been simulated using a 7 qubit IBMQ simulator.

\section{Conclusion}
\label{sec:conclusion}
Image processing is indispensable in a plethora of everyday applications. Some quantum image processing algorithms have been studied in theory, which show polynomial and exponential speed-up over their classical analogues. There exists a number of quantum image encoding methods, one of which (QPIE) encodes the pixel positions and pixel values as the basis vectors and corresponding amplitudes of a quantum state. This naturally implies that swap test, which calculates the closeness between two quantum states, is a plausible tool to identify similar patterns in quantum images. In this paper, we investigated the efficiency of swap test for this task in real quantum processors. Our results showed that for the typical values of the gate errors present in IBMQ quantum machines, the output of swap test becomes completely noisy corresponding to three or more qubits. For lower dimensional states, it can still faithfully identify the pattern closest to a given pattern. To mitigate the effect of noise, we propose to exploit the destructive swap test, which showed considerable improvement for the three-qubit states. With access to larger quantum processors, it will be interesting to see whether the destructive swap test still remains successful against the noise for higher dimensional states. Considering the fact that a simple circuit such as swap test becomes completely noisy for three qubits, it is questionable how much practically useful the larger IBMQ processors will be for moderately complex algorithms. A scaling analysis with respect to dimension of the image states can be performed to investigate this question. To apply destructive swap test to larger images, we divided them into smaller segments which can be encoded using the 5 or 7 qubit quantum devices. The segment-wise overlaps were aggregated by defining an average overlap, which though not being equivalent to the actual distance between full quantum image states, remained consistent while distinguishing between two very similar or very different ones. The above findings remain true for both binary and greyscale images. We tested the destructive swap test for identifying a cavity in the binary image of human blood-vessel obtained by MRI, which again showed consistency, thus implying possible applications of this protocol in biomedical fields. As observed from the results involving simple and complex binary patterns, and greyscale MNIST images, the efficiency of average overlap may diminish with increasing range of pixel values and the complexity of the pattern. However, if it is possible to encode the full quantum states using the real IBMQ processors with suitably controlled noise, the destructive swap test is without any doubt a potentially useful tool for pattern recognition, as clear from our results in this work. Besides IBMQ devices, we also used the nitrogen vacancy centre in diamond to exhibit the destructive swap test for single qubit states, which shows good agreement with theoretical results. A future aim will be to find a suitable function of the segment-wise overlaps, which is equivalent to the distance between the full images. Also an important objective will be to find whether the machine learning tools can be used to counteract the noise in the outputs. We believe that the destructive swap test based protocol can be used for pattern recognition in all the relevant fields and for quantum machine learning algorithms.

\acknowledgments

The work was financially supported by the European Union's Horizon 2020 research and innovation programme under FET-OPEN Grant Agreement No. 828946 (PATHOS). We acknowledge the use of IBM Quantum services. The views expressed are those of the authors, and do not reflect the official policy or position of IBM or the IBM Quantum team.

\bibliography{main}{}

\begin{thebibliography}{74}%
\makeatletter
\providecommand \@ifxundefined [1]{%
 \@ifx{#1\undefined}
}%
\providecommand \@ifnum [1]{%
 \ifnum #1\expandafter \@firstoftwo
 \else \expandafter \@secondoftwo
 \fi
}%
\providecommand \@ifx [1]{%
 \ifx #1\expandafter \@firstoftwo
 \else \expandafter \@secondoftwo
 \fi
}%
\providecommand \natexlab [1]{#1}%
\providecommand \enquote  [1]{``#1''}%
\providecommand \bibnamefont  [1]{#1}%
\providecommand \bibfnamefont [1]{#1}%
\providecommand \citenamefont [1]{#1}%
\providecommand \href@noop [0]{\@secondoftwo}%
\providecommand \href [0]{\begingroup \@sanitize@url \@href}%
\providecommand \@href[1]{\@@startlink{#1}\@@href}%
\providecommand \@@href[1]{\endgroup#1\@@endlink}%
\providecommand \@sanitize@url [0]{\catcode `\\12\catcode `\$12\catcode
  `\&12\catcode `\#12\catcode `\^12\catcode `\_12\catcode `\%12\relax}%
\providecommand \@@startlink[1]{}%
\providecommand \@@endlink[0]{}%
\providecommand \url  [0]{\begingroup\@sanitize@url \@url }%
\providecommand \@url [1]{\endgroup\@href {#1}{\urlprefix }}%
\providecommand \urlprefix  [0]{URL }%
\providecommand \Eprint [0]{\href }%
\providecommand \doibase [0]{http://dx.doi.org/}%
\providecommand \selectlanguage [0]{\@gobble}%
\providecommand \bibinfo  [0]{\@secondoftwo}%
\providecommand \bibfield  [0]{\@secondoftwo}%
\providecommand \translation [1]{[#1]}%
\providecommand \BibitemOpen [0]{}%
\providecommand \bibitemStop [0]{}%
\providecommand \bibitemNoStop [0]{.\EOS\space}%
\providecommand \EOS [0]{\spacefactor3000\relax}%
\providecommand \BibitemShut  [1]{\csname bibitem#1\endcsname}%
\let\auto@bib@innerbib\@empty
\bibitem [{\citenamefont {Bennett}\ \emph {et~al.}(1993)\citenamefont
  {Bennett}, \citenamefont {Brassard}, \citenamefont {Cr\'epeau}, \citenamefont
  {Jozsa}, \citenamefont {Peres},\ and\ \citenamefont
  {Wootters}}]{teleportation}%
  \BibitemOpen
  \bibfield  {author} {\bibinfo {author} {\bibfnamefont {C.~H.}\ \bibnamefont
  {Bennett}}, \bibinfo {author} {\bibfnamefont {G.}~\bibnamefont {Brassard}},
  \bibinfo {author} {\bibfnamefont {C.}~\bibnamefont {Cr\'epeau}}, \bibinfo
  {author} {\bibfnamefont {R.}~\bibnamefont {Jozsa}}, \bibinfo {author}
  {\bibfnamefont {A.}~\bibnamefont {Peres}}, \ and\ \bibinfo {author}
  {\bibfnamefont {W.~K.}\ \bibnamefont {Wootters}},\ }\href {\doibase
  10.1103/PhysRevLett.70.1895} {\bibfield  {journal} {\bibinfo  {journal}
  {Phys. Rev. Lett.}\ }\textbf {\bibinfo {volume} {70}},\ \bibinfo {pages}
  {1895} (\bibinfo {year} {1993})}\BibitemShut {NoStop}%
\bibitem [{\citenamefont {Bennett}\ and\ \citenamefont
  {Wiesner}(1992)}]{densecoding}%
  \BibitemOpen
  \bibfield  {author} {\bibinfo {author} {\bibfnamefont {C.~H.}\ \bibnamefont
  {Bennett}}\ and\ \bibinfo {author} {\bibfnamefont {S.~J.}\ \bibnamefont
  {Wiesner}},\ }\href {\doibase 10.1103/PhysRevLett.69.2881} {\bibfield
  {journal} {\bibinfo  {journal} {Phys. Rev. Lett.}\ }\textbf {\bibinfo
  {volume} {69}},\ \bibinfo {pages} {2881} (\bibinfo {year}
  {1992})}\BibitemShut {NoStop}%
\bibitem [{\citenamefont {Shor}(1997)}]{shor}%
  \BibitemOpen
  \bibfield  {author} {\bibinfo {author} {\bibfnamefont {P.~W.}\ \bibnamefont
  {Shor}},\ }\href {\doibase 10.1137/S0097539795293172} {\bibfield  {journal}
  {\bibinfo  {journal} {SIAM Journal on Computing}\ }\textbf {\bibinfo {volume}
  {26}},\ \bibinfo {pages} {1484} (\bibinfo {year} {1997})}\BibitemShut
  {NoStop}%
\bibitem [{\citenamefont {Grover}(1997)}]{grover}%
  \BibitemOpen
  \bibfield  {author} {\bibinfo {author} {\bibfnamefont {L.~K.}\ \bibnamefont
  {Grover}},\ }\href {\doibase 10.1103/PhysRevLett.79.325} {\bibfield
  {journal} {\bibinfo  {journal} {Phys. Rev. Lett.}\ }\textbf {\bibinfo
  {volume} {79}},\ \bibinfo {pages} {325} (\bibinfo {year} {1997})}\BibitemShut
  {NoStop}%
\bibitem [{\citenamefont {Lloyd}(1996)}]{simulation1}%
  \BibitemOpen
  \bibfield  {author} {\bibinfo {author} {\bibfnamefont {S.}~\bibnamefont
  {Lloyd}},\ }\href {\doibase 10.1126/science.273.5278.1073} {\bibfield
  {journal} {\bibinfo  {journal} {Science}\ }\textbf {\bibinfo {volume}
  {273}},\ \bibinfo {pages} {1073} (\bibinfo {year} {1996})}\BibitemShut
  {NoStop}%
\bibitem [{\citenamefont {Peng}\ \emph {et~al.}(2009)\citenamefont {Peng},
  \citenamefont {Zhang}, \citenamefont {Du},\ and\ \citenamefont
  {Suter}}]{simulation2}%
  \BibitemOpen
  \bibfield  {author} {\bibinfo {author} {\bibfnamefont {X.}~\bibnamefont
  {Peng}}, \bibinfo {author} {\bibfnamefont {J.}~\bibnamefont {Zhang}},
  \bibinfo {author} {\bibfnamefont {J.}~\bibnamefont {Du}}, \ and\ \bibinfo
  {author} {\bibfnamefont {D.}~\bibnamefont {Suter}},\ }\href {\doibase
  10.1103/PhysRevLett.103.140501} {\bibfield  {journal} {\bibinfo  {journal}
  {Phys. Rev. Lett.}\ }\textbf {\bibinfo {volume} {103}},\ \bibinfo {pages}
  {140501} (\bibinfo {year} {2009})}\BibitemShut {NoStop}%
\bibitem [{\citenamefont {Peng}\ and\ \citenamefont
  {Suter}(2010)}]{simulation3}%
  \BibitemOpen
  \bibfield  {author} {\bibinfo {author} {\bibfnamefont {X.-h.}\ \bibnamefont
  {Peng}}\ and\ \bibinfo {author} {\bibfnamefont {D.}~\bibnamefont {Suter}},\
  }\href {\doibase 10.1007/s11467-009-0067-x} {\bibfield  {journal} {\bibinfo
  {journal} {Frontiers of Physics in China}\ }\textbf {\bibinfo {volume} {5}},\
  \bibinfo {pages} {1} (\bibinfo {year} {2010})}\BibitemShut {NoStop}%
\bibitem [{\citenamefont {\'Alvarez}\ and\ \citenamefont
  {Suter}(2010)}]{simulation4}%
  \BibitemOpen
  \bibfield  {author} {\bibinfo {author} {\bibfnamefont {G.~A.}\ \bibnamefont
  {\'Alvarez}}\ and\ \bibinfo {author} {\bibfnamefont {D.}~\bibnamefont
  {Suter}},\ }\href {\doibase 10.1103/PhysRevLett.104.230403} {\bibfield
  {journal} {\bibinfo  {journal} {Phys. Rev. Lett.}\ }\textbf {\bibinfo
  {volume} {104}},\ \bibinfo {pages} {230403} (\bibinfo {year}
  {2010})}\BibitemShut {NoStop}%
\bibitem [{\citenamefont {Peng}\ \emph {et~al.}(2014)\citenamefont {Peng},
  \citenamefont {Luo}, \citenamefont {Zheng}, \citenamefont {Kou},
  \citenamefont {Suter},\ and\ \citenamefont {Du}}]{simulation5}%
  \BibitemOpen
  \bibfield  {author} {\bibinfo {author} {\bibfnamefont {X.}~\bibnamefont
  {Peng}}, \bibinfo {author} {\bibfnamefont {Z.}~\bibnamefont {Luo}}, \bibinfo
  {author} {\bibfnamefont {W.}~\bibnamefont {Zheng}}, \bibinfo {author}
  {\bibfnamefont {S.}~\bibnamefont {Kou}}, \bibinfo {author} {\bibfnamefont
  {D.}~\bibnamefont {Suter}}, \ and\ \bibinfo {author} {\bibfnamefont
  {J.}~\bibnamefont {Du}},\ }\href {\doibase 10.1103/PhysRevLett.113.080404}
  {\bibfield  {journal} {\bibinfo  {journal} {Phys. Rev. Lett.}\ }\textbf
  {\bibinfo {volume} {113}},\ \bibinfo {pages} {080404} (\bibinfo {year}
  {2014})}\BibitemShut {NoStop}%
\bibitem [{\citenamefont {Álvarez}\ \emph {et~al.}(2015)\citenamefont
  {Álvarez}, \citenamefont {Suter},\ and\ \citenamefont
  {Kaiser}}]{simulation6}%
  \BibitemOpen
  \bibfield  {author} {\bibinfo {author} {\bibfnamefont {G.~A.}\ \bibnamefont
  {Álvarez}}, \bibinfo {author} {\bibfnamefont {D.}~\bibnamefont {Suter}}, \
  and\ \bibinfo {author} {\bibfnamefont {R.}~\bibnamefont {Kaiser}},\ }\href
  {\doibase 10.1126/science.1261160} {\bibfield  {journal} {\bibinfo  {journal}
  {Science}\ }\textbf {\bibinfo {volume} {349}},\ \bibinfo {pages} {846}
  (\bibinfo {year} {2015})},\ \Eprint
  {http://arxiv.org/abs/https://www.science.org/doi/pdf/10.1126/science.1261160}
  {https://www.science.org/doi/pdf/10.1126/science.1261160} \BibitemShut
  {NoStop}%
\bibitem [{\citenamefont {Chen}\ \emph {et~al.}(2016)\citenamefont {Chen},
  \citenamefont {Wu}, \citenamefont {Su}, \citenamefont {Cai}, \citenamefont
  {Wang}, \citenamefont {Yang}, \citenamefont {Li}, \citenamefont {Liu},
  \citenamefont {Lu},\ and\ \citenamefont {Pan}}]{simulation7}%
  \BibitemOpen
  \bibfield  {author} {\bibinfo {author} {\bibfnamefont {M.-C.}\ \bibnamefont
  {Chen}}, \bibinfo {author} {\bibfnamefont {D.}~\bibnamefont {Wu}}, \bibinfo
  {author} {\bibfnamefont {Z.-E.}\ \bibnamefont {Su}}, \bibinfo {author}
  {\bibfnamefont {X.-D.}\ \bibnamefont {Cai}}, \bibinfo {author} {\bibfnamefont
  {X.-L.}\ \bibnamefont {Wang}}, \bibinfo {author} {\bibfnamefont
  {T.}~\bibnamefont {Yang}}, \bibinfo {author} {\bibfnamefont {L.}~\bibnamefont
  {Li}}, \bibinfo {author} {\bibfnamefont {N.-L.}\ \bibnamefont {Liu}},
  \bibinfo {author} {\bibfnamefont {C.-Y.}\ \bibnamefont {Lu}}, \ and\ \bibinfo
  {author} {\bibfnamefont {J.-W.}\ \bibnamefont {Pan}},\ }\href {\doibase
  10.1103/PhysRevLett.116.070502} {\bibfield  {journal} {\bibinfo  {journal}
  {Phys. Rev. Lett.}\ }\textbf {\bibinfo {volume} {116}},\ \bibinfo {pages}
  {070502} (\bibinfo {year} {2016})}\BibitemShut {NoStop}%
\bibitem [{\citenamefont {Li}\ \emph {et~al.}(2017)\citenamefont {Li},
  \citenamefont {Fan}, \citenamefont {Wang}, \citenamefont {Ye}, \citenamefont
  {Zeng}, \citenamefont {Zhai}, \citenamefont {Peng},\ and\ \citenamefont
  {Du}}]{simulation8}%
  \BibitemOpen
  \bibfield  {author} {\bibinfo {author} {\bibfnamefont {J.}~\bibnamefont
  {Li}}, \bibinfo {author} {\bibfnamefont {R.}~\bibnamefont {Fan}}, \bibinfo
  {author} {\bibfnamefont {H.}~\bibnamefont {Wang}}, \bibinfo {author}
  {\bibfnamefont {B.}~\bibnamefont {Ye}}, \bibinfo {author} {\bibfnamefont
  {B.}~\bibnamefont {Zeng}}, \bibinfo {author} {\bibfnamefont {H.}~\bibnamefont
  {Zhai}}, \bibinfo {author} {\bibfnamefont {X.}~\bibnamefont {Peng}}, \ and\
  \bibinfo {author} {\bibfnamefont {J.}~\bibnamefont {Du}},\ }\href {\doibase
  10.1103/PhysRevX.7.031011} {\bibfield  {journal} {\bibinfo  {journal} {Phys.
  Rev. X}\ }\textbf {\bibinfo {volume} {7}},\ \bibinfo {pages} {031011}
  (\bibinfo {year} {2017})}\BibitemShut {NoStop}%
\bibitem [{\citenamefont {Aaronson}\ and\ \citenamefont
  {Arkhipov}(2011)}]{bsampling1}%
  \BibitemOpen
  \bibfield  {author} {\bibinfo {author} {\bibfnamefont {S.}~\bibnamefont
  {Aaronson}}\ and\ \bibinfo {author} {\bibfnamefont {A.}~\bibnamefont
  {Arkhipov}},\ }in\ \href {\doibase 10.1145/1993636.1993682} {\emph {\bibinfo
  {booktitle} {Proceedings of the Forty-Third Annual ACM Symposium on Theory of
  Computing}}},\ \bibinfo {series and number} {STOC '11}\ (\bibinfo
  {publisher} {Association for Computing Machinery},\ \bibinfo {address} {New
  York, NY, USA},\ \bibinfo {year} {2011})\ p.\ \bibinfo {pages}
  {333–342}\BibitemShut {NoStop}%
\bibitem [{\citenamefont {Broome}\ \emph {et~al.}(2013)\citenamefont {Broome},
  \citenamefont {Fedrizzi}, \citenamefont {Rahimi-Keshari}, \citenamefont
  {Dove}, \citenamefont {Aaronson}, \citenamefont {Ralph},\ and\ \citenamefont
  {White}}]{bsampling2}%
  \BibitemOpen
  \bibfield  {author} {\bibinfo {author} {\bibfnamefont {M.~A.}\ \bibnamefont
  {Broome}}, \bibinfo {author} {\bibfnamefont {A.}~\bibnamefont {Fedrizzi}},
  \bibinfo {author} {\bibfnamefont {S.}~\bibnamefont {Rahimi-Keshari}},
  \bibinfo {author} {\bibfnamefont {J.}~\bibnamefont {Dove}}, \bibinfo {author}
  {\bibfnamefont {S.}~\bibnamefont {Aaronson}}, \bibinfo {author}
  {\bibfnamefont {T.~C.}\ \bibnamefont {Ralph}}, \ and\ \bibinfo {author}
  {\bibfnamefont {A.~G.}\ \bibnamefont {White}},\ }\href {\doibase
  10.1126/science.1231440} {\bibfield  {journal} {\bibinfo  {journal}
  {Science}\ }\textbf {\bibinfo {volume} {339}},\ \bibinfo {pages} {794}
  (\bibinfo {year} {2013})},\ \Eprint
  {http://arxiv.org/abs/https://www.science.org/doi/pdf/10.1126/science.1231440}
  {https://www.science.org/doi/pdf/10.1126/science.1231440} \BibitemShut
  {NoStop}%
\bibitem [{\citenamefont {Spring}\ \emph {et~al.}(2013)\citenamefont {Spring},
  \citenamefont {Metcalf}, \citenamefont {Humphreys}, \citenamefont
  {Kolthammer}, \citenamefont {Jin}, \citenamefont {Barbieri}, \citenamefont
  {Datta}, \citenamefont {Thomas-Peter}, \citenamefont {Langford},
  \citenamefont {Kundys}, \citenamefont {Gates}, \citenamefont {Smith},
  \citenamefont {Smith},\ and\ \citenamefont {Walmsley}}]{bsampling3}%
  \BibitemOpen
  \bibfield  {author} {\bibinfo {author} {\bibfnamefont {J.~B.}\ \bibnamefont
  {Spring}}, \bibinfo {author} {\bibfnamefont {B.~J.}\ \bibnamefont {Metcalf}},
  \bibinfo {author} {\bibfnamefont {P.~C.}\ \bibnamefont {Humphreys}}, \bibinfo
  {author} {\bibfnamefont {W.~S.}\ \bibnamefont {Kolthammer}}, \bibinfo
  {author} {\bibfnamefont {X.-M.}\ \bibnamefont {Jin}}, \bibinfo {author}
  {\bibfnamefont {M.}~\bibnamefont {Barbieri}}, \bibinfo {author}
  {\bibfnamefont {A.}~\bibnamefont {Datta}}, \bibinfo {author} {\bibfnamefont
  {N.}~\bibnamefont {Thomas-Peter}}, \bibinfo {author} {\bibfnamefont {N.~K.}\
  \bibnamefont {Langford}}, \bibinfo {author} {\bibfnamefont {D.}~\bibnamefont
  {Kundys}}, \bibinfo {author} {\bibfnamefont {J.~C.}\ \bibnamefont {Gates}},
  \bibinfo {author} {\bibfnamefont {B.~J.}\ \bibnamefont {Smith}}, \bibinfo
  {author} {\bibfnamefont {P.~G.~R.}\ \bibnamefont {Smith}}, \ and\ \bibinfo
  {author} {\bibfnamefont {I.~A.}\ \bibnamefont {Walmsley}},\ }\href {\doibase
  10.1126/science.1231692} {\bibfield  {journal} {\bibinfo  {journal}
  {Science}\ }\textbf {\bibinfo {volume} {339}},\ \bibinfo {pages} {798}
  (\bibinfo {year} {2013})},\ \Eprint
  {http://arxiv.org/abs/https://www.science.org/doi/pdf/10.1126/science.1231692}
  {https://www.science.org/doi/pdf/10.1126/science.1231692} \BibitemShut
  {NoStop}%
\bibitem [{\citenamefont {Tillmann}\ \emph {et~al.}(2013)\citenamefont
  {Tillmann}, \citenamefont {Daki{\'c}}, \citenamefont {Heilmann},
  \citenamefont {Nolte}, \citenamefont {Szameit},\ and\ \citenamefont
  {Walther}}]{bsampling4}%
  \BibitemOpen
  \bibfield  {author} {\bibinfo {author} {\bibfnamefont {M.}~\bibnamefont
  {Tillmann}}, \bibinfo {author} {\bibfnamefont {B.}~\bibnamefont {Daki{\'c}}},
  \bibinfo {author} {\bibfnamefont {R.}~\bibnamefont {Heilmann}}, \bibinfo
  {author} {\bibfnamefont {S.}~\bibnamefont {Nolte}}, \bibinfo {author}
  {\bibfnamefont {A.}~\bibnamefont {Szameit}}, \ and\ \bibinfo {author}
  {\bibfnamefont {P.}~\bibnamefont {Walther}},\ }\href {\doibase
  10.1038/nphoton.2013.102} {\bibfield  {journal} {\bibinfo  {journal} {Nature
  Photonics}\ }\textbf {\bibinfo {volume} {7}},\ \bibinfo {pages} {540}
  (\bibinfo {year} {2013})}\BibitemShut {NoStop}%
\bibitem [{\citenamefont {Crespi}\ \emph {et~al.}(2013)\citenamefont {Crespi},
  \citenamefont {Osellame}, \citenamefont {Ramponi}, \citenamefont {Brod},
  \citenamefont {Galv{\~a}o}, \citenamefont {Spagnolo}, \citenamefont
  {Vitelli}, \citenamefont {Maiorino}, \citenamefont {Mataloni},\ and\
  \citenamefont {Sciarrino}}]{bsampling5}%
  \BibitemOpen
  \bibfield  {author} {\bibinfo {author} {\bibfnamefont {A.}~\bibnamefont
  {Crespi}}, \bibinfo {author} {\bibfnamefont {R.}~\bibnamefont {Osellame}},
  \bibinfo {author} {\bibfnamefont {R.}~\bibnamefont {Ramponi}}, \bibinfo
  {author} {\bibfnamefont {D.~J.}\ \bibnamefont {Brod}}, \bibinfo {author}
  {\bibfnamefont {E.~F.}\ \bibnamefont {Galv{\~a}o}}, \bibinfo {author}
  {\bibfnamefont {N.}~\bibnamefont {Spagnolo}}, \bibinfo {author}
  {\bibfnamefont {C.}~\bibnamefont {Vitelli}}, \bibinfo {author} {\bibfnamefont
  {E.}~\bibnamefont {Maiorino}}, \bibinfo {author} {\bibfnamefont
  {P.}~\bibnamefont {Mataloni}}, \ and\ \bibinfo {author} {\bibfnamefont
  {F.}~\bibnamefont {Sciarrino}},\ }\href {\doibase 10.1038/nphoton.2013.112}
  {\bibfield  {journal} {\bibinfo  {journal} {Nature Photonics}\ }\textbf
  {\bibinfo {volume} {7}},\ \bibinfo {pages} {545} (\bibinfo {year}
  {2013})}\BibitemShut {NoStop}%
\bibitem [{\citenamefont {Wang}\ \emph {et~al.}(2017)\citenamefont {Wang},
  \citenamefont {He}, \citenamefont {Li}, \citenamefont {Su}, \citenamefont
  {Li}, \citenamefont {Huang}, \citenamefont {Ding}, \citenamefont {Chen},
  \citenamefont {Liu}, \citenamefont {Qin}, \citenamefont {Li}, \citenamefont
  {He}, \citenamefont {Schneider}, \citenamefont {Kamp}, \citenamefont {Peng},
  \citenamefont {H{\"o}fling}, \citenamefont {Lu},\ and\ \citenamefont
  {Pan}}]{bsampling6}%
  \BibitemOpen
  \bibfield  {author} {\bibinfo {author} {\bibfnamefont {H.}~\bibnamefont
  {Wang}}, \bibinfo {author} {\bibfnamefont {Y.}~\bibnamefont {He}}, \bibinfo
  {author} {\bibfnamefont {Y.-H.}\ \bibnamefont {Li}}, \bibinfo {author}
  {\bibfnamefont {Z.-E.}\ \bibnamefont {Su}}, \bibinfo {author} {\bibfnamefont
  {B.}~\bibnamefont {Li}}, \bibinfo {author} {\bibfnamefont {H.-L.}\
  \bibnamefont {Huang}}, \bibinfo {author} {\bibfnamefont {X.}~\bibnamefont
  {Ding}}, \bibinfo {author} {\bibfnamefont {M.-C.}\ \bibnamefont {Chen}},
  \bibinfo {author} {\bibfnamefont {C.}~\bibnamefont {Liu}}, \bibinfo {author}
  {\bibfnamefont {J.}~\bibnamefont {Qin}}, \bibinfo {author} {\bibfnamefont
  {J.-P.}\ \bibnamefont {Li}}, \bibinfo {author} {\bibfnamefont {Y.-M.}\
  \bibnamefont {He}}, \bibinfo {author} {\bibfnamefont {C.}~\bibnamefont
  {Schneider}}, \bibinfo {author} {\bibfnamefont {M.}~\bibnamefont {Kamp}},
  \bibinfo {author} {\bibfnamefont {C.-Z.}\ \bibnamefont {Peng}}, \bibinfo
  {author} {\bibfnamefont {S.}~\bibnamefont {H{\"o}fling}}, \bibinfo {author}
  {\bibfnamefont {C.-Y.}\ \bibnamefont {Lu}}, \ and\ \bibinfo {author}
  {\bibfnamefont {J.-W.}\ \bibnamefont {Pan}},\ }\href {\doibase
  10.1038/nphoton.2017.63} {\bibfield  {journal} {\bibinfo  {journal} {Nature
  Photonics}\ }\textbf {\bibinfo {volume} {11}},\ \bibinfo {pages} {361}
  (\bibinfo {year} {2017})}\BibitemShut {NoStop}%
\bibitem [{\citenamefont {Harrow}\ \emph {et~al.}(2009)\citenamefont {Harrow},
  \citenamefont {Hassidim},\ and\ \citenamefont {Lloyd}}]{harrow_2009}%
  \BibitemOpen
  \bibfield  {author} {\bibinfo {author} {\bibfnamefont {A.~W.}\ \bibnamefont
  {Harrow}}, \bibinfo {author} {\bibfnamefont {A.}~\bibnamefont {Hassidim}}, \
  and\ \bibinfo {author} {\bibfnamefont {S.}~\bibnamefont {Lloyd}},\ }\href
  {\doibase 10.1103/PhysRevLett.103.150502} {\bibfield  {journal} {\bibinfo
  {journal} {Phys. Rev. Lett.}\ }\textbf {\bibinfo {volume} {103}},\ \bibinfo
  {pages} {150502} (\bibinfo {year} {2009})}\BibitemShut {NoStop}%
\bibitem [{\citenamefont {Cai}\ \emph {et~al.}(2013)\citenamefont {Cai},
  \citenamefont {Weedbrook}, \citenamefont {Su}, \citenamefont {Chen},
  \citenamefont {Gu}, \citenamefont {Zhu}, \citenamefont {Li}, \citenamefont
  {Liu}, \citenamefont {Lu},\ and\ \citenamefont {Pan}}]{Cai_2013}%
  \BibitemOpen
  \bibfield  {author} {\bibinfo {author} {\bibfnamefont {X.-D.}\ \bibnamefont
  {Cai}}, \bibinfo {author} {\bibfnamefont {C.}~\bibnamefont {Weedbrook}},
  \bibinfo {author} {\bibfnamefont {Z.-E.}\ \bibnamefont {Su}}, \bibinfo
  {author} {\bibfnamefont {M.-C.}\ \bibnamefont {Chen}}, \bibinfo {author}
  {\bibfnamefont {M.}~\bibnamefont {Gu}}, \bibinfo {author} {\bibfnamefont
  {M.-J.}\ \bibnamefont {Zhu}}, \bibinfo {author} {\bibfnamefont
  {L.}~\bibnamefont {Li}}, \bibinfo {author} {\bibfnamefont {N.-L.}\
  \bibnamefont {Liu}}, \bibinfo {author} {\bibfnamefont {C.-Y.}\ \bibnamefont
  {Lu}}, \ and\ \bibinfo {author} {\bibfnamefont {J.-W.}\ \bibnamefont {Pan}},\
  }\href {\doibase 10.1103/PhysRevLett.110.230501} {\bibfield  {journal}
  {\bibinfo  {journal} {Phys. Rev. Lett.}\ }\textbf {\bibinfo {volume} {110}},\
  \bibinfo {pages} {230501} (\bibinfo {year} {2013})}\BibitemShut {NoStop}%
\bibitem [{\citenamefont {Pan}\ \emph {et~al.}(2014)\citenamefont {Pan},
  \citenamefont {Cao}, \citenamefont {Yao}, \citenamefont {Li}, \citenamefont
  {Ju}, \citenamefont {Chen}, \citenamefont {Peng}, \citenamefont {Kais},\ and\
  \citenamefont {Du}}]{Pan_2014}%
  \BibitemOpen
  \bibfield  {author} {\bibinfo {author} {\bibfnamefont {J.}~\bibnamefont
  {Pan}}, \bibinfo {author} {\bibfnamefont {Y.}~\bibnamefont {Cao}}, \bibinfo
  {author} {\bibfnamefont {X.}~\bibnamefont {Yao}}, \bibinfo {author}
  {\bibfnamefont {Z.}~\bibnamefont {Li}}, \bibinfo {author} {\bibfnamefont
  {C.}~\bibnamefont {Ju}}, \bibinfo {author} {\bibfnamefont {H.}~\bibnamefont
  {Chen}}, \bibinfo {author} {\bibfnamefont {X.}~\bibnamefont {Peng}}, \bibinfo
  {author} {\bibfnamefont {S.}~\bibnamefont {Kais}}, \ and\ \bibinfo {author}
  {\bibfnamefont {J.}~\bibnamefont {Du}},\ }\href {\doibase
  10.1103/PhysRevA.89.022313} {\bibfield  {journal} {\bibinfo  {journal} {Phys.
  Rev. A}\ }\textbf {\bibinfo {volume} {89}},\ \bibinfo {pages} {022313}
  (\bibinfo {year} {2014})}\BibitemShut {NoStop}%
\bibitem [{\citenamefont {Barz}\ \emph {et~al.}(2017)\citenamefont {Barz},
  \citenamefont {Kassal}, \citenamefont {Ringbauer}, \citenamefont {Lipp},
  \citenamefont {Daki{\'c}}, \citenamefont {Aspuru-Guzik},\ and\ \citenamefont
  {Walther}}]{barz_2017}%
  \BibitemOpen
  \bibfield  {author} {\bibinfo {author} {\bibfnamefont {S.}~\bibnamefont
  {Barz}}, \bibinfo {author} {\bibfnamefont {I.}~\bibnamefont {Kassal}},
  \bibinfo {author} {\bibfnamefont {M.}~\bibnamefont {Ringbauer}}, \bibinfo
  {author} {\bibfnamefont {Y.~O.}\ \bibnamefont {Lipp}}, \bibinfo {author}
  {\bibfnamefont {B.}~\bibnamefont {Daki{\'c}}}, \bibinfo {author}
  {\bibfnamefont {A.}~\bibnamefont {Aspuru-Guzik}}, \ and\ \bibinfo {author}
  {\bibfnamefont {P.}~\bibnamefont {Walther}},\ }\href {\doibase
  10.1038/srep42653} {\bibfield  {journal} {\bibinfo  {journal} {Scientific
  reports}\ }\textbf {\bibinfo {volume} {7}},\ \bibinfo {pages} {42653}
  (\bibinfo {year} {2017})}\BibitemShut {NoStop}%
\bibitem [{\citenamefont {Zheng}\ \emph {et~al.}(2017)\citenamefont {Zheng},
  \citenamefont {Song}, \citenamefont {Chen}, \citenamefont {Xia},
  \citenamefont {Liu}, \citenamefont {Guo}, \citenamefont {Zhang},
  \citenamefont {Xu}, \citenamefont {Deng}, \citenamefont {Huang},
  \citenamefont {Wu}, \citenamefont {Yan}, \citenamefont {Zheng}, \citenamefont
  {Lu}, \citenamefont {Pan}, \citenamefont {Wang}, \citenamefont {Lu},\ and\
  \citenamefont {Zhu}}]{zheng_2017}%
  \BibitemOpen
  \bibfield  {author} {\bibinfo {author} {\bibfnamefont {Y.}~\bibnamefont
  {Zheng}}, \bibinfo {author} {\bibfnamefont {C.}~\bibnamefont {Song}},
  \bibinfo {author} {\bibfnamefont {M.-C.}\ \bibnamefont {Chen}}, \bibinfo
  {author} {\bibfnamefont {B.}~\bibnamefont {Xia}}, \bibinfo {author}
  {\bibfnamefont {W.}~\bibnamefont {Liu}}, \bibinfo {author} {\bibfnamefont
  {Q.}~\bibnamefont {Guo}}, \bibinfo {author} {\bibfnamefont {L.}~\bibnamefont
  {Zhang}}, \bibinfo {author} {\bibfnamefont {D.}~\bibnamefont {Xu}}, \bibinfo
  {author} {\bibfnamefont {H.}~\bibnamefont {Deng}}, \bibinfo {author}
  {\bibfnamefont {K.}~\bibnamefont {Huang}}, \bibinfo {author} {\bibfnamefont
  {Y.}~\bibnamefont {Wu}}, \bibinfo {author} {\bibfnamefont {Z.}~\bibnamefont
  {Yan}}, \bibinfo {author} {\bibfnamefont {D.}~\bibnamefont {Zheng}}, \bibinfo
  {author} {\bibfnamefont {L.}~\bibnamefont {Lu}}, \bibinfo {author}
  {\bibfnamefont {J.-W.}\ \bibnamefont {Pan}}, \bibinfo {author} {\bibfnamefont
  {H.}~\bibnamefont {Wang}}, \bibinfo {author} {\bibfnamefont {C.-Y.}\
  \bibnamefont {Lu}}, \ and\ \bibinfo {author} {\bibfnamefont {X.}~\bibnamefont
  {Zhu}},\ }\href {\doibase 10.1103/PhysRevLett.118.210504} {\bibfield
  {journal} {\bibinfo  {journal} {Phys. Rev. Lett.}\ }\textbf {\bibinfo
  {volume} {118}},\ \bibinfo {pages} {210504} (\bibinfo {year}
  {2017})}\BibitemShut {NoStop}%
\bibitem [{\citenamefont {Manzano}\ \emph {et~al.}(2009)\citenamefont
  {Manzano}, \citenamefont {Paw{\l}owski},\ and\ \citenamefont
  {Brukner}}]{Manzano_2009}%
  \BibitemOpen
  \bibfield  {author} {\bibinfo {author} {\bibfnamefont {D.}~\bibnamefont
  {Manzano}}, \bibinfo {author} {\bibfnamefont {M.}~\bibnamefont
  {Paw{\l}owski}}, \ and\ \bibinfo {author} {\bibfnamefont
  {{\v{C}}.}~\bibnamefont {Brukner}},\ }\href {\doibase
  10.1088/1367-2630/11/11/113018} {\bibfield  {journal} {\bibinfo  {journal}
  {New Journal of Physics}\ }\textbf {\bibinfo {volume} {11}},\ \bibinfo
  {pages} {113018} (\bibinfo {year} {2009})}\BibitemShut {NoStop}%
\bibitem [{\citenamefont {Lloyd}\ \emph {et~al.}(2013)\citenamefont {Lloyd},
  \citenamefont {Mohseni},\ and\ \citenamefont
  {Rebentrost}}]{lloyd2013quantum}%
  \BibitemOpen
  \bibfield  {author} {\bibinfo {author} {\bibfnamefont {S.}~\bibnamefont
  {Lloyd}}, \bibinfo {author} {\bibfnamefont {M.}~\bibnamefont {Mohseni}}, \
  and\ \bibinfo {author} {\bibfnamefont {P.}~\bibnamefont {Rebentrost}},\
  }\href@noop {} {\enquote {\bibinfo {title} {Quantum algorithms for supervised
  and unsupervised machine learning},}\ } (\bibinfo {year} {2013}),\ \Eprint
  {http://arxiv.org/abs/1307.0411} {arXiv:1307.0411 [quant-ph]} \BibitemShut
  {NoStop}%
\bibitem [{\citenamefont {Rebentrost}\ \emph {et~al.}(2014)\citenamefont
  {Rebentrost}, \citenamefont {Mohseni},\ and\ \citenamefont
  {Lloyd}}]{rebentrost_2014}%
  \BibitemOpen
  \bibfield  {author} {\bibinfo {author} {\bibfnamefont {P.}~\bibnamefont
  {Rebentrost}}, \bibinfo {author} {\bibfnamefont {M.}~\bibnamefont {Mohseni}},
  \ and\ \bibinfo {author} {\bibfnamefont {S.}~\bibnamefont {Lloyd}},\ }\href
  {\doibase 10.1103/PhysRevLett.113.130503} {\bibfield  {journal} {\bibinfo
  {journal} {Phys. Rev. Lett.}\ }\textbf {\bibinfo {volume} {113}},\ \bibinfo
  {pages} {130503} (\bibinfo {year} {2014})}\BibitemShut {NoStop}%
\bibitem [{\citenamefont {Cai}\ \emph {et~al.}(2015)\citenamefont {Cai},
  \citenamefont {Wu}, \citenamefont {Su}, \citenamefont {Chen}, \citenamefont
  {Wang}, \citenamefont {Li}, \citenamefont {Liu}, \citenamefont {Lu},\ and\
  \citenamefont {Pan}}]{qmachine1}%
  \BibitemOpen
  \bibfield  {author} {\bibinfo {author} {\bibfnamefont {X.-D.}\ \bibnamefont
  {Cai}}, \bibinfo {author} {\bibfnamefont {D.}~\bibnamefont {Wu}}, \bibinfo
  {author} {\bibfnamefont {Z.-E.}\ \bibnamefont {Su}}, \bibinfo {author}
  {\bibfnamefont {M.-C.}\ \bibnamefont {Chen}}, \bibinfo {author}
  {\bibfnamefont {X.-L.}\ \bibnamefont {Wang}}, \bibinfo {author}
  {\bibfnamefont {L.}~\bibnamefont {Li}}, \bibinfo {author} {\bibfnamefont
  {N.-L.}\ \bibnamefont {Liu}}, \bibinfo {author} {\bibfnamefont {C.-Y.}\
  \bibnamefont {Lu}}, \ and\ \bibinfo {author} {\bibfnamefont {J.-W.}\
  \bibnamefont {Pan}},\ }\href {\doibase 10.1103/PhysRevLett.114.110504}
  {\bibfield  {journal} {\bibinfo  {journal} {Phys. Rev. Lett.}\ }\textbf
  {\bibinfo {volume} {114}},\ \bibinfo {pages} {110504} (\bibinfo {year}
  {2015})}\BibitemShut {NoStop}%
\bibitem [{\citenamefont {Li}\ \emph {et~al.}(2015)\citenamefont {Li},
  \citenamefont {Liu}, \citenamefont {Xu},\ and\ \citenamefont
  {Du}}]{qmachine2}%
  \BibitemOpen
  \bibfield  {author} {\bibinfo {author} {\bibfnamefont {Z.}~\bibnamefont
  {Li}}, \bibinfo {author} {\bibfnamefont {X.}~\bibnamefont {Liu}}, \bibinfo
  {author} {\bibfnamefont {N.}~\bibnamefont {Xu}}, \ and\ \bibinfo {author}
  {\bibfnamefont {J.}~\bibnamefont {Du}},\ }\href {\doibase
  10.1103/PhysRevLett.114.140504} {\bibfield  {journal} {\bibinfo  {journal}
  {Phys. Rev. Lett.}\ }\textbf {\bibinfo {volume} {114}},\ \bibinfo {pages}
  {140504} (\bibinfo {year} {2015})}\BibitemShut {NoStop}%
\bibitem [{\citenamefont {Rist{\`e}}\ \emph {et~al.}(2017)\citenamefont
  {Rist{\`e}}, \citenamefont {da~Silva}, \citenamefont {Ryan}, \citenamefont
  {Cross}, \citenamefont {C{\'o}rcoles}, \citenamefont {Smolin}, \citenamefont
  {Gambetta}, \citenamefont {Chow},\ and\ \citenamefont {Johnson}}]{qmachine3}%
  \BibitemOpen
  \bibfield  {author} {\bibinfo {author} {\bibfnamefont {D.}~\bibnamefont
  {Rist{\`e}}}, \bibinfo {author} {\bibfnamefont {M.~P.}\ \bibnamefont
  {da~Silva}}, \bibinfo {author} {\bibfnamefont {C.~A.}\ \bibnamefont {Ryan}},
  \bibinfo {author} {\bibfnamefont {A.~W.}\ \bibnamefont {Cross}}, \bibinfo
  {author} {\bibfnamefont {A.~D.}\ \bibnamefont {C{\'o}rcoles}}, \bibinfo
  {author} {\bibfnamefont {J.~A.}\ \bibnamefont {Smolin}}, \bibinfo {author}
  {\bibfnamefont {J.~M.}\ \bibnamefont {Gambetta}}, \bibinfo {author}
  {\bibfnamefont {J.~M.}\ \bibnamefont {Chow}}, \ and\ \bibinfo {author}
  {\bibfnamefont {B.~R.}\ \bibnamefont {Johnson}},\ }\href {\doibase
  10.1038/s41534-017-0017-3} {\bibfield  {journal} {\bibinfo  {journal} {npj
  Quantum Information}\ }\textbf {\bibinfo {volume} {3}},\ \bibinfo {pages}
  {16} (\bibinfo {year} {2017})}\BibitemShut {NoStop}%
\bibitem [{\citenamefont {Dunjko}\ \emph {et~al.}(2016)\citenamefont {Dunjko},
  \citenamefont {Taylor},\ and\ \citenamefont {Briegel}}]{qmachine4}%
  \BibitemOpen
  \bibfield  {author} {\bibinfo {author} {\bibfnamefont {V.}~\bibnamefont
  {Dunjko}}, \bibinfo {author} {\bibfnamefont {J.~M.}\ \bibnamefont {Taylor}},
  \ and\ \bibinfo {author} {\bibfnamefont {H.~J.}\ \bibnamefont {Briegel}},\
  }\href {\doibase 10.1103/PhysRevLett.117.130501} {\bibfield  {journal}
  {\bibinfo  {journal} {Phys. Rev. Lett.}\ }\textbf {\bibinfo {volume} {117}},\
  \bibinfo {pages} {130501} (\bibinfo {year} {2016})}\BibitemShut {NoStop}%
\bibitem [{\citenamefont {Buffoni}\ and\ \citenamefont
  {Caruso}(2020)}]{buffoni_2020}%
  \BibitemOpen
  \bibfield  {author} {\bibinfo {author} {\bibfnamefont {L.}~\bibnamefont
  {Buffoni}}\ and\ \bibinfo {author} {\bibfnamefont {F.}~\bibnamefont
  {Caruso}},\ }\href {\doibase 10.1209/0295-5075/132/60004} {\bibfield
  {journal} {\bibinfo  {journal} {EPL}\ }\textbf {\bibinfo {volume} {132}},\
  \bibinfo {pages} {60004} (\bibinfo {year} {2020})}\BibitemShut {NoStop}%
\bibitem [{\citenamefont {Arute~et al.}(2019)}]{qsupremacy_google}%
  \BibitemOpen
  \bibfield  {author} {\bibinfo {author} {\bibfnamefont {F.}~\bibnamefont
  {Arute~et al.}},\ }\href {\doibase 10.1038/s41586-019-1666-5} {\bibfield
  {journal} {\bibinfo  {journal} {Nature}\ }\textbf {\bibinfo {volume} {574}},\
  \bibinfo {pages} {505} (\bibinfo {year} {2019})}\BibitemShut {NoStop}%
\bibitem [{\citenamefont {et~al.}(2020)}]{china_supremacy1}%
  \BibitemOpen
  \bibfield  {author} {\bibinfo {author} {\bibfnamefont {H.-S.~Z.}\
  \bibnamefont {et~al.}},\ }\href {\doibase 10.1126/science.abe8770} {\bibfield
   {journal} {\bibinfo  {journal} {Science}\ }\textbf {\bibinfo {volume}
  {370}},\ \bibinfo {pages} {1460} (\bibinfo {year} {2020})}\BibitemShut
  {NoStop}%
\bibitem [{\citenamefont {Zhong~et al.}(2021)}]{china_supremacy2}%
  \BibitemOpen
  \bibfield  {author} {\bibinfo {author} {\bibfnamefont {H.-S.}\ \bibnamefont
  {Zhong~et al.}},\ }\href {\doibase 10.1103/PhysRevLett.127.180502} {\bibfield
   {journal} {\bibinfo  {journal} {Phys. Rev. Lett.}\ }\textbf {\bibinfo
  {volume} {127}},\ \bibinfo {pages} {180502} (\bibinfo {year}
  {2021})}\BibitemShut {NoStop}%
\bibitem [{\citenamefont {Wu~et al.}(2021)}]{china_supremacy3}%
  \BibitemOpen
  \bibfield  {author} {\bibinfo {author} {\bibfnamefont {Y.}~\bibnamefont
  {Wu~et al.}},\ }\href {\doibase 10.1103/PhysRevLett.127.180501} {\bibfield
  {journal} {\bibinfo  {journal} {Phys. Rev. Lett.}\ }\textbf {\bibinfo
  {volume} {127}},\ \bibinfo {pages} {180501} (\bibinfo {year}
  {2021})}\BibitemShut {NoStop}%
\bibitem [{ibm()}]{ibmq}%
  \BibitemOpen
  \href@noop {} {}\bibinfo {howpublished}
  {\url{https://www.ibm.com/quantum-computing/}}\BibitemShut {NoStop}%
\bibitem [{\citenamefont {Preskill}(2018)}]{Preskill2018quantumcomputingin}%
  \BibitemOpen
  \bibfield  {author} {\bibinfo {author} {\bibfnamefont {J.}~\bibnamefont
  {Preskill}},\ }\href {\doibase 10.22331/q-2018-08-06-79} {\bibfield
  {journal} {\bibinfo  {journal} {{Quantum}}\ }\textbf {\bibinfo {volume}
  {2}},\ \bibinfo {pages} {79} (\bibinfo {year} {2018})}\BibitemShut {NoStop}%
\bibitem [{\citenamefont {Zhang}\ \emph
  {et~al.}(2015{\natexlab{a}})\citenamefont {Zhang}, \citenamefont {Lu},\ and\
  \citenamefont {Gao}}]{qsobel_2015}%
  \BibitemOpen
  \bibfield  {author} {\bibinfo {author} {\bibfnamefont {Y.}~\bibnamefont
  {Zhang}}, \bibinfo {author} {\bibfnamefont {K.}~\bibnamefont {Lu}}, \ and\
  \bibinfo {author} {\bibfnamefont {Y.}~\bibnamefont {Gao}},\ }\href {\doibase
  10.1007/s11432-014-5158-9} {\bibfield  {journal} {\bibinfo  {journal}
  {Science China Information Sciences}\ }\textbf {\bibinfo {volume} {58}},\
  \bibinfo {pages} {1} (\bibinfo {year} {2015}{\natexlab{a}})}\BibitemShut
  {NoStop}%
\bibitem [{\citenamefont {Yao}\ \emph {et~al.}(2017)\citenamefont {Yao},
  \citenamefont {Wang}, \citenamefont {Liao}, \citenamefont {Chen},
  \citenamefont {Pan}, \citenamefont {Li}, \citenamefont {Zhang}, \citenamefont
  {Lin}, \citenamefont {Wang}, \citenamefont {Luo}, \citenamefont {Zheng},
  \citenamefont {Li}, \citenamefont {Zhao}, \citenamefont {Peng},\ and\
  \citenamefont {Suter}}]{suter_2017}%
  \BibitemOpen
  \bibfield  {author} {\bibinfo {author} {\bibfnamefont {X.-W.}\ \bibnamefont
  {Yao}}, \bibinfo {author} {\bibfnamefont {H.}~\bibnamefont {Wang}}, \bibinfo
  {author} {\bibfnamefont {Z.}~\bibnamefont {Liao}}, \bibinfo {author}
  {\bibfnamefont {M.-C.}\ \bibnamefont {Chen}}, \bibinfo {author}
  {\bibfnamefont {J.}~\bibnamefont {Pan}}, \bibinfo {author} {\bibfnamefont
  {J.}~\bibnamefont {Li}}, \bibinfo {author} {\bibfnamefont {K.}~\bibnamefont
  {Zhang}}, \bibinfo {author} {\bibfnamefont {X.}~\bibnamefont {Lin}}, \bibinfo
  {author} {\bibfnamefont {Z.}~\bibnamefont {Wang}}, \bibinfo {author}
  {\bibfnamefont {Z.}~\bibnamefont {Luo}}, \bibinfo {author} {\bibfnamefont
  {W.}~\bibnamefont {Zheng}}, \bibinfo {author} {\bibfnamefont
  {J.}~\bibnamefont {Li}}, \bibinfo {author} {\bibfnamefont {M.}~\bibnamefont
  {Zhao}}, \bibinfo {author} {\bibfnamefont {X.}~\bibnamefont {Peng}}, \ and\
  \bibinfo {author} {\bibfnamefont {D.}~\bibnamefont {Suter}},\ }\href
  {\doibase 10.1103/PhysRevX.7.031041} {\bibfield  {journal} {\bibinfo
  {journal} {Phys. Rev. X}\ }\textbf {\bibinfo {volume} {7}},\ \bibinfo {pages}
  {031041} (\bibinfo {year} {2017})}\BibitemShut {NoStop}%
\bibitem [{\citenamefont {Cavalieri}\ and\ \citenamefont
  {Maio}(2020)}]{Cavalieri2020AQE}%
  \BibitemOpen
  \bibfield  {author} {\bibinfo {author} {\bibfnamefont {G.}~\bibnamefont
  {Cavalieri}}\ and\ \bibinfo {author} {\bibfnamefont {D.}~\bibnamefont
  {Maio}},\ }\href@noop {} {\enquote {\bibinfo {title} {A quantum edge
  detection algorithm},}\ } (\bibinfo {year} {2020}),\ \Eprint
  {http://arxiv.org/abs/2012.11036} {arXiv:2012.11036 [quant-ph]} \BibitemShut
  {NoStop}%
\bibitem [{\citenamefont {Xu}\ \emph {et~al.}(2020)\citenamefont {Xu},
  \citenamefont {He}, \citenamefont {Qiu},\ and\ \citenamefont {Ma}}]{Xu:20}%
  \BibitemOpen
  \bibfield  {author} {\bibinfo {author} {\bibfnamefont {P.}~\bibnamefont
  {Xu}}, \bibinfo {author} {\bibfnamefont {Z.}~\bibnamefont {He}}, \bibinfo
  {author} {\bibfnamefont {T.}~\bibnamefont {Qiu}}, \ and\ \bibinfo {author}
  {\bibfnamefont {H.}~\bibnamefont {Ma}},\ }\href {\doibase 10.1364/OE.386283}
  {\bibfield  {journal} {\bibinfo  {journal} {Opt. Express}\ }\textbf {\bibinfo
  {volume} {28}},\ \bibinfo {pages} {12508} (\bibinfo {year}
  {2020})}\BibitemShut {NoStop}%
\bibitem [{\citenamefont {Trugenberger}(2002)}]{Trugenberger2002}%
  \BibitemOpen
  \bibfield  {author} {\bibinfo {author} {\bibfnamefont {C.~A.}\ \bibnamefont
  {Trugenberger}},\ }\href@noop {} {\bibfield  {journal} {\bibinfo  {journal}
  {Quantum Information Processing}\ }\textbf {\bibinfo {volume} {1}},\ \bibinfo
  {pages} {471} (\bibinfo {year} {2002})}\BibitemShut {NoStop}%
\bibitem [{\citenamefont {Sch\"utzhold}(2003)}]{schutzhold_2003}%
  \BibitemOpen
  \bibfield  {author} {\bibinfo {author} {\bibfnamefont {R.}~\bibnamefont
  {Sch\"utzhold}},\ }\href {\doibase 10.1103/PhysRevA.67.062311} {\bibfield
  {journal} {\bibinfo  {journal} {Phys. Rev. A}\ }\textbf {\bibinfo {volume}
  {67}},\ \bibinfo {pages} {062311} (\bibinfo {year} {2003})}\BibitemShut
  {NoStop}%
\bibitem [{\citenamefont {Pham}\ and\ \citenamefont
  {Park}(2014)}]{petruccione_book2014}%
  \BibitemOpen
  \bibfield  {author} {\bibinfo {author} {\bibfnamefont {D.}~\bibnamefont
  {Pham}}\ and\ \bibinfo {author} {\bibfnamefont {S.}~\bibnamefont {Park}},\
  }\href {https://books.google.it/books?id=1vdWBQAAQBAJ} {\emph {\bibinfo
  {title} {PRICAI 2014: Trends in Artificial Intelligence: 13th Pacific Rim
  International Conference on Artificial Intelligence, PRICAI 2014, Gold Coast,
  QLD, Australia, December 1-5, 2014, Proceedings}}},\ Lecture Notes in
  Computer Science\ (\bibinfo  {publisher} {Springer International
  Publishing},\ \bibinfo {year} {2014})\BibitemShut {NoStop}%
\bibitem [{\citenamefont {Prousalis}\ and\ \citenamefont
  {Konofaos}(2019)}]{prousalis_2019}%
  \BibitemOpen
  \bibfield  {author} {\bibinfo {author} {\bibfnamefont {K.}~\bibnamefont
  {Prousalis}}\ and\ \bibinfo {author} {\bibfnamefont {N.}~\bibnamefont
  {Konofaos}},\ }\href {\doibase 10.1038/s41598-019-43697-3} {\bibfield
  {journal} {\bibinfo  {journal} {Scientific Reports}\ }\textbf {\bibinfo
  {volume} {9}},\ \bibinfo {pages} {7226} (\bibinfo {year} {2019})}\BibitemShut
  {NoStop}%
\bibitem [{\citenamefont {Banchi}\ \emph {et~al.}(2020)\citenamefont {Banchi},
  \citenamefont {Zhuang},\ and\ \citenamefont {Pirandola}}]{banchi2020}%
  \BibitemOpen
  \bibfield  {author} {\bibinfo {author} {\bibfnamefont {L.}~\bibnamefont
  {Banchi}}, \bibinfo {author} {\bibfnamefont {Q.}~\bibnamefont {Zhuang}}, \
  and\ \bibinfo {author} {\bibfnamefont {S.}~\bibnamefont {Pirandola}},\ }\href
  {\doibase 10.1103/PhysRevApplied.14.064026} {\bibfield  {journal} {\bibinfo
  {journal} {Phys. Rev. Applied}\ }\textbf {\bibinfo {volume} {14}},\ \bibinfo
  {pages} {064026} (\bibinfo {year} {2020})}\BibitemShut {NoStop}%
\bibitem [{\citenamefont {Neigovzen}\ \emph {et~al.}(2009)\citenamefont
  {Neigovzen}, \citenamefont {Neves}, \citenamefont {Sollacher},\ and\
  \citenamefont {Glaser}}]{neigovzen_2009}%
  \BibitemOpen
  \bibfield  {author} {\bibinfo {author} {\bibfnamefont {R.}~\bibnamefont
  {Neigovzen}}, \bibinfo {author} {\bibfnamefont {J.~L.}\ \bibnamefont
  {Neves}}, \bibinfo {author} {\bibfnamefont {R.}~\bibnamefont {Sollacher}}, \
  and\ \bibinfo {author} {\bibfnamefont {S.~J.}\ \bibnamefont {Glaser}},\
  }\href {\doibase 10.1103/PhysRevA.79.042321} {\bibfield  {journal} {\bibinfo
  {journal} {Phys. Rev. A}\ }\textbf {\bibinfo {volume} {79}},\ \bibinfo
  {pages} {042321} (\bibinfo {year} {2009})}\BibitemShut {NoStop}%
\bibitem [{\citenamefont {Montanaro}(2015)}]{montanaro2015}%
  \BibitemOpen
  \bibfield  {author} {\bibinfo {author} {\bibfnamefont {A.}~\bibnamefont
  {Montanaro}},\ }\href@noop {} {\enquote {\bibinfo {title} {Quantum pattern
  matching fast on average},}\ } (\bibinfo {year} {2015}),\ \Eprint
  {http://arxiv.org/abs/1408.1816} {arXiv:1408.1816 [quant-ph]} \BibitemShut
  {NoStop}%
\bibitem [{\citenamefont {Zhang}\ \emph
  {et~al.}(2015{\natexlab{b}})\citenamefont {Zhang}, \citenamefont {Lu},
  \citenamefont {Xu}, \citenamefont {Gao},\ and\ \citenamefont
  {Wilson}}]{Zhang2015LocalFP}%
  \BibitemOpen
  \bibfield  {author} {\bibinfo {author} {\bibfnamefont {Y.}~\bibnamefont
  {Zhang}}, \bibinfo {author} {\bibfnamefont {K.}~\bibnamefont {Lu}}, \bibinfo
  {author} {\bibfnamefont {K.}~\bibnamefont {Xu}}, \bibinfo {author}
  {\bibfnamefont {Y.}~\bibnamefont {Gao}}, \ and\ \bibinfo {author}
  {\bibfnamefont {R.~C.}\ \bibnamefont {Wilson}},\ }\href@noop {} {\bibfield
  {journal} {\bibinfo  {journal} {Quantum Information Processing}\ }\textbf
  {\bibinfo {volume} {14}},\ \bibinfo {pages} {1573} (\bibinfo {year}
  {2015}{\natexlab{b}})}\BibitemShut {NoStop}%
\bibitem [{\citenamefont {Jiang}\ \emph {et~al.}(2016)\citenamefont {Jiang},
  \citenamefont {Dang},\ and\ \citenamefont {Wang}}]{jiang_2016}%
  \BibitemOpen
  \bibfield  {author} {\bibinfo {author} {\bibfnamefont {N.}~\bibnamefont
  {Jiang}}, \bibinfo {author} {\bibfnamefont {Y.}~\bibnamefont {Dang}}, \ and\
  \bibinfo {author} {\bibfnamefont {J.}~\bibnamefont {Wang}},\ }\href {\doibase
  10.1007/s11128-016-1364-2} {\bibfield  {journal} {\bibinfo  {journal}
  {Quantum Information Processing}\ }\textbf {\bibinfo {volume} {15}},\
  \bibinfo {pages} {3543} (\bibinfo {year} {2016})}\BibitemShut {NoStop}%
\bibitem [{\citenamefont {Soni}\ and\ \citenamefont
  {Rasool}(2020)}]{soni_2020}%
  \BibitemOpen
  \bibfield  {author} {\bibinfo {author} {\bibfnamefont {K.~K.}\ \bibnamefont
  {Soni}}\ and\ \bibinfo {author} {\bibfnamefont {A.}~\bibnamefont {Rasool}},\
  }\href {\doibase https://doi.org/10.1016/j.procs.2020.03.230} {\bibfield
  {journal} {\bibinfo  {journal} {Procedia Computer Science}\ }\textbf
  {\bibinfo {volume} {167}},\ \bibinfo {pages} {1991} (\bibinfo {year}
  {2020})},\ \bibinfo {note} {international Conference on Computational
  Intelligence and Data Science}\BibitemShut {NoStop}%
\bibitem [{\citenamefont {Tezuka}\ \emph {et~al.}(2022)\citenamefont {Tezuka},
  \citenamefont {Nakaji}, \citenamefont {Satoh},\ and\ \citenamefont
  {Yamamoto}}]{tezuka_2022}%
  \BibitemOpen
  \bibfield  {author} {\bibinfo {author} {\bibfnamefont {H.}~\bibnamefont
  {Tezuka}}, \bibinfo {author} {\bibfnamefont {K.}~\bibnamefont {Nakaji}},
  \bibinfo {author} {\bibfnamefont {T.}~\bibnamefont {Satoh}}, \ and\ \bibinfo
  {author} {\bibfnamefont {N.}~\bibnamefont {Yamamoto}},\ }\href {\doibase
  10.1103/PhysRevA.105.032440} {\bibfield  {journal} {\bibinfo  {journal}
  {Phys. Rev. A}\ }\textbf {\bibinfo {volume} {105}},\ \bibinfo {pages}
  {032440} (\bibinfo {year} {2022})}\BibitemShut {NoStop}%
\bibitem [{\citenamefont {Kapoor}\ \emph {et~al.}(2016)\citenamefont {Kapoor},
  \citenamefont {Wiebe},\ and\ \citenamefont {Svore}}]{qmachinepattern1}%
  \BibitemOpen
  \bibfield  {author} {\bibinfo {author} {\bibfnamefont {A.}~\bibnamefont
  {Kapoor}}, \bibinfo {author} {\bibfnamefont {N.}~\bibnamefont {Wiebe}}, \
  and\ \bibinfo {author} {\bibfnamefont {K.}~\bibnamefont {Svore}},\ }in\ \href
  {https://proceedings.neurips.cc/paper/2016/file/d47268e9db2e9aa3827bba3afb7ff94a-Paper.pdf}
  {\emph {\bibinfo {booktitle} {Advances in Neural Information Processing
  Systems}}},\ Vol.~\bibinfo {volume} {29},\ \bibinfo {editor} {edited by\
  \bibinfo {editor} {\bibfnamefont {D.}~\bibnamefont {Lee}}, \bibinfo {editor}
  {\bibfnamefont {M.}~\bibnamefont {Sugiyama}}, \bibinfo {editor}
  {\bibfnamefont {U.}~\bibnamefont {Luxburg}}, \bibinfo {editor} {\bibfnamefont
  {I.}~\bibnamefont {Guyon}}, \ and\ \bibinfo {editor} {\bibfnamefont
  {R.}~\bibnamefont {Garnett}}}\ (\bibinfo  {publisher} {Curran Associates,
  Inc.},\ \bibinfo {year} {2016})\BibitemShut {NoStop}%
\bibitem [{\citenamefont {Benedetti}\ \emph {et~al.}(2017)\citenamefont
  {Benedetti}, \citenamefont {Realpe-G\'omez}, \citenamefont {Biswas},\ and\
  \citenamefont {Perdomo-Ortiz}}]{qmachinepattern2}%
  \BibitemOpen
  \bibfield  {author} {\bibinfo {author} {\bibfnamefont {M.}~\bibnamefont
  {Benedetti}}, \bibinfo {author} {\bibfnamefont {J.}~\bibnamefont
  {Realpe-G\'omez}}, \bibinfo {author} {\bibfnamefont {R.}~\bibnamefont
  {Biswas}}, \ and\ \bibinfo {author} {\bibfnamefont {A.}~\bibnamefont
  {Perdomo-Ortiz}},\ }\href {\doibase 10.1103/PhysRevX.7.041052} {\bibfield
  {journal} {\bibinfo  {journal} {Phys. Rev. X}\ }\textbf {\bibinfo {volume}
  {7}},\ \bibinfo {pages} {041052} (\bibinfo {year} {2017})}\BibitemShut
  {NoStop}%
\bibitem [{\citenamefont {Denchev}\ \emph {et~al.}(2012)\citenamefont
  {Denchev}, \citenamefont {Ding}, \citenamefont {Vishwanathan},\ and\
  \citenamefont {Neven}}]{qmachinepattern3}%
  \BibitemOpen
  \bibfield  {author} {\bibinfo {author} {\bibfnamefont {V.~S.}\ \bibnamefont
  {Denchev}}, \bibinfo {author} {\bibfnamefont {N.}~\bibnamefont {Ding}},
  \bibinfo {author} {\bibfnamefont {S.~V.~N.}\ \bibnamefont {Vishwanathan}}, \
  and\ \bibinfo {author} {\bibfnamefont {H.}~\bibnamefont {Neven}},\ }in\
  \href@noop {} {\emph {\bibinfo {booktitle} {ICML}}}\ (\bibinfo {year}
  {2012})\BibitemShut {NoStop}%
\bibitem [{\citenamefont {Schuld}\ \emph {et~al.}(2016)\citenamefont {Schuld},
  \citenamefont {Sinayskiy},\ and\ \citenamefont
  {Petruccione}}]{qmachinepattern4}%
  \BibitemOpen
  \bibfield  {author} {\bibinfo {author} {\bibfnamefont {M.}~\bibnamefont
  {Schuld}}, \bibinfo {author} {\bibfnamefont {I.}~\bibnamefont {Sinayskiy}}, \
  and\ \bibinfo {author} {\bibfnamefont {F.}~\bibnamefont {Petruccione}},\
  }\href {\doibase 10.1103/PhysRevA.94.022342} {\bibfield  {journal} {\bibinfo
  {journal} {Phys. Rev. A}\ }\textbf {\bibinfo {volume} {94}},\ \bibinfo
  {pages} {022342} (\bibinfo {year} {2016})}\BibitemShut {NoStop}%
\bibitem [{\citenamefont {Schuld}\ \emph {et~al.}(2015)\citenamefont {Schuld},
  \citenamefont {Sinayskiy},\ and\ \citenamefont
  {Petruccione}}]{schuld_qmachine}%
  \BibitemOpen
  \bibfield  {author} {\bibinfo {author} {\bibfnamefont {M.}~\bibnamefont
  {Schuld}}, \bibinfo {author} {\bibfnamefont {I.}~\bibnamefont {Sinayskiy}}, \
  and\ \bibinfo {author} {\bibfnamefont {F.}~\bibnamefont {Petruccione}},\
  }\href {\doibase 10.1080/00107514.2014.964942} {\bibfield  {journal}
  {\bibinfo  {journal} {Contemporary Physics}\ }\textbf {\bibinfo {volume}
  {56}},\ \bibinfo {pages} {172} (\bibinfo {year} {2015})}\BibitemShut
  {NoStop}%
\bibitem [{\citenamefont {Garcia-Escartin}\ and\ \citenamefont
  {Chamorro-Posada}(2013)}]{garcia_2013}%
  \BibitemOpen
  \bibfield  {author} {\bibinfo {author} {\bibfnamefont {J.~C.}\ \bibnamefont
  {Garcia-Escartin}}\ and\ \bibinfo {author} {\bibfnamefont {P.}~\bibnamefont
  {Chamorro-Posada}},\ }\href {\doibase 10.1103/PhysRevA.87.052330} {\bibfield
  {journal} {\bibinfo  {journal} {Phys. Rev. A}\ }\textbf {\bibinfo {volume}
  {87}},\ \bibinfo {pages} {052330} (\bibinfo {year} {2013})}\BibitemShut
  {NoStop}%
\bibitem [{\citenamefont {Le}\ \emph {et~al.}(2011)\citenamefont {Le},
  \citenamefont {Dong},\ and\ \citenamefont {Hirota}}]{frqi}%
  \BibitemOpen
  \bibfield  {author} {\bibinfo {author} {\bibfnamefont {P.~Q.}\ \bibnamefont
  {Le}}, \bibinfo {author} {\bibfnamefont {F.}~\bibnamefont {Dong}}, \ and\
  \bibinfo {author} {\bibfnamefont {K.}~\bibnamefont {Hirota}},\ }\href
  {\doibase 10.1007/s11128-010-0177-y} {\bibfield  {journal} {\bibinfo
  {journal} {Quantum Information Processing}\ }\textbf {\bibinfo {volume}
  {10}},\ \bibinfo {pages} {63} (\bibinfo {year} {2011})}\BibitemShut {NoStop}%
\bibitem [{\citenamefont {Zhang}\ \emph {et~al.}(2013)\citenamefont {Zhang},
  \citenamefont {Lu}, \citenamefont {Gao},\ and\ \citenamefont {Wang}}]{neqr}%
  \BibitemOpen
  \bibfield  {author} {\bibinfo {author} {\bibfnamefont {Y.}~\bibnamefont
  {Zhang}}, \bibinfo {author} {\bibfnamefont {K.}~\bibnamefont {Lu}}, \bibinfo
  {author} {\bibfnamefont {Y.}~\bibnamefont {Gao}}, \ and\ \bibinfo {author}
  {\bibfnamefont {M.}~\bibnamefont {Wang}},\ }\href {\doibase
  10.1007/s11128-013-0567-z} {\bibfield  {journal} {\bibinfo  {journal}
  {Quantum Information Processing}\ }\textbf {\bibinfo {volume} {12}},\
  \bibinfo {pages} {2833} (\bibinfo {year} {2013})}\BibitemShut {NoStop}%
\bibitem [{\citenamefont {Buhrman}\ \emph {et~al.}(2001)\citenamefont
  {Buhrman}, \citenamefont {Cleve}, \citenamefont {Watrous},\ and\
  \citenamefont {de~Wolf}}]{buhrman2001}%
  \BibitemOpen
  \bibfield  {author} {\bibinfo {author} {\bibfnamefont {H.}~\bibnamefont
  {Buhrman}}, \bibinfo {author} {\bibfnamefont {R.}~\bibnamefont {Cleve}},
  \bibinfo {author} {\bibfnamefont {J.}~\bibnamefont {Watrous}}, \ and\
  \bibinfo {author} {\bibfnamefont {R.}~\bibnamefont {de~Wolf}},\ }\href
  {\doibase 10.1103/PhysRevLett.87.167902} {\bibfield  {journal} {\bibinfo
  {journal} {Phys. Rev. Lett.}\ }\textbf {\bibinfo {volume} {87}},\ \bibinfo
  {pages} {167902} (\bibinfo {year} {2001})}\BibitemShut {NoStop}%
\bibitem [{\citenamefont {Gottesman}\ and\ \citenamefont
  {Chuang}(2001)}]{gottesman_2001}%
  \BibitemOpen
  \bibfield  {author} {\bibinfo {author} {\bibfnamefont {D.}~\bibnamefont
  {Gottesman}}\ and\ \bibinfo {author} {\bibfnamefont {I.}~\bibnamefont
  {Chuang}},\ }\href@noop {} {\enquote {\bibinfo {title} {Quantum digital
  signatures},}\ } (\bibinfo {year} {2001}),\ \Eprint
  {http://arxiv.org/abs/quant-ph/0105032} {arXiv:quant-ph/0105032} \BibitemShut
  {NoStop}%
\bibitem [{\citenamefont {Kang}\ \emph {et~al.}(2019)\citenamefont {Kang},
  \citenamefont {Heo}, \citenamefont {Choi}, \citenamefont {Moon},\ and\
  \citenamefont {Han}}]{kang2019}%
  \BibitemOpen
  \bibfield  {author} {\bibinfo {author} {\bibfnamefont {M.-S.}\ \bibnamefont
  {Kang}}, \bibinfo {author} {\bibfnamefont {J.}~\bibnamefont {Heo}}, \bibinfo
  {author} {\bibfnamefont {S.-G.}\ \bibnamefont {Choi}}, \bibinfo {author}
  {\bibfnamefont {S.}~\bibnamefont {Moon}}, \ and\ \bibinfo {author}
  {\bibfnamefont {S.-W.}\ \bibnamefont {Han}},\ }\href {\doibase
  10.1038/s41598-019-42662-4} {\bibfield  {journal} {\bibinfo  {journal}
  {Scientific Reports}\ }\textbf {\bibinfo {volume} {9}},\ \bibinfo {pages}
  {6167} (\bibinfo {year} {2019})}\BibitemShut {NoStop}%
\bibitem [{\citenamefont {{Jozsa}}(1994)}]{Jozsa_1994}%
  \BibitemOpen
  \bibfield  {author} {\bibinfo {author} {\bibfnamefont {R.}~\bibnamefont
  {{Jozsa}}},\ }\href {\doibase 10.1080/09500349414552171} {\bibfield
  {journal} {\bibinfo  {journal} {Journal of Modern Optics}\ }\textbf {\bibinfo
  {volume} {41}},\ \bibinfo {pages} {2315} (\bibinfo {year}
  {1994})}\BibitemShut {NoStop}%
\bibitem [{\citenamefont {Schumacher}(1995)}]{schumacher_1995}%
  \BibitemOpen
  \bibfield  {author} {\bibinfo {author} {\bibfnamefont {B.}~\bibnamefont
  {Schumacher}},\ }\href {\doibase 10.1103/PhysRevA.51.2738} {\bibfield
  {journal} {\bibinfo  {journal} {Phys. Rev. A}\ }\textbf {\bibinfo {volume}
  {51}},\ \bibinfo {pages} {2738} (\bibinfo {year} {1995})}\BibitemShut
  {NoStop}%
\bibitem [{\citenamefont {Cincio}\ \emph {et~al.}(2018)\citenamefont {Cincio},
  \citenamefont {Suba{\c{s}}{\i}}, \citenamefont {Sornborger},\ and\
  \citenamefont {Coles}}]{cincio_2018}%
  \BibitemOpen
  \bibfield  {author} {\bibinfo {author} {\bibfnamefont {L.}~\bibnamefont
  {Cincio}}, \bibinfo {author} {\bibfnamefont {Y.}~\bibnamefont
  {Suba{\c{s}}{\i}}}, \bibinfo {author} {\bibfnamefont {A.~T.}\ \bibnamefont
  {Sornborger}}, \ and\ \bibinfo {author} {\bibfnamefont {P.~J.}\ \bibnamefont
  {Coles}},\ }\href {\doibase 10.1088/1367-2630/aae94a} {\ \textbf {\bibinfo
  {volume} {20}},\ \bibinfo {pages} {113022} (\bibinfo {year}
  {2018})}\BibitemShut {NoStop}%
\bibitem [{\citenamefont {Deng}(2012)}]{deng2012mnist}%
  \BibitemOpen
  \bibfield  {author} {\bibinfo {author} {\bibfnamefont {L.}~\bibnamefont
  {Deng}},\ }\href@noop {} {\bibfield  {journal} {\bibinfo  {journal} {IEEE
  Signal Processing Magazine}\ }\textbf {\bibinfo {volume} {29}},\ \bibinfo
  {pages} {141} (\bibinfo {year} {2012})}\BibitemShut {NoStop}%
\bibitem [{\citenamefont {Wrachtrup}\ and\ \citenamefont
  {Jelezko}(2006)}]{6991}%
  \BibitemOpen
  \bibfield  {author} {\bibinfo {author} {\bibfnamefont {J.}~\bibnamefont
  {Wrachtrup}}\ and\ \bibinfo {author} {\bibfnamefont {F.}~\bibnamefont
  {Jelezko}},\ }\href@noop {} {\bibfield  {journal} {\bibinfo  {journal} {J.
  Phys.: Condens. Matter}\ }\textbf {\bibinfo {volume} {18}},\ \bibinfo {pages}
  {S807} (\bibinfo {year} {2006})}\BibitemShut {NoStop}%
\bibitem [{\citenamefont {Suter}\ and\ \citenamefont
  {Jelezko}(2017)}]{Suter201750}%
  \BibitemOpen
  \bibfield  {author} {\bibinfo {author} {\bibfnamefont {D.}~\bibnamefont
  {Suter}}\ and\ \bibinfo {author} {\bibfnamefont {F.}~\bibnamefont
  {Jelezko}},\ }\href@noop {} {\bibfield  {journal} {\bibinfo  {journal}
  {Progress in Nuclear Magnetic Resonance Spectroscopy}\ }\textbf {\bibinfo
  {volume} {98-99}},\ \bibinfo {pages} {50 } (\bibinfo {year}
  {2017})}\BibitemShut {NoStop}%
\bibitem [{\citenamefont {Zhang}\ \emph {et~al.}(2019)\citenamefont {Zhang},
  \citenamefont {Hegde},\ and\ \citenamefont {Suter}}]{zhang18}%
  \BibitemOpen
  \bibfield  {author} {\bibinfo {author} {\bibfnamefont {J.}~\bibnamefont
  {Zhang}}, \bibinfo {author} {\bibfnamefont {S.~S.}\ \bibnamefont {Hegde}}, \
  and\ \bibinfo {author} {\bibfnamefont {D.}~\bibnamefont {Suter}},\ }\href
  {\doibase 10.1103/PhysRevApplied.12.064047} {\bibfield  {journal} {\bibinfo
  {journal} {Phys. Rev. Applied}\ }\textbf {\bibinfo {volume} {12}},\ \bibinfo
  {pages} {064047} (\bibinfo {year} {2019})}\BibitemShut {NoStop}%
\bibitem [{\citenamefont {Zhang}\ \emph {et~al.}(2020)\citenamefont {Zhang},
  \citenamefont {Hegde},\ and\ \citenamefont {Suter}}]{zhang19}%
  \BibitemOpen
  \bibfield  {author} {\bibinfo {author} {\bibfnamefont {J.}~\bibnamefont
  {Zhang}}, \bibinfo {author} {\bibfnamefont {S.~S.}\ \bibnamefont {Hegde}}, \
  and\ \bibinfo {author} {\bibfnamefont {D.}~\bibnamefont {Suter}},\ }\href
  {\doibase 10.1103/PhysRevLett.125.030501} {\bibfield  {journal} {\bibinfo
  {journal} {Phys. Rev. Lett.}\ }\textbf {\bibinfo {volume} {125}},\ \bibinfo
  {pages} {030501} (\bibinfo {year} {2020})}\BibitemShut {NoStop}%
\bibitem [{\citenamefont {Hegde}\ \emph {et~al.}(2020)\citenamefont {Hegde},
  \citenamefont {Zhang},\ and\ \citenamefont {Suter}}]{swathi19}%
  \BibitemOpen
  \bibfield  {author} {\bibinfo {author} {\bibfnamefont {S.~S.}\ \bibnamefont
  {Hegde}}, \bibinfo {author} {\bibfnamefont {J.}~\bibnamefont {Zhang}}, \ and\
  \bibinfo {author} {\bibfnamefont {D.}~\bibnamefont {Suter}},\ }\href
  {\doibase 10.1103/PhysRevLett.124.220501} {\bibfield  {journal} {\bibinfo
  {journal} {Phys. Rev. Lett.}\ }\textbf {\bibinfo {volume} {124}},\ \bibinfo
  {pages} {220501} (\bibinfo {year} {2020})}\BibitemShut {NoStop}%
\bibitem [{\citenamefont {Zhang}\ \emph {et~al.}(2021)\citenamefont {Zhang},
  \citenamefont {Hegde},\ and\ \citenamefont {Suter}}]{arXiv:2108.13738}%
  \BibitemOpen
  \bibfield  {author} {\bibinfo {author} {\bibfnamefont {J.}~\bibnamefont
  {Zhang}}, \bibinfo {author} {\bibfnamefont {S.~S.}\ \bibnamefont {Hegde}}, \
  and\ \bibinfo {author} {\bibfnamefont {D.}~\bibnamefont {Suter}},\
  }\href@noop {} {\bibfield  {journal} {\bibinfo  {journal} {arXiv:2108.13738
  [quant-ph]}\ } (\bibinfo {year} {2021})}\BibitemShut {NoStop}%
\bibitem [{\citenamefont {M{\"u}ller}\ \emph {et~al.}(2012)\citenamefont
  {M{\"u}ller}, \citenamefont {Reinhardt},\ and\ \citenamefont
  {Strickland}}]{muller2012neural}%
  \BibitemOpen
  \bibfield  {author} {\bibinfo {author} {\bibfnamefont {B.}~\bibnamefont
  {M{\"u}ller}}, \bibinfo {author} {\bibfnamefont {J.}~\bibnamefont
  {Reinhardt}}, \ and\ \bibinfo {author} {\bibfnamefont {M.}~\bibnamefont
  {Strickland}},\ }\href {https://books.google.it/books?id=on0QBwAAQBAJ} {\emph
  {\bibinfo {title} {Neural Networks: An Introduction}}},\ Physics of Neural
  Networks\ (\bibinfo  {publisher} {Springer Berlin Heidelberg},\ \bibinfo
  {year} {2012})\BibitemShut {NoStop}%
\end{thebibliography}%
\bibliographystyle{apsrev4-1}
\end{document}